\documentclass[a4paper, 11pt, amsmath, amssymb]{article}
\pdfoutput=1
\usepackage{jheppub}

\usepackage{amsmath}
\usepackage{amssymb}
\usepackage{graphicx}
\usepackage{dcolumn}
\usepackage{bm}
\usepackage{hyperref}
\usepackage{epsfig}
\usepackage{epstopdf}
\usepackage{mathtools}

\newcommand{\bc}{\begin{center}}
\newcommand{\ec}{\end{center}} 
\newcommand{\be}{\begin{equation}}
\newcommand{\ee}{\end{equation}}

\newcommand{\bk}{\mathbf{k}}

\newcommand{\bq}{\mathbf{q}}

\newcommand{\bz}{\mathbf{z}}

\newcommand{\vz}{{\bf z}}

\newcommand{\lp}{\left(}
\newcommand{\rp}{\right)}

\title{Bridging Soft-Hard Transport Properties of Quark-Gluon Plasmas with CUJET3.0}

\author[a]{Jiechen Xu}
\author[b,c]{Jinfeng Liao}
\author[a]{Miklos Gyulassy}
\affiliation[a]{Department of Physics, Columbia University,\\538 West 120th Street, New York, NY 10027, USA}
\affiliation[b]{Physics Department and Center for Exploration of Energy and Matter, Indiana University,\\2401 North Milo B. Sampson Lane, Bloomington, IN 47408, USA}
\affiliation[c]{RIKEN BNL Research Center,\\Building 510A, Brookhaven National Laboratory, Upton, NY 11973, USA}

\emailAdd{xjc@phys.columbia.edu}
\emailAdd{liaoji@indiana.edu}
\emailAdd{gyulassy@phys.columbia.edu}

\date{\today}

\abstract{A new model (CUJET3.0) of jet quenching in nuclear collisions
coupled to bulk data constrained
  (VISH2+1D) viscous hydrodynamic backgrounds is constructed by
  generalizing the 
perturbative QCD based (CUJET2.0) model to 
  include two complementary  non-perturbative chromodynamical features of the QCD
  confinement cross-over phase transition near $T_c\approx 160$ MeV: (1) the suppression of quark and
  gluon chromo-electric-charged (cec) degrees of freedom and (2) the emergence of
  chromo-magnetic-monopole (cmm) degrees of freedom.  Such a 
 semi Quark Gluon Monopole Plasma (sQGMP) microscopic scenario is tested by
  comparing predictions of the leading hadron nuclear modification
  factors, $R^h_{AA}(p_T>10{\rm GeV/c},\sqrt{s})$, and their azimuthal elliptic
  asymmetry $v^h_2(p_T>10{\rm GeV/c},\sqrt{s})$ with available data on $h=\pi,D,B$ jet fragments from nuclear
  collisions at RHIC($\sqrt{s}=0.2$ ATeV) and LHC($\sqrt{s}$=2.76
  ATeV).  
The cmm degrees of freedom in the sQGMP model near $T_c$ are shown to
  solve robustly the long standing $R_{AA}$ vs $v_2$ puzzle by
  predicting a maximum of the jet quenching parameter field
  $\hat{q}(E,T)/T^3$ near $T_c$. The robustness of CUJET3.0 model to a number of  theoretical uncertainties  is critically tested. 
Moreover the  consistency 
of jet quenching with observed bulk perfect fluidity is  
demonstrated by 
extrapolating the sQGMP $\hat{q}$
  down to thermal energy $E\sim 3 T$ scales and showing that the sQGMP
  shear viscosity to entropy density ratio $\eta/s \approx T^3/\hat{q}$ 
  falls  close to the unitarity bound, $1/4\pi$, in the range $(1-2)T_c$. 
Detailed comparisons of the CUJET2.0 and CUJET3.0 models reveal   
  the fact that remarkably different $\hat{q}(T)$ dependence could be consistent with the same $R_{AA}$ data and could only be distinguished by anisotropy observables. These findings demonstrate clearly the inadequacy of focusing on the jet path averaged  quantity $\left\langle\hat{q}\right\rangle$ as the only relevant medium property to characterize jet quenching, and  point to  the crucial roles of other essential factors beyond just the $\left\langle\hat{q}\right\rangle $, such as the chromo electric and magnetic composition of the plasma, the screening masses  and the running couplings at multiple scales which all strongly influence jet energy loss.  
}


\begin{document}

\maketitle  

\flushbottom

\section{Introduction}
\label{sec:intro}

In ultrarelativistic heavy-ion collisions at the BNL Relativistic
Heavy Ion Collider (RHIC)
\cite{Adare:2008qa,Adare:2010sp,Adare:2012wg,Abelev:2009wx} and the
CERN Large Hadron Collider (LHC)
\cite{Abelev:2012di,Abelev:2012hxa,ATLAS:2011ah,CMS:2012aa,Chatrchyan:2012xq,ALICE:2012ab,Abelev:2014ipa,CMS:2012vxa},
strongly-coupled quark-gluon plasmas (sQGP) are created with deconfined color
degrees of freedom under extremely hot conditions at
sufficiently high beam energies
\cite{Gyulassy:2004zy,Shuryak:2004cy,Muller:2012zq}. This new phase of
Quantum Chromodynamics (QCD) matter provides a unique environment  that contains an abundance of information about both
the perturbative and nonperturbative aspects of QCD. In such
collisions, large transverse momentum ($p_T$) partons are produced in
the pre-thermal stage and subsequently 
traverse the entire
medium. They scatter strongly with the dynamical degrees of freedom in the thermal
medium, and undergo both radiative and collisional energy loss
\cite{Baier:2000mf,Gyulassy:2003mc,Kovner:2003zj,Jacobs:2004qv,Armesto:2011ht,CasalderreySolana:2011us,Burke:2013yra}. This
``jet quenching'' effect leads to suppressed yields of high $p_T$
hadrons in nuclei-nuclei (A+A) collisions compared with the yields in
scaled proton-proton (p+p) collisions at the same center of mass
energy. It has been discovered experimentally at both RHIC and LHC and serves as one
of the key evidences for the formation of sQGP in heavy-ion
collisions \cite{Gyulassy:2004zy,Shuryak:2004cy,Muller:2012zq}. 
Jet quenching observables probe the color composition and 
chromo field fluctuations in high density QCD
matter that can provide insight into the novel dynamical mechanisms responsible
for the surprising ``perfect fluidity'' property discovered through
the systematics of 
bulk collective low $p_T$ azimuthal harmonics at RHIC and LHC.

Among a variety of jet quenching observables, two primary informative  ones (at single hadron level) 
\footnote{In the present study we focus on single hadron observables, and the current CUJET implementation considers energy loss of single partons that are subsequently mapped to hadrons. This should be distinguished from studies that focus on full jet evolution for describing reconstructed jet observables. The extension of CUJET framework toward full jet study would be a future project.}
are the nuclear modification factor $R_{AA}$ which is defined as the ratio of the A+A spectrum to the  p+p spectrum scaled by the number of binary collisions,
\begin{eqnarray}
R_{AA}(p_T,y;\sqrt{s},b) = \frac{{dN_{AA}}/{dyp_Tdp_T}}{N_{\rm bin}\;{dN_{pp}}/{dyp_Tdp_T}}\;,
\label{RAAdef}
\end{eqnarray}
and the azimuthal elliptical harmonics $v_2$ which is the second Fourier coefficient in
\begin{eqnarray}
\frac{ dN}{dyp_T dp_T d\phi}(p_T,\phi,y;\sqrt{s},b)= 
\frac{1}{2\pi} \frac{dN}{dyp_Tdp_T}\times\left[   1+2\sum_{n=1}^{\infty} v_n\cos \lp n ( \phi - \Psi_n) \rp \right].
\label{v2def}
\end{eqnarray}
They characterize the overall magnitude and the angular dependence of
jet quenching in heavy-ion collisions, imposing stringent constraints
on the mechanism of parton-medium interactions in jet energy loss
models. However, most perturbative QCD (pQCD) based jet
quenching frameworks have been found to fail to 
describe {\em simultaneously}
~\footnote{It should be pointed out that recent studies on full jet observables (see e.g. [23]) provide quite reasonable descriptions of reconstructed jet $R_{AA}$ and $v_{2}$ measurements at the LHC. The different situation for single hadron versus full jet results may be due to different sensitivities of different types of observables, or may also plausibly hint at systematic uncertainty in various studies of single hadron observables due to limited understanding of hadronization processes.} 
the high $p_T$
light hadrons' and open heavy flavors' $R_{AA}$ and $v_2$ data at RHIC
and LHC \cite{Molnar:2001ux,Noronha:2010zc, Molnar:2013eqa, Betz:2013caa,Betz:2014cza}.

One of the main problems with the conventional perturbative QCD
picture of  the (HTL) quasiparticle degrees of freedom in 
Quark Gluon Plasmas 
is that  leading
order (LO) pQCD estimates of the QGP's shear viscosity to entropy
density ratio
\cite{Hosoya:1983xm,Danielewicz:1984ww,Thoma:1991em,Hirano:2005wx},
\begin{eqnarray}
\frac{\eta}{s}\approx \frac{0.071}{\alpha_s^2 \log (1/\alpha_s)}\;,
\label{pQCDetas}
\end{eqnarray}
predict that this ratio should be of the order unity, which is an order of magnitude larger than the
unitarity $\eta/s=1/4\pi$ lower bound\cite{Danielewicz:1984ww, Kovtun:2004de}
that was found to  be required to explain
the ``perfect fluidity''property of QGP produced in high energy nuclear collisions\cite{Lee:2005gw,Gyulassy:2004zy,Shuryak:2004cy,Luzum:2008cw, Noronha:2010zc, Gale:2012rq}. 
While the  factor of $\sim 5$ quenching of hard leading hadrons
observed in central collisions with 
$R_{AA}\sim 0.2$, was well predicted \cite{Gyulassy:2003mc} even with perturbative 
QCD jet medium coupling,
the collective bulk azimuthal flow moments observed at RHIC and LHC
appear  to require much stronger 
interactions such as those assumed, e.g., in AdS/CFT black hole modeling
of the sQGP to account for perfect fluidity \cite{ Kovtun:2004de, Noronha:2010zc}
 The long-standing ``jet'' $R_{AA}(p_T>5~{\rm GeV})$ vs ``bulk'' $v_2(p_T<2~{\rm GeV})$~\cite{Molnar:2001ux, Noronha:2010zc} as well as the ``jet'' azimuthal $v_2(p_T>10~{\rm GeV})$ 
puzzles \cite{Shuryak:2001me, Betz:2013caa,Betz:2014cza,Xu:2014ica}
 continue to raise critical questions about our understanding of 
the dynamics and composition of the
QGP medium produced in high energy nuclear collisions
and especially the consistency of information derived from 
high $p_T>10$ GeV jet quenching observables and low $p_T<2$ GeV bulk collective flow observables at RHIC
and LHC. 
While the soft hadrons' $v_2$ (originating  from collective flow of the nearly perfect bulk fluid) and the hard hadrons' $v_2$ (due to geometric anisotropy of in-medium path lengths) are phenomenolgocially via different mechanism, they are both generated by {\it the same underlying QGP medium} for which a consistent microscopic model should account for  both the soft and the hard transport properties. Many critical questions need to be addressed here.   
 How do the effective degrees of freedom of nonperturbative QCD origin blend interpolate between the confined Hadron Resonance Gas (HRG) world
at low energy energy density into
an asymptotically free quark gluon plasma at extreme densities? Are there effective quasi-particles in the highly non-perturbative non-conformal temperature
range near the deconfinement transition temperature
$T_c\sim 160$ MeV? How do those ``emergent'' degrees of freedom near $T_c$ 
affect high $p_T >10$ GeV jet flavor observables? Can an effective quasi-particle description be constructed that is consistent with lattice QCD thermodynamic data and simultaneously  could reconcile the apparent  inconsistency between the
bulk ``perfect fluid'', minimally viscous hydrodynamics and the success (modulo $v_2$) of pQCD based 
jet quenching phenomenology?
Can the combined set of soft plus hard observables be used to  
elucidate the mechanism of color confinement?
The goal of this paper is to demonstrate in detail the existence
 of at least one possible model that involves partially suppressed cec together
with emergent cmm effective degrees of freedom (the  
semi-Quark-Gluon-Monopole-Plasma model) that via the CUJET3.0 numerical 
framework allows us to constrain the $\hat{q}(E,T)$ and $\eta/s(T)$ fields
over a much wider range than previously possible.   

We begin by recalling a possible mechanism of color confinement based
on a ``dual superconductor'' picture as proposed by Mandelstam, Nambu,
Polyakov, 't Hooft
\cite{Mandelstam:1974pi,Nambu:1974zg,Polyakov:1976fu,'tHooft:1981ht}
--- It is understood that in type II superconductors the condensate of
cooper pairs generates a ``Meissner Effect'' that repels magnetic field
lines and squeezes monopole pair fields into flux tubes. In models 
possessing electric-magnetic duality, e.g. the Seiberg-Witten solution of the
$\mathcal{N}=2$ supersymmetric gauge theory
\cite{Seiberg:1994rs,Seiberg:1994aj}, a ``dual superconductor''
phase does emerge in the strong coupling regime. Stable magnetic monopoles condensates can be generated leading to ``Dual Meissner
Effect'' that forces the chromo-electric fields sourced by cec pairs
to form  flux tubes
that gives rise to linearly rising potential and confines quark-antiquark pairs. 

Partially motivated by the ``dual superconductivity'' picture of color
confinement, a magnetic scenario for the near $T_c$ QCD plasma was
proposed in
\cite{Liao:2006ry,Liao:2007mj,Liao:2008vj,Liao:2008jg,Liao:2008dk,Liao:2012tw}. This scenario 
emphasizes the change in chromo degrees of freedom with the gauge coupling, and recasts the QCD phase diagram into
electrically and magnetically dominated regimes. For example, focusing on  increasing temperature (at zero baryonic density ),  a particular 
temperature $T_{E=M}$ may be identified as  a new phase boundary   where the coupling strength of electric (E) and
magnetic (M) interactions are equal and satisfy the Dirac quantization
condition \cite{Dirac:1931kp}. Interestingly,
a novel post-confinement non-conformal non-perturbative 
region emerges at $T_c<T<T_{E=M}\sim 1.4T_c$ where
chromo magnetic monopoles (cmm) are the lightest degrees of freedom (DOFs) in the system
while chromo electrically charged (cec) quasi-particles 
are strongly correlated and
connected by flux tubes. Phenomenologically it has been qualitatively demonstrated that with the inclusion of such emergent monopoles
near $T_c$,  the scatterings in both the soft and hard sectors are strongly enhanced and thus help explaining the small $\eta/s$ of the bulk sQGP as well as  leading to  significant $v_2$ of high $p_T$ hadrons~\cite{Liao:2006ry,Ratti:2008jz,Liao:2008dk,Zhang:2012ie,Zhang:2012ha,Zhang:2013oca}.

However, a quantitative and phenomenologically robust modeling
framework for the QCD matter in the near $T_c$ regime has not been
built previously. Such a framework has to couple the hard probes with the
state-of-the-art bulk constrained viscous hydrodynamic
$(T(x,t),u^\mu(x,t))$ fields. It also needs to have a sophisticated
implementation of the microscopic nonperturbative physics for both the
cec and cmm degrees of freedom.  For that, we need to constrain how
the E and M quasi-particles DOF vary with temperature in a way that is
consistent with available lattice QCD data
\cite{Bazavov:2014pvz,Borsanyi:2013bia,Nakamura:2003pu,Bazavov:2009zn,Borsanyi:2010bp,Bazavov:2012jq,Borsanyi:2011sw}
on thermodynamic pressure, entropy density, energy density as well as
the Polyakov loop and quark susceptibilities.  Specifically for the
nonperturbative dynamics of the cec near $T_c$, the ``semi-QGP'' model
\cite{Pisarski:2006hz,Hidaka:2008dr,Hidaka:2009hs,Hidaka:2009ma,Lin:2013qu,Lin:2013efa}
was previously proposed to build in the color suppression effect in
the deconfinement transition region. In that model the Polyakov loop
is the relevant parameter that controls confinement/deconfinement of
color electric charge DOFs. When temperature drops towards $T_c$, the
excitation of cec quarks and gluons are quenched by powers of the
loop, resulting in a number of novel phenomenological effects. For
example it was found in the semi-QGP that there is a mild enhancement
and a strong suppression for the production of thermal dileptons and
hard photons \cite{Gale:2014dfa,Hidaka:2015ima}.

The above considerations of the nonperturbative medium near $T_c$ have motivated
us to propose and study a nonperturbative
semi-Quark-Gluon-Monopole-Plasma (sQGMP) model for the ``perfect chromo fluid''   
near the deconfinement
transition range. In \cite{Xu:2014tda}, we summarized the results
of our CUJET3.0 extension of a pQCD based
energy loss model CUJET2.0 discussed\cite{Xu:2014ica,Xu:2014wua} which
integrates local parton energy loss  over (2+1)D viscous hydrodynamic 
flows and and models jet medium interactions via the sQGMP 
quasi-particle model picture of the chromo structure of 
the fluid that include specific non-pertubative features 
related to confinement in
the vicinity of $T_c$.  In CUJET3.0 all thermodynamic
properties are constrained by lattice QCD data. At very high temperature $T\gg T_c$ the
model by construction would smoothly reduce to CUJET2.0 corresponding to a perturbative Hard Thermal Loop
(HTL) picture of the QGP. As T approaches $T_c$, the chromo-electric charge (cec) degrees of freedom  are
suppressed as powers of the Polyakov loop as in the semi-QGP
framework while chromo-magnetic monopoles emerge to account
for the total lattice QCD pressure or the total entropy density. 
Jet quenching observables of high $p_T$
light hadrons' and open heavy flavors' $R_{AA}$ and $v_2$ at RHIC and
LHC are computed using CUJET3.0 and are shown to be in agreement with all data
simultaneously withing present experimental uncertainties. 
This model therefore provides a semi-quantitative  bridge  
between local equilibrium bulk ``perfect fluidity'' and 
high energy far from equilibrium jet transport phenomena.

Following up the  condensed summary of the CUJET3.0 results reported in 
\cite{Xu:2014tda}, we will present in this paper the theoretical
details of the CUJET3.0 framework and concentrate on the robustness
and consistency of its components as well on its phenomenological applications 
that successfully account for a large set of leading-hadron-suppression data. We address a number of key questions concerning the
theoretical robustness of the underlying sQGMP microscopic scenario 
and report new results that help in estimating 
theoretical uncertainties in our present 
understanding of jet energy loss phenomena via
the sQGMP model. Such questions include: (1) Can an effective quasi-particle
chromodynamic model be formulated with sufficient ab-initio lattice QCD and experimental data constraints to explain simultaneously 
both long and short wavelength observables in high energy A+A reactions? 
(2) How do theoretical uncertainties 
on Quark  and Glue cec 
quasi-particle number densities $\rho_Q(T)$ and $\rho_G(T)$ 
and chromo Magnetic monopole cmm  density $\rho_M(T)$ of the sQGMP
propagate to the observables? 
(3) How do uncertainties in the chromo electric and magnetic
 screening masses ($\mu_{E,M}(T\sim T_c)$) near $T_c$
 effect the observables? (4) Can we 
constrain in the sQGMP model the temperature dependence of the  jet quenching
parameter ($\hat{q}(E,T)$) well enough to predict the shear viscosity to entropy
ratio ($\eta/s(T)$) by extrapolating down to $E\sim 3 T$ thermal scales? (5) 
How does the   effective path length dependence of light and
heavy quark energy loss depend on the detailed cec and cmc 
composition of sQGMP matter, and how do these compare to perturbative 
QCD HTL path length dependences?  

The organization of this paper is as follows: in section
\ref{sec:CUJET3.0}, we briefly review the CUJET2.0 pQCD energy loss
kernel -- the dynamical Djordjevic-Gyulassy-Levai-Vitev (DGLV) opacity
series with multi-scale running strong couplings, the semi-QGP model,
the magnetic scenario of the near $T_c$ QCD matter, as well as how the
perturbative and non-perturbative ingredients are integrated in the
CUJET3.0 framework. In section \ref{sec:L}, we discuss two different
scenarios for the deconfinement of color electric charge (cec)
degrees of freedom  near $T_c$, and explore
how the  jet quenching observables computed from three schemes  with
different $c_m$ would vary in CUJET3.0 framework.  We then investigate the influence of
color composition of the fluid on key transport properties, the $\hat{q}/T^3$ and $\eta/s$, and propose 
schematic
strategies for constraining $\eta/s(T)$ from high $p_T$ jet quenching
data and vice versa. 
In section \ref{sec:PL}, we present a systematic study of the
path length dependence of parton energy loss in the sQGMP, including
both the light quark and the heavy flavor. We summarize and propose
possible future studies in section \ref{sec:conclusion}. 
We discuss in appendix \ref{appx:flow} the effects
of relativistic hydrodynamic flows fields on anisotropic jet suppressions in
the sQGMP, and study in appendix \ref{appx:hybrid} which single 
ingredient is most significant to generate the simultaneous satisfactory 
description of high $p_T$ $R_{AA}$ and
$v_{2}$ within CUJET3.0 framework.

\section{The CUJET3.0 framework}
\label{sec:CUJET3.0}

The CUJET3.0 jet quenching framework 
generalizes the CUJET2.0 perturbative
non-Abelian gluon bremsstrahlung kernel of the DGLV opacity expansion
theory (section \ref{sec:DGLV}) of 
elastic radiative jet energy
loss to incorporate lattice QCD
constraints on the nonperturbative semi-QGP suppression
of color DOFs (section \ref{sec:semi-QGP}), and 
emergent nonperturbative chromo-magnetic monopole DOF near 
the crossover QCD temperature regime $T\sim T_c$ regime (section
\ref{sec:monopoles}). CUJET3.0 incorporates lattice QCD data 
on the QCD pressure equation of state (EOS) $P(T)$, nonperturbative
chromo electric and magnetic  screening masses, $m_E(T),m_M(T)$,
,and  the Polyakov loop $L(T)$ (section \ref{sec:pQCD+sQGMP}). In the following
subsections, we present comprehensive discussions of the details
of these ingredients and study robustness to 
systematic theoretical uncertainties
\ref{sec:NL-FL}  
associated with ``slow'' quark liberation as suggested by $L(T)$ 
data compared to ``fast'' quark liberation as 
suggested by the light quark susceptibility, $\chi_u(T)$, data.

\subsection{Perturbative QCD jet quenching: DGLV opacity expansion}
\label{sec:DGLV}

In the pQCD paradigm, radiative processes dominate the jet-medium interactions for a highly energetic parton. Based on different physical assumptions and approximations about the virtuality and branching of the hard parton, the nature of the medium through which the energetic parton propagates, and the kinematics for the interaction between
medium and projectile parton \cite{Armesto:2011ht}, energy loss models like AMY \cite{Arnold:2002ja,Arnold:2002zm,Arnold:2003zc}, ASW \cite{Wiedemann:2000za,Salgado:2003gb,Armesto:2003jh,Armesto:2004pt,Armesto:2005iq}, BDMPS-Z \cite{Baier:1996kr,Zakharov:1997uu,Baier:1998kq}, Higher Twist \cite{Guo:2000nz,Wang:2001ifa,Majumder:2007ae}, (D)GLV \cite{Gyulassy:1993hr,Gyulassy:1999zd,Gyulassy:2000er,Gyulassy:2002yv,Vitev:2002pf,Djordjevic:2003zk,Gyulassy:2003mc,WHDG,Buzzatti:2011vt} have been built and developed. In the dynamical DGLV opacity expansion theory \cite{Gyulassy:1993hr,Gyulassy:2000er,Djordjevic:2003zk,Djordjevic:2008iz}, the inclusive single gluon emission spectrum in the $n=1$ opacity series with multi-scale running strong couplings \cite{Buzzatti:2012dy} reads \cite{Xu:2014ica,Xu:2014wua,Xu:2014tda}:
\begin{eqnarray}
 x_E\frac{dN_g^{n=1}}{dx_E}&=& \frac{18 C_R}{\pi^2} \frac{4+N_f}{16+9N_f} \int{d\tau}\; \rho(\bz) \Gamma(\vz)\;\int{d^2k_{\perp}} \;\nonumber\\
& \times&\;\alpha_s( \frac{\bk_{\perp}^2}{x_+ (1-x_+)} )\;\int{d^2q}\frac{\alpha_s^2(\bq_{\perp}^2)}{\mu^2(\vz)}\frac{f_E^2\mu^2(\vz)}{\bq_{\perp}^2(\bq_{\perp}^2+f_E^2\mu^2(\vz))}\nonumber\\
& \times&\;\frac{-2(\bk_{\perp}-\bq_{\perp})}{(\bk_{\perp}-\bq_{\perp})^2+\chi^2(\bz)} \left[ \frac{\bk_{\perp}}{\bk_{\perp}^2+\chi^2(\bz)} - \frac{(\bk_{\perp}-\bq_{\perp})}{(\bk_{\perp}-\bq_{\perp})^2+\chi^2(\bz)} \right]\nonumber\\
& \times&\;{\left[1-\cos\lp\frac{(\bk_{\perp}-\bq_{\perp})^2+\chi^2(\bz)}{2 x_+ E } \tau\rp\right]}\left(\frac{x_E}{x_+}\right)\left| \frac{dx_+}{dx_E} \right| \;\;.
\label{rcDGLV}
\end{eqnarray} 
$C_R=4/3$ or $3$ is the quadratic Casimir of the quark or gluon; the transverse coordinate of the hard parton is denoted by $ \vz=\lp x_0+\tau\cos\phi,y_0+\tau\sin\phi; \tau\rp$; $\rho(\vz)$ and $T(\vz)$ is the particle number density and the medium temperature in the local rest frame. In the presence of hydrodynamical 4-velocity fields $u_f^{\mu}(\vz) $, boosting back to the lab frame, one should take into account a relativistic correction $\Gamma(\bz)=u^{\mu}_fn_{\mu}$ \cite{Liu:2006he,Baier:2006pt}, where the flow 4-velocity $u^{\mu}_f=\gamma_f(1,\vec{\beta}_f)$ and null hard parton 4-velocity 
$n^\mu=(1,\vec{\beta}_{j})$. E is the energy of the hard parton in the lab frame, $\bf{k}_{\perp}$ ($|{\bf k}_{\perp}|\leq x_EE\cdot\Gamma(\bz)$) and $\bf{q}_{\perp}$ ($|{\bf q}_{\perp}|\leq 6T(\bz)E\cdot\Gamma(\bz)$) is the local transverse momentum of the radiated gluon and the local transverse momentum transfer respectively. The Debye screening mass $\mu(\vz)$ is determined from solving the self-consistent equation 
\begin{eqnarray}
\mu^2(\vz) = \sqrt{4\pi\alpha_s(\mu^2(\vz))}T(\vz)\sqrt{1+N_f/6}
\label{DebyeMass}
\end{eqnarray}
as in \cite{Peshier:2006ah}; $\chi^2(\vz)=M^2 x_+^2+m_g^2(\vz)(1-x_+)$ regulates the soft collinear divergences in the color antennae and controls the Landau-Pomeranchuk-Migdal (LPM) phase, the gluon plasmon mass $ m_g(\vz)=f_E\mu(\vz) / \sqrt{2} $, and $f_E$ is the HTL chromo-electric deformation parameter, with $f_E=1$ by default \cite{Djordjevic:2008iz}. The gluon fractional energy $x_E$ and fractional plus-momentum $x_+$ are connected by $x_+(x_E)=x_E[1+\sqrt{1-(k_\perp/x_EE)^2}]/2$. 

In the CUJET2.0 model, Zakharov's 1-loop pQCD running scheme is used \cite{Zakharov:2008kt,Zakharov:2007pj,Xu:2014ica}. This running is cutoff in the infrared when the strong coupling strength reaches a maximum value $\alpha_{max}$ for $Q\le Q_{min}$: 
\be
\alpha_s(Q^2) = \begin{cases}
\alpha_{max} & \mbox{if } Q \le Q_{min}\;, \\
\dfrac{4\pi}{9\log(Q^2/\Lambda_{QCD}^2)}  & \mbox{if } Q > Q_{min}\;.
\end{cases}
\label{AlphaRunMax}
\ee
where the minimum running scale $Q_{min}$ is fixed by $\alpha_{max}$ via $Q_{min}=\Lambda_{QCD}\exp\left\lbrace {2\pi}/{9\alpha_{max}}\right\rbrace\;$, with $\Lambda_{QCD}=200$ MeV. Note that the one power of $\alpha_s$ originating from gluon radiation vertex runs with the virtuality $k_\perp^2/[x_+(1-x_+)]$, while the two powers of $\alpha_s$ originating from parton-medium scatterings run with the transverse momentum exchange $q_\perp^2$.

For collisional processes, TG elastic energy loss formula \cite{Thoma:1990fm,Bjorken:1982tu,Peigne:2008nd} with Peign\'{e}-Peshier running coupling prescription \cite{Peigne:2008nd} is used \cite{Xu:2014ica,Xu:2014wua}:
\be\label{rcCUJETElastic}
\begin{split}
\frac{dE(\bz)}{d\tau}= & - C_R \pi \left[ \alpha(\mu(\bz))\alpha( 6 E(\bz)\Gamma(\bz) T(\bz)) \right] T(\bz)^2 \lp 1+\frac{N_f}{6} \rp \\
& \times \log \left[ \frac{6T(\bz)\sqrt{E(\bz)^2\Gamma(\bz)^2-M^2}}{\lp E(\bz)\Gamma(\bz)-\sqrt{E(\bz)^2\Gamma(\bz)^2-M^2}+6T(\bz)\rp\mu(\bz)} \right],
\end{split}
\ee
and
\be
\bar{N_c} = \int_{0}^{\tau_{max}} d\tau \left[ \frac{\alpha(\mu(\bz)) \alpha(6 E(\bz)\Gamma(\bz) T(\bz))}{\mu(\bz)^2} \right] \left[ \frac{\Gamma(\bz)}{\gamma_f} \frac{18 \zeta(3)}{\pi} (4+N_f) T(\bz)^3 \right]\;\;.
\label{rcNumOfColl}
\ee
respectively. Note that the calculation of the energy loss and the average number of collisions $\bar{N_c}$ involves recursively solving the $E(\bz)$ integral equation.

In Eq.~\eqref{rcDGLV}\eqref{rcCUJETElastic}\eqref{rcNumOfColl}, the bulk evolution profiles $(T(\vz),\rho(\vz),u_f^{\mu}(\vz) )$ are generated from the VISH2+1 code \cite{Song:2008si,Shen:2010uy,Renk:2010qx} with MC-Glauber initial condition, $\tau_0=0.6$ fm/c, s95p-PCE Equation of State (EOS), $\eta/s=0.08$, and Cooper-Frye freeze-out temperature 120 MeV \cite{Song:2010mg,Majumder:2011uk,Qiu:2011hf,Shen:2011eg,Shen:2012vn,Shen:2014vra}. Event-averaged smooth profiles are embedded, and the path integrations $\int d\tau$ for jets initially produced at transverse coordinates $({\bf x}_0,\phi)$ are cutoff at dynamical $T(\vz({\bf x}_0,\phi,\tau))|_{\tau_{max}}\equiv T_{cut} =160$ MeV hypersurfaces \cite{Xu:2014ica}.

\begin{figure*}[!t]
\bc
\includegraphics[width=0.49\textwidth]{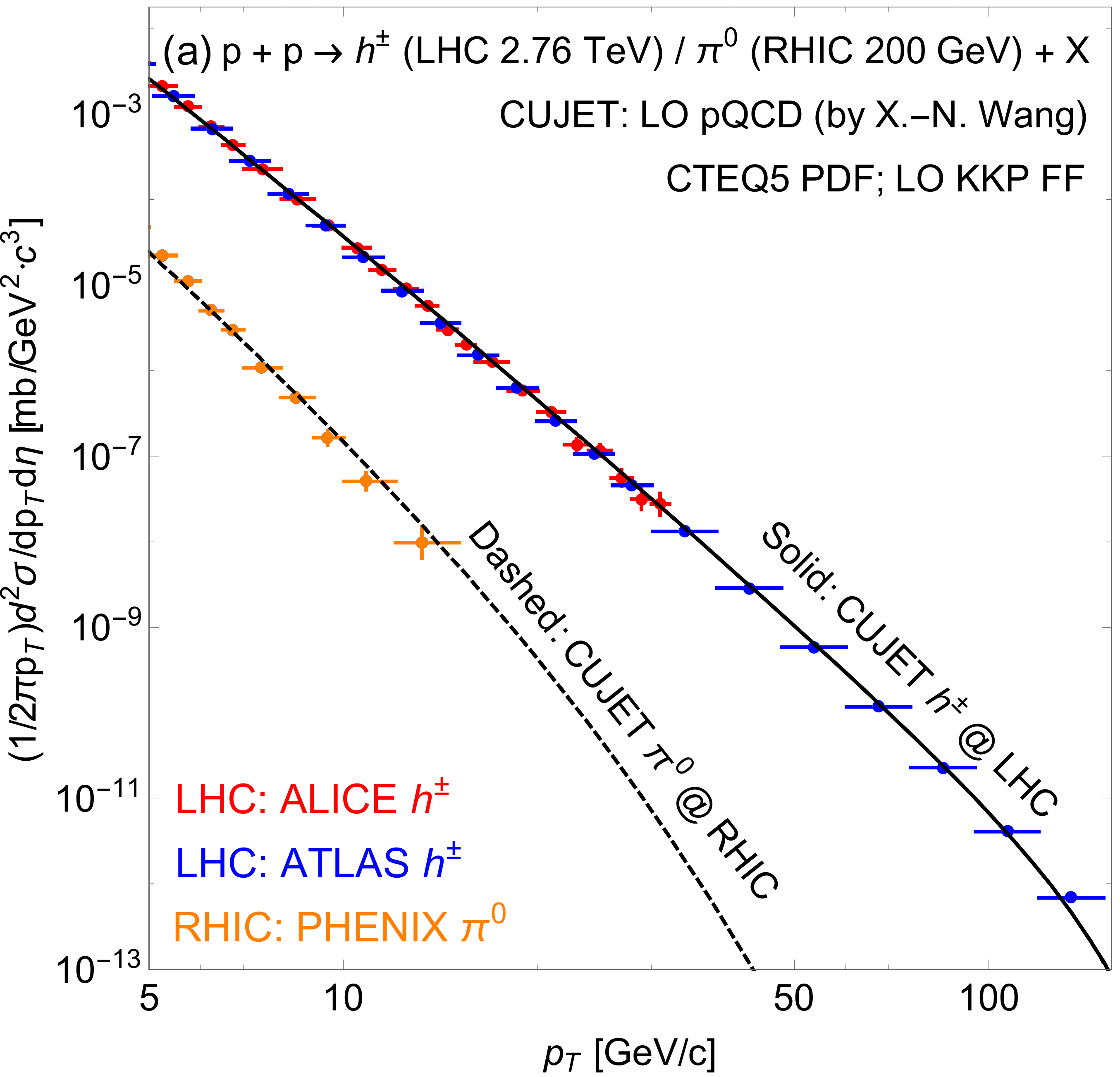}
\includegraphics[width=0.49\textwidth]{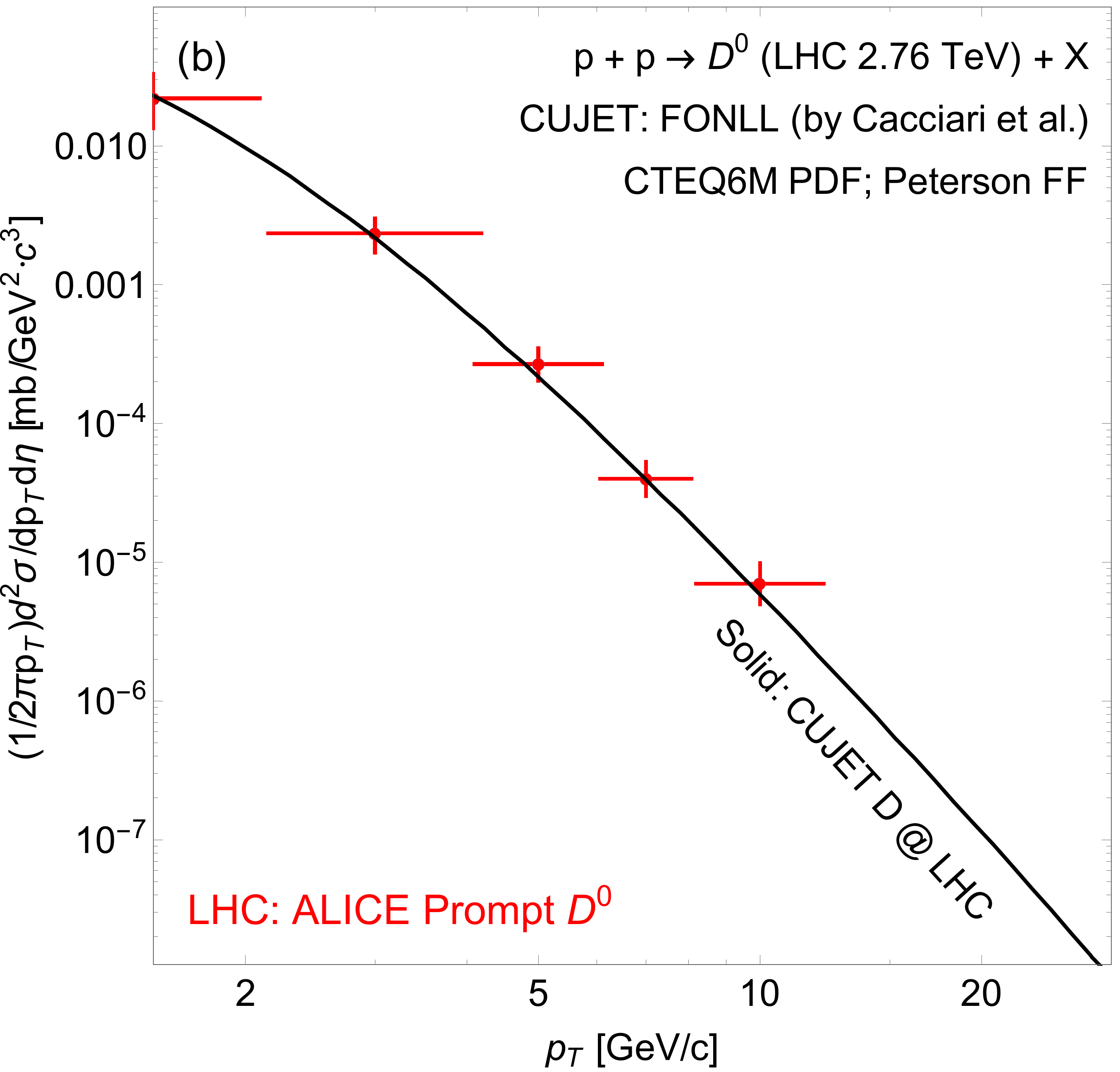}
\caption{\label{fig:pp}
(Color online)
(a) The measured invariant differential cross sections of inclusive charged particles in the mid-rapidity regime in p+p $\sqrt{s_{NN}}=2.76$ TeV collisions at the LHC (ALICE (red, \cite{Abelev:2013ala}), ATLAS (blue, \cite{Aad:2015wga})), and those of neutral pions in p+p $\sqrt{s_{NN}}=200$ GeV collisions at RHIC (PHENIX (orange, \cite{Adler:2003pb})), compared with CUJET pp references that are generated from LO pQCD calculations \cite{Wang:private} with CTEQ5 Parton Distribution Functions (PDFs) and LO KKP Fragmentation Functions (FFs). (b) Data of prompt ${\rm D}^0$ invariant differential cross sections (mid-rapidity) in p+p $\sqrt{s_{NN}}=2.76$ TeV collisions at the LHC (ALICE (red, \cite{Abelev:2012vra})) compared with CUJET pp references generated from FONLL calculations \cite{VOGT} with CTEQ6M PDFs and Peterson FFs.
}
\ec
\end{figure*}

Fluctuations about the mean radiative and elastic energy loss are taken into account in the following approximations:
Poisson multiple gluon emission is assumed in the radiative sector;
Gaussian fluctuations are assumed in the elastic
sector. The total energy loss probability distribution is constructed from the
convolution of the radiative and the elastic sector. In order to get the quenched leading hadron spectra at high $p_T$, this parton-level distribution is convoluted 
with LO
pQCD pp spectra (for gluon and ligh quark) \cite{Wang:private} or FONLL pp spectra (for charm and
bottom) \cite{VOGT}, folded over the
 MC-Glauber A+A initial hard scattering probability distribution
\cite{Glauber:1970jm,Shen:2010uy,Song:2011hk,Heinz:2013bua,Shen:2015qta},
and hadronized with parton fragmentation functions (LO KKP \cite{KKP} for gluon and light quark\footnote{There are systematic studies that confront current NLO parton fragmentation functions with inclusive charged-particle spectra at hadron colliders and compare the varied consistencies between NLO FFs and data, cf. e.g. \cite{d'Enterria:2013vba}. The light-hadron pp references in CUJET are generated from LO pQCD calculations with CTEQ5 PDFs plus LO KKP FFs. The results are consistent with available data, as shown in Fig.~\ref{fig:pp}(a).}, Peterson \cite{PETERSON} for charm and bottom).

The pp baselines embedded in CUJET are plotted in Fig.~\ref{fig:pp}.
For light hadrons, data from corresponding RHIC and LHC measurements \cite{Abelev:2013ala,Aad:2015wga,Adler:2003pb} are well described by the LO pQCD results with a K-factor of 1.13 (fitting to the PHENIX datum at $p_T=$ 6.24 GeV) at RHIC and of 1.76 (fitting to the ATLAS datum at $p_T=$ 6.26 GeV) at the LHC. In the heavy flavor sector, the FONLL results with a K-factor of 0.43 (fitting to the ALICE datum at $p_T=$ 7 GeV) are in agreements with data from ALICE measurements \cite{Abelev:2012vra}.  
The most important message demonstrated by the comparison in Fig.~\ref{fig:pp}, is that the {\em $p_T$ spectra shapes} used in CUJET are consistent with RHIC and LHC data. The fitted LO  K-factors (that are theoretically predictable only with NLO and beyond) drop out of the single hadron observables $R_{AA}$ and $v_2$ which are only dependent on the spectra shapes and independent of absolute normalization.

In the CUJET2.0 framework, only $a=q,g$ (where $a$ is the quasi-parton type, $q$ stands for quark while $g$ for gluon) HTL 
components are included. 
While in the CUJET3.0 framework, $a=m$ ($m$ is the particle monopole) is also included 
and running coupling elastic and radiative energy loss
are modified as we discuss in the next section 
to incorporate nonconformal nonperturbative lattice QCD 
data to constrain the composition and screening properties of the sQGMP.

\subsection{Nonperturbative QCD matter near $T_c$}
\label{sec:near-Tc}

In the vicinity of the QCD deconfinement transition temperature $T_c\sim \Lambda_{QCD}$,   the strength of the chromo-electric coupling becomes very strong. Novel nonperturbative effects should enter and modify the properties of the QGP in this  regime. Based on first-principle lattice QCD calculations and reasonable theoretical assumptions,  in the CUJET3.0 framework, we model  the near $T_c$ QCD matter as a semi-Quark-Gluon-Monopole-Plasma (sQGMP) that  includes two ingredients with nonperturbative origins  -- the semi-QGP suppression of color electric DOFs and the emergent chromo magnetic monopoles.

\subsubsection{The semi-quark-gluon plasma model}
\label{sec:semi-QGP}

The semi-QGP model was developed and discussed in detail in \cite{Pisarski:2006hz,Hidaka:2008dr,Hidaka:2009hs,Hidaka:2009ma,Lin:2013qu,Lin:2013efa,Gale:2014dfa,Hidaka:2015ima}. It is constructed to describe QCD as temperature $T\rightarrow T^+_c$, where both the naive perturbative methods and the hadronic models are not applicable. A main emphasis is on the ``mismatch'' seen from lattice data between the liberation of thermal excitations (as indicated by e.g. rapid increase of entropy density around $T_c$) and the liberation of ``color'' (as indicated by the rather slow increase of Polyakov loop toward fully deconfined limit). This observation indicates at a region above $T_c$ where significant nonperturbative suppression of color charge is still present.    
In the semi-QGP, how color is suppressed is quantified by the decrease of the expectation value of the Polyakov loop $\langle \ell \rangle$, which is the trace of a straight Wilson line in imaginary time. Properly normalized $\langle \ell \rangle$ is near unity in the perturbative QGP, but in the near $T_c$ regime, it is smaller than 1 (from lattice, $\langle \ell \rangle \sim 0.1$ at $T_c$). This implies a non-trivial distribution for the eigenvalues of the Wilson line and a nontrivial background field for the time-like component of the gauge field $A_0$. In the presence of a nonzero $A_0$, as $T\rightarrow T_c^+$, the colored excitations are suppressed by powers of the Polyakov loop.

Let us briefly review how this suppression works in the semi-QGP following \cite{Hidaka:2009hs,Hidaka:2009ma}. Adopting the double line basis for color factors, fundamental quarks carry a single index in the fundamental representation, $a=1,\cdots,N_c$, and adjoint gluons carry a pair of fundamental indices, ab. In an $SU(N_c)$ gauge theory, under mean field approximation, we take the temporal component of the gluon field to be a constant, diagonal, traceless matrix
\begin{eqnarray}
(A_0^{cl})^{ab}=\delta^{ab} Q^a/g
\label{A0background}
\end{eqnarray}
where g is the coupling constant for the $SU(N_c)$ gauge theory. For the spatial components of the gluon vector potential, $A_i$, there is no background field.
In the Euclidean spacetime, the Wilson line in the temporal direction is
\begin{eqnarray}
L(\vec{x})\equiv \mathcal{P} \exp \left( ig \int_0^{1/T} d\tau A_0(\tau,\vec{x}) \right) .
\label{WilsonLine}
\end{eqnarray}
We neglect fluctuations in $A_0$ to leading order in the coupling constant under mean field approximation. Gauge invariant Polyakov loops are traces of powers of the Wilson line:
\begin{eqnarray}
\ell_n(Q) \equiv \langle tr L^n\rangle / N_c = \sum_{a=1}^{N_c} e^{inQ^a/T} / N_c.
\label{PolyakovLoops}
\end{eqnarray}
We define $\ell$ as the first Polyakov loop $\ell\equiv \ell_1$. Physically, one can think of the Polyakov loop as measuring the extra free energy $F$ which arises from adding a colored heavy quark to a thermal bath, $\langle \ell \rangle\sim\exp(-F/T)$. In the perturbative QGP, all $Q^a$'s vanish and $\ell_n=1$. In the confined phase of a pure gauge theory, eigenvalues of $Q$ are uniformly distributed on a circle of radius $T$, and $\ell_n=0$ if $n\neq [N_c/2]N_c$. Dynamical quarks act as a background $Z(N)$ field, if they are present, there is no rigorous definition of a confined phase, and all Polyakov loops are nonzero at nonzero temperatures \cite{Hidaka:2009hs}. Lattice simulations find that $\ell$ is small ($\langle\ell \rangle \sim 0.1$) in the phase transition regime for $N_c=3$ and $N_f=3$ \cite{Hidaka:2009ma}.


In the imaginary time formalism, the Euclidean four
momentum is $P_\mu=(p_0,{\mathbf p})$, where $p_0$ is an even/odd multiple of $\pi T$ for bosons/fermions. Expanding around the background field in Eq.~\eqref{A0background}, the 4-momentum of a quark becomes $P_\mu^a=(p_0+Q^a,\mathbf{p})$, and the 4-momentum of a gluon becomes $P_\mu^{ab}=(p_0+Q^{ab},\mathbf{p})$ ($Q^{ab}\equiv Q^a-Q^b$). Each $Q^a$ is typically a non-integral multiple of $2\pi T$, in the space of diagonal generators, it is like an imaginary chemical potential for color charges. To analytically continue from Euclidean to Minkowski spacetime, one continues the entire Euclidean energy to $-iE$, where $E$ is a continuous energy variable. For quarks, the usual Fermi-Dirac statistical distribution function $\widetilde{n}(E)$ becomes
\begin{eqnarray}
\widetilde{n}_a(E)=\frac{1}{e^{(E-iQ^a)/T}+1}=\sum_{j=1}^{\infty}(-1)^{j+1} e^{-j(E-iQ^a)/T}\;.
\end{eqnarray}
The first term represents the Boltzmann approximation to the quantum distribution function, and is accompanied by $\exp(iQ^a/T )$.
Consider the trace of the quark propagator which enters e.g. in the calculation of the pressure at leading order. The sum is
\begin{eqnarray}
\frac{1}{N_c}\sum_{a}\widetilde{n}_a(E)=\sum_{j=1}^{\infty}(-1)^{j+1} e^{-jE/T} \frac{1}{N_c} tr L^j\;.
\end{eqnarray}
Denote $\langle \cdots \rangle_Q$ as the average over the Q distribution and an integration over the particles three momenta. At $T\rightarrow T^+_c$ where $\ell$ is nonzero but small,
\begin{eqnarray}
\langle \sum_{a} \widetilde{n}_a \rangle_Q \sim N_c T^3 \ell\;.
\end{eqnarray}
This means the distribution function for a quark field vanishes as a single power of the loop, i.e. $n_q\sim\ell$.
For gluons, the usual Bose-Einstein statistical distribution function ${n}(E)$ becomes
\begin{eqnarray}
{n}_{ab}(E)=\frac{1}{e^{(E-i(Q^a-Q^b))/T}-1}=\sum_{j=1}^{\infty} e^{-j(E-i(Q^a-Q^b))/T}\;.
\end{eqnarray}
Consider summing over the color indices of the gluon propagator, to avoid taking the trace which is part of the $1/N_c$ correction, one sums separately over a and b. Since $\sum_a e^{iQ^a/T}\sum_b e^{-iQ^b/T}\;=\;trL\;trL^\dagger$, we have
\begin{eqnarray}
\langle \sum_{ab} {n}_{ab} \rangle_Q \sim N_c^2 T^3 \ell^2\;.
\end{eqnarray}
This means in the near $T_c$ regime, the density of gluons vanishes as the square of the loop, i.e. $n_g\sim\ell^2$. In the perturbative regime, the density of massless fields is necessarily a pure number times $T^3$ therefore such a suppression is not present. Note that we always perform a global color rotation to enforce that the expectation value of the Polyakov loop $\ell$ is real.

\subsubsection{The magnetic scenario for the near $T_c$ QCD plasma}
\label{sec:monopoles}

The magnetic scenario for the near $T_c$ QCD matter was proposed and discussed in \cite{Liao:2006ry,Liao:2007mj,Liao:2008vj,Liao:2008jg,Liao:2008dk,Liao:2012tw}, and there have since been extensive studies of the magnetic component of the plasma using different approaches \cite{Chernodub:2006gu,Ratti:2008jz,Zakharov:2014bca,Jiang:2015cva}. In this scenario, the QGP not too far above $T_c$ contains not only electrically charged quasi-particles (cec), quarks and gluons, but also magnetically charged quasiparticles (cmc), monopoles and dyons. 

This approach is different from many traditional discussions, which focus on the thermodynamic transition    and divide the temperature regimes into the hadronic phase at $T <T_c$ and the   QGP phase at $T >T_c$. Rather, the emphasis is on  the competition between EQPs and MQPs, based on which one may divide the phases of QCD matter   into the €œmagnetically dominated region at $T<T_{E=M}\sim 1.4 T_c $ and the â€electrically dominated€  region at $T>T_{E=M}$.  This picture is largely motivated by  analogy with electric-magnetic duality in supersymmetric Yang-Mills theories. The key aspect of the physics involved is the coupling strength of the electric (e) and magnetic (g) interaction, which can lead to different dominance of dynamical degrees of freedom in different regimes. Under Dirac quantization condition \cite{Dirac:1931kp}, $e\cdot g=n/2$, and magnetic objects are in the adjoint color representation if $n=2$. In a so-called E/M-equilibrium region, the couplings are equal,  i.e. $e=g$, densities as well as masses of both EQPs and MQPs are comparable. Then depending on the change of these couplings in different physical regimes, the ``balance'' between E and M sectors would shift one way or the other, giving rise to distinctive phases. 

Let us start with the QGP at very high temperature $T\gg T_c$ where the electric coupling is weak.  This regime is well described by perturbative EQPs with small quark and gluon effective masses.  The monopoles in this case are heavy, dilute and strongly coupled, but they  play a minor role and contribute little to the overall bulk properties. They do manifest themselves through nonperturbative contributions to certain observables at the soft magnetic scale.

On the other hand, as T goes down and approaches the confinement transition $T \rightarrow T_c$, the converse is expected to happen: the electric coupling becomes very strong and the EQPs, i.e. quarks and gluons, are getting heavier and gradually suppressed dynamically. The emergent MQPs gradually become light, abundant, and  dominate the system at $T<T_{E=M}$. With further decrease of temperature toward $T_c$ the thermal monopoles will eventually reach the Bose-Einstein condensation,  forming a 't Hooft-Mandelstam ``dual superconductor'' \cite{Mandelstam:1974pi,'tHooft:1981ht} that enforces color confinement.  In the   post-confinement region at $T_c < T < T_{E=M}$ EQPs are still strongly correlated and connected by the electric flux tubes, but MQPs are the dominant DOFs and they serve as an effective description of the strong nonperturbative gauge dynamics. In \cite{Liao:2008jg}, the authors showed that gauge theory monopoles in a deconfined phase behave as magnetic charges in a Coulomb plasmas.  At $T \approx 1.3T_c$ where lattice potentials indicate flux tubes dissolve, an estimate of total density of magnetic quasi-particles is $n_{cmc} \approx 4.4-6.6\;{\rm fm}^{−3}$ \cite{Liao:2007mj}.From an analysis of the lattice monopole-(anti)monopole correlators, they showed that the temperature dependence of the magnetic couplings in gauge theories is indeed the inverse of the electric one as per the electric-magnetic duality arguments. More specifically, the magnetic part of the QGP at $T\sim 1-3T_c$ possesses an effective plasma parameter in the ``good liquid'' domain, thus in consistency with   the ``nearly perfect liquid'' property observed at RHIC and LHC.



\subsection{Jet suppression in semi-Quark-Gluon-Monopole-Plasmas}
\label{sec:pQCD+sQGMP}

Having discussed the foundations of the sQGMP, let us integrate it into the jet energy loss kernel in section \ref{sec:DGLV}. The critical component in Eq.~\eqref{rcDGLV} is the 1-HTL dynamical scattering potential,  
\be
x\frac{dN}{dx}\propto{...} \int d^2{q} \left[  \frac{\rho \, \alpha_s^2(\bq_\perp^2)\, f_E^2}{\bq_\perp^2 (\bq_\perp^2 + f_E^2 \mu^2)} \right]  ...\;\;.
\label{Potential}
\ee
Since the sQGMP contains both chromo electrically charged quasi-particles (cec) and chromo magnetically charged quasiparticles (cmc), when jets propagate through the medium near $T_c$, scattering channels of $E+E$ and $E+M$ exist simultaneously. One way to generalize Eq.~\eqref{Potential} is to symmetrize it with respect to the E and M components of the kernel based on demanding electric-magnetic duality as illustrated in e.g. the celebrated Seiberg-Witten solution of the ${\cal N}=2$ super-Yang-Mills theory. This leads to the following modified form of the kernel:
\begin{eqnarray}
x\frac{dN}{dx}\propto{...} \int d^2{q} \left[ \frac{\rho_E \left(\alpha_s(\bq_\perp^2)\alpha_s(\bq_\perp^2)\right) f_E^2}{\bq_\perp^2 (\bq_\perp^2 + f_E^2 \mu^2)} + \frac{ \rho_M \left(\alpha_E(\bq_\perp^2)\alpha_M(\bq_\perp^2)\right)  f_M^2}{\bq_\perp^2 (\bq_\perp^2 + f_M^2 \mu^2)}\;\right]...\;\;.
\label{EMPotential}
\end{eqnarray}
Where $\alpha_s\equiv\alpha_E$, and $\alpha_E\cdot\alpha_M=1$ at any scale by Dirac quantization condition \cite{Dirac:1931kp,Liao:2008jg}. The total quasi-particle number density $\rho$ is divided into EQPs with fraction $\chi_T = \rho_E / \rho$ and MQPs with fraction $1-\chi_T=\rho_M/\rho$. The parameter $f_E$ and $f_M$ is defined via $f_E\equiv\mu_E/\mu$ and $f_M\equiv\mu_M/\mu$, with $\mu_E$ and $\mu_M$ being the electric and magnetic screening mass respectively. We emphasize that Eq.(\ref{EMPotential}) is a nonperturbative sQGMP model ansatz that differs substantially from other generalization of HTL, see e.g.~\cite{Djordjevic:2008iz,Djordjevic:2011dd}. 

To determine $\chi_T$, one notices that: (1) when temperature is high, $\chi_T$ should reach unity, i.e. $\chi_T(T\gg T_c) \to 1$; (2) in the vicinity of the regime $T\sim (1-3)T_c$, the renormalized expectation value of the Polyakov loop L (let us redefine $L\equiv \ell = \langle tr \mathcal{P} \exp\lbrace ig\int_{0}^{1/T} d\tau A_0\rbrace  \rangle /N_c$) deviates significantly from unity, implying the suppression $\sim L$ for quarks and $\sim L^2$ for gluons in the semi-QGP model \cite{Hidaka:2008dr,Hidaka:2009ma,Dumitru:2010mj,Lin:2013efa}. In the regime the quark and gluon density drop much faster than the thermodynamic quantities. This points to ``missing'' DOFs, in the magnetic scenario~\cite{Liao:2006ry,Liao:2008jg}, they are identified as chromo-magnetic monopoles who emerge in gauge theories at strong coupling and are thermal excitations of the vacuum magnetic condensate as in the ``dual superconductivity'' picture of color confinement~\cite{Bali:2000gf,Ripka:2003vv,Kondo:2014sta}. For the cec component fraction, we use the semi-QGP ansatz: 
\begin{eqnarray}
\chi_T(T) = c_q \, L(T) + c_g \, L^2(T)\;.
\label{chiTL}
\end{eqnarray}
For the respective fraction of quarks and gluons, where we take the Stefan-Boltzmann (SB) fraction coefficients, $c_q = (10.5 N_f )/(10.5 N_f + 16)$ and $c_g = 16/(10.5 N_f + 16)$. To be consistent with lattice data, we fit the temperature dependent Polyakov loop $L(T)$ ($T$ in GeV) with 
\begin{eqnarray}
L(T) = \left[\frac{1}{2}+\frac{1}{2}{\rm Tanh}[7.69(T-0.0726)]\right]^{10} 
\label{PolyakovLoop}.
\end{eqnarray} 
Eq.~\eqref{PolyakovLoop} adequately fits both the HotQCD \cite{Bazavov:2009zn} and Wuppertal-Budapest \cite{Borsanyi:2010bp} Collaboration results, c.f. Fig.~\ref{fig:FL}(a). With $\chi_T$ and $(1-\chi_T)$,  $\rho_E/\rho$ and $\rho_M/\rho$ are completely fixed. 

To specify the electric and magnetic screening mass ($\mu_{E,M}=f_{E,M}\, \mu$), we recall that at very high temperature, one expects (1) $f_E \to 1$, i.e. $\mu_E \sim g T$ from
HTL results and (2) $f_M \sim g$, i.e. $\mu_M \sim g^2 T$, from magnetic scaling in dimensional reduction. Assuming E-M duality, the screening masses are expected to scale as
\begin{eqnarray}
\mu_{E,M}^2 \sim \alpha_{E,M}\,\rho_{E,M}/T\;.
\end{eqnarray} 
The extrapolation to lower temperature thus gives 
\begin{eqnarray}
\mu_E^2\sim \alpha_E\,(\chi_T\rho)/T  \sim \chi_T \mu^2\;,
\end{eqnarray}
and we expect the electric screening mass to be suppressed as $\sqrt{\chi_T(T)}$ in the near $T_c$ regime but approach the HTL $\mu(T)$ at high T limit. Regarding the magnetic screening mass, since we have $n_M\sim(\alpha_E T)^3$ following the magnetic scaling, then
\begin{eqnarray}
\mu_M^2\sim \alpha_M\,(\alpha_E T)^3/T\sim \alpha_E^2 T^2 \sim g^2\,g^2  T^2 \sim g^2 \mu^2\;.
\end{eqnarray}
This prescription is supported by lattice data \cite{Nakamura:2003pu}. Therefore, we assume the following
local temperature dependent screening masses in the CUJET3.0 model:
\begin{eqnarray}
f_E(T(\vz))  = \sqrt{\chi_T(T(\vz))}  \quad ,  \quad f_M(T(\vz)) = c_m \, g(T(\vz))\, .
\label{f_EM}
\end{eqnarray} 
To be consistent   
with previous treatments in Eq.~\eqref{rcDGLV} and \eqref{DebyeMass},   
the local electric ``coupling'' $g(T(\vz))$ is defined via
\begin{eqnarray}
g(T(\vz))=\sqrt{4\pi\alpha_s(\mu^2(T(\vz)))}=\frac{\mu(T(\vz))}{T(\vz)\sqrt{1+N_f/6}}.
\end{eqnarray}
Note that $c_m$ is a constant parameter that can be constrained by lattice data on the magnetic screening. Fig.~\ref{fig:FL}(b) illustrates the agreement between this prescription of $\mu_{E,M}$ and lattice extracted values \cite{Nakamura:2003pu}. 

Finally, in the CUJET3.0 framework, the energy loss kernel Eq.~\eqref{rcDGLV} is generalized to
\begin{eqnarray}
 x_E\frac{dN_g^{n=1}}{dx_E}&=& \frac{18 C_R}{\pi^2} \frac{4+N_f}{16+9N_f} \int{d\tau}\; \rho(\bz) \Gamma(\vz)\;\int{d^2k_{\perp}} \alpha_s( \frac{\bk_{\perp}^2}{x_+ (1-x_+)} )\;\nonumber\\
& \times&\;\int{d^2q}\frac{ \alpha_s^2(\bq_\perp^2)\left (f_E^2+ \frac{f_E^2 f_M^2 \mu^2(\bz)}{\bq_\perp^2} \right )\chi_T+\left (f_M^2+ \frac{f_E^2 f_M^2 \mu^2(\bz)}{\bq_\perp^2} \right) (1- \chi_T) }{(\bq_\perp^2 + f_E^2 \mu^2(\bz))(\bq_\perp^2 + f_M^2 \mu^2(\bz))}\nonumber\\
& \times&\;\frac{-2(\bk_{\perp}-\bq_{\perp})}{(\bk_{\perp}-\bq_{\perp})^2+\chi^2(\bz)} \left[ \frac{\bk_{\perp}}{\bk_{\perp}^2+\chi^2(\bz)} - \frac{(\bk_{\perp}-\bq_{\perp})}{(\bk_{\perp}-\bq_{\perp})^2+\chi^2(\bz)} \right]\nonumber\\
& \times&\;{\left[1-\cos\lp\frac{(\bk_{\perp}-\bq_{\perp})^2+\chi^2(\bz)}{2 x_+ E } \tau\rp\right]}\left(\frac{x_E}{x_+}\right)\left| \frac{dx_+}{dx_E} \right| \;\;,
\label{emEnergyLoss}
\end{eqnarray} 
where $\chi_T$ and $f_{E,M}$ follows Eq.~\eqref{chiTL} and Eq.~\eqref{f_EM}. We note that in the temperature range $T\sim T_c$, the coupling $\alpha_s$ becomes non-perturbative~\cite{Liao:2006ry,Liao:2008jg,Zakharov:2008kt,Randall:1998ra}.  Analysis of  lattice data \cite{Liao:2008jg} suggests the following thermal running coupling form:
\be
\alpha_s(Q^2)=\dfrac{\alpha_c}{1+\frac{9\alpha_c}{4\pi} \log(\frac{Q^2}{T_c^2})}\;,
\label{TcEnhancement}
\ee
with $T_c=160$ MeV. Note that at large $Q^2$, Eq.~\eqref{TcEnhancement} converges to vacuum running $\alpha_s(Q^2)=\frac{4\pi}{9\log(Q^2/\Lambda^2)}$; while at $Q=T_c$, $\alpha_s(T^2_c)=\alpha_c$. 

\section{Liberation schemes for color degrees of freedom}
\label{sec:L}

As discussed in section \ref{sec:near-Tc}, within the semi-QGP model, the expectation value of the Polyakov loop $L$ (note that we have redefined $L\equiv \ell$ in section \ref{sec:pQCD+sQGMP}, and we use this notation for the rest of this paper) is the only relevant parameter for the confinement/deconfinement transition, upon proper renormalizations, $L$ serves as a suppression factor for the colored excitations as $T\rightarrow T_c^+$. However, it is questionable whether or not $L$ is an \emph{order parameter} for the phase transition. Besides the fact that lattice calculations point to a $L\sim 0.1$ at $T_c$, Eq.~\eqref{A0background}\eqref{WilsonLine}\eqref{PolyakovLoops} also indicate that to a certain degree the loop physically characterizes the free energy of an \emph{infinitely massive static quark}. Since (1) in the perturbative QGP phase dynamical light quarks dominate the medium transport properties; and (2) to boost $v_2$ in line with data, a strongly enhanced jet scattering near $T_c$ makes decisive contributions \cite{Xu:2014tda}; then, the nonperturbative property of the sQGMP, in particular, the rate at which fractional chromo-electric DOFs are liberated (defined as $r_d(T)\equiv d\chi_T/dT$) in the near $T_c$ regime will play a significant role in computing jet quenching observables within the CUJET3.0 framework and should be studied more systematically.

\subsection{Polyakov loop versus quark number susceptibility}
\label{sec:NL-FL}

Another useful measure of the nonperturbative suppression of the color electric DOF is provided by the quark number susceptibilities~\cite{McLerran:1987pz,Gottlieb:1988cq,Gavai:1989ce,Gottlieb:1987ac}. Such susceptibilities quantify the quark number fluctuations that can be obtained from the QCD partition function at vanishing chemical potentials. Denote $u$, $d$, $s$ as up, down, strange quark whose numbers are conserved charges in QCD. Starting from the pressure,
\begin{eqnarray}
\frac{p}{T^4}=\frac{1}{VT^3}{\rm ln}Z(V,T,\mu_u,\mu_d,\mu_s)\;,
\end{eqnarray}
moments of charge fluctuations are defined as follows,
\begin{eqnarray}
\chi_{ijk}^{uds}=\frac{\partial^{i+j+k}{p/T^4}}{\partial(\mu_u/T)^i\partial(\mu_d/T)^j\partial(\mu_s/T)^k}\;.
\end{eqnarray}
Concentrate on the quadratic fluctuation,
\begin{eqnarray}
\chi_{2}^{u,d,s}=\frac{1}{VT^3}\langle N_{u,d,s}^2\rangle.
\end{eqnarray}
And $\chi_{2}^{u,d,s}$ is the diagonal susceptibility of $u$, $d$, $s$ quark number density. Singlet susceptibilities of other conserved charges in QCD such as baryon number B, strangeness S and electric charge Q can be obtained from the above quark number susceptibilities \cite{Borsanyi:2011sw}.

The diagonal susceptibility is proposed as part of the order parameter for chiral symmetry breaking/restoration in \cite{McLerran:1987pz}. Considering a gas of free quarks, if the quark mass $m$ is small, then $\chi_2$ is expected to be large since it is relatively easy to create an additional quark. For instances, if $m\ll T$, then in the continuum limit, $\chi_2\sim N_f T$.
If $m$ is large, then it will be difficult to create a quark or antiquark, the susceptibilities will be suppressed by ${\rm exp}(-m/T)$. Realistically, in the high T phase, though strongly interacting, if the fundamental excitations of the system are low-$m$ objects with the quantum numbers of quarks, then $\chi_2$ is still expected to be large. Meanwhile, in the
low T phase, $\chi_2$ will be small since quarks are confined and the nonzero quark number states have large masses. Thus in the chirally symmetric phase, the quark number susceptibility is large, which is consistent with a plasma of light quarks; while in the chiral symmetry broken phase, the quark number susceptibility is small, as expected from quark confinement. It however may be noted that in the parton-hadron boundary regime, various bound states like baryons and mesons (and even other exotic composite objects) carry conserved charges and contribute to the susceptibilities. As previous studies have shown~\cite{Liao:2005pa,Kim:2009uu,Shi:2013zxa}, such contributions are mostly important for the higher-order susceptibilities as well as for the off-diagonal ones. The leading order diagonal susceptibilities could therefore serve as a reasonable measure for the counting of quark degrees of freedom in the plasma. 

Therefore, besides interpolating the renormalized lattice Polyakov loop as in Eq.~\eqref{PolyakovLoop}, we parametrize the lattice diagonal susceptibility of u quark number density as
\begin{eqnarray}
\chi_2^u(T) = 0.91 \times \left[ \frac{1}{2} \left\lbrace 1+{\rm Tanh}[15.65(T-0.0607)] \right\rbrace  \right]^{10} 
\label{chi2u}. 
\end{eqnarray} 
Where T is the temperature in the unit of GeV. Note that at extremely high temperature, the $\chi_2^u(T)$ is not unity, so we renormalize the susceptibility by its value at $T\rightarrow \infty$ and define a new quantity $\tilde{\chi}_2^u(T)$ as
\begin{eqnarray}
\tilde{\chi}_2^u(T) = \left[ \frac{1}{2} \left\lbrace 1+{\rm Tanh}[15.65(T-0.0607)] \right\rbrace  \right]^{10}
\label{chi2u_}.
\end{eqnarray} 
The $\tilde{\chi}_2^u(T)$ plays a similar role as properly renormalized $L$ for quark DOFs. Let us denote the original liberation scheme, c.f. Eq.~\eqref{chiTL}, that follows the power law of the Polyakov loop as in the semi-QGP model, as $\chi_T^L$ ($\chi_T^L\equiv \chi_T$ in Eq.~\eqref{chiTL}); and the new deconfinement scheme where the diagonal susceptibility of light quark number density dominates the transition, as $\chi_T^u$ ($\chi_T^u=\rho_E/\rho$):
\begin{eqnarray}
\chi_T^u = c_q \, \tilde{\chi}_2^u + c_g \, L^2\;\;.
\label{chiTu}
\end{eqnarray}
Note that in this scheme, the magnetically charged quasi-particles, i.e. chromo-magnetic monopoles, consist a density fraction of $1-\chi_T^u=\rho_M/\rho$.

\begin{figure*}[!t]
\bc
\includegraphics[width=0.49\textwidth]{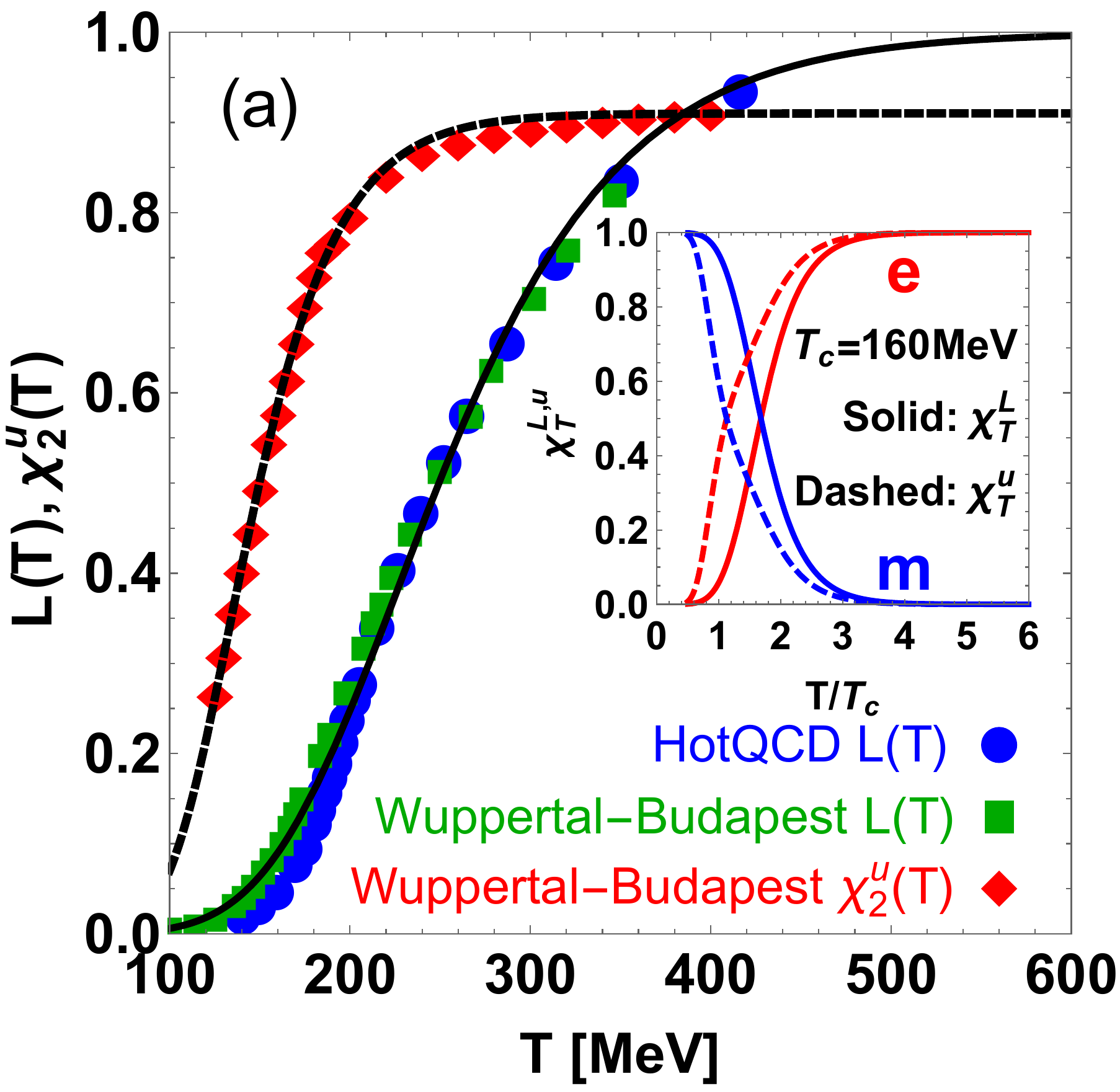}
\includegraphics[width=0.45\textwidth]{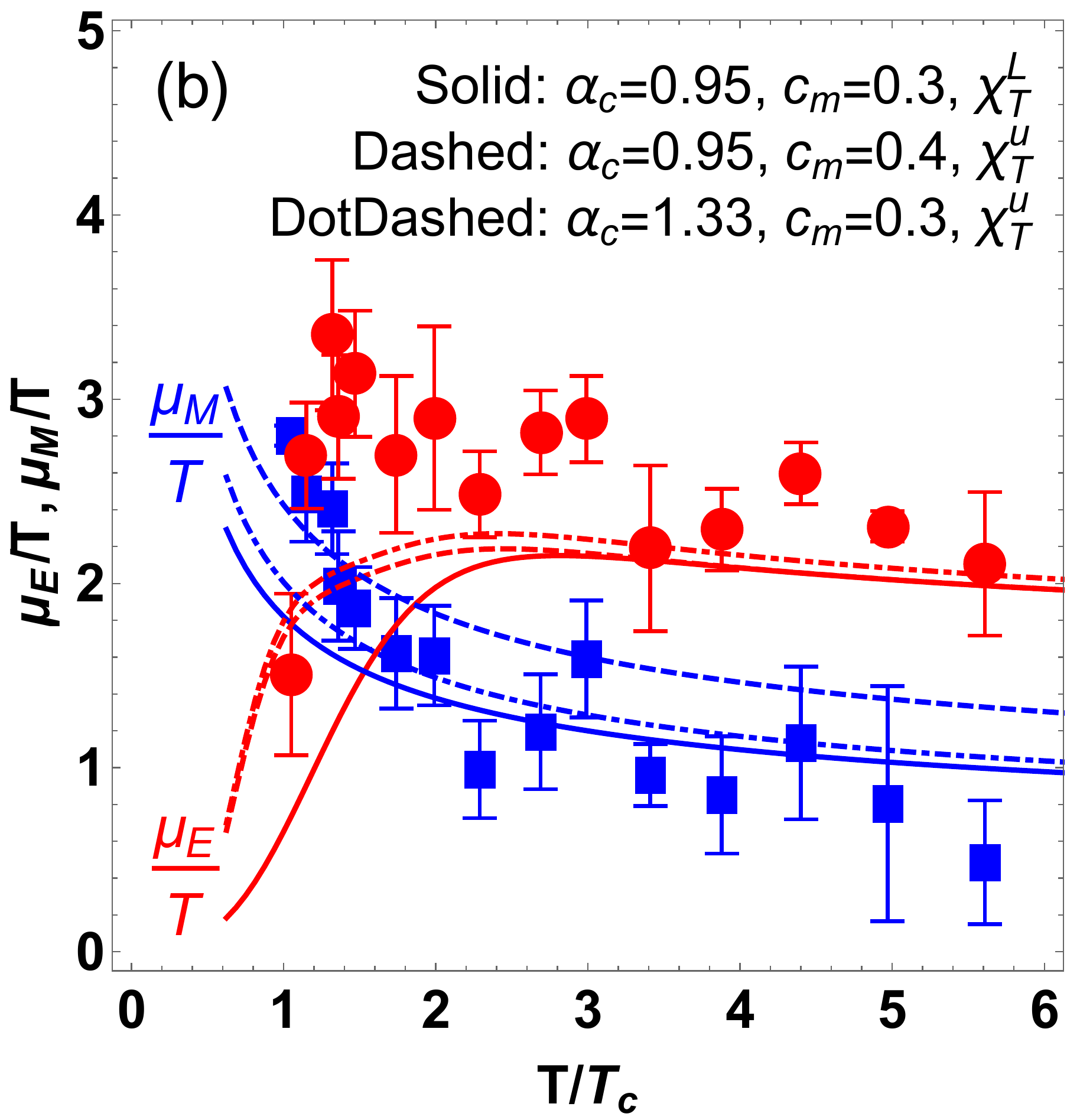}
\caption{\label{fig:FL}
(Color online)
(a) Renormalized Polyakov loop $L(T)$ (blue circle: \cite{Bazavov:2009zn}, green square: \cite{Borsanyi:2010bp}) and diagonal susceptibility of light quark number density $\chi_2^u(T)$ (red diamond: \cite{Borsanyi:2011sw}) computed from lattice QCD, fitted with the parametrization of Eq.~\eqref{PolyakovLoop} and \eqref{chi2u}. The inset shows the the density fraction of color electric DOFs (red, $\chi_T=\rho_E/\rho$) and color magnetic DOFs (blue, $1-\chi_T=\rho_M/\rho$) within the liberation scheme $\chi_T^L$ (solid) and $\chi_T^u$ (dashed), in the temperature range $T\sim0.6-6.0\;T_c$, where $T_c=160$ MeV. Notice that in $\chi_T^L$, $\rho_E\approx \rho_M$ at $T\sim 1.7\,T_c$; in $\chi_T^u$, $\rho_E\approx \rho_M$ at $T\sim 1.1\,T_c$; and these temperatures are where $r_d(T)\equiv d\chi_T/dT$ should peak in $\chi_T^{L,u}$.
(b) The dimensionless electric (red) and magnetic (blue) screening mass $\mu_{E,M}/T$ in the CUJET3.0 model i.e. Eq.~\eqref{f_EM}, for scheme (i) \eqref{SchI}, (ii) \eqref{SchII}, and (iii) \eqref{SchIII}, compared with results from Lattice QCD \cite{Nakamura:2003pu}. Note that the $\alpha_c$ and $c_m$ parameters in (i)(ii)(iii) are chosen such that the high-$p_T$ reference $R_{AA}$ datum can be well-fitted, c.f. Fig.~\ref{fig:FL-RAA-v2}(a). Note that for (i) and (ii), $\mu_E/T\approx\mu_M/T$ at around the same temperature, i.e. $T\sim 1.5-1.6\,T_c$, while (iii)'s $\mu_E/T$ and $\mu_M/T$ intersect at $T\sim 1.1\,T_c$. In the near $T_c$ regime, (i) and (ii)'s $\mu_E-\mu_M$ are approximately identical, both are less than (iii)'s.
}
\ec
\end{figure*}

Fig.~\ref{fig:FL}(a) shows the lattice QCD data on renormalized Polyakov loop and diagonal light quark susceptibility from the HotQCD \cite{Bazavov:2009zn} and Wuppertal-Budapest \cite{Borsanyi:2010bp,Borsanyi:2011sw} Collaboration, as well as the parametrization fit Eq.~\eqref{PolyakovLoop} and Eq.~\eqref{chi2u}. The E and M quasi-particle density fraction in both the $\chi_T^L$ and $\chi_T^u$ scheme are plotted in the inset of Fig.~\ref{fig:FL}(a). Note that $\rho_E/\rho=\chi_T^{L,u}$ and $\rho_M/\rho=1-\chi_T^{L,u}$. The two different schemes, for the rate of ``quark liberation'', with $\chi_T^L$ the ``slow'' and $\chi_T^u$ the ``fast'', provide useful estimates of theoretical systematic uncertainties associated with the quark component of the sQGMP model. Note that in the inset of Fig.~\ref{fig:FL}(a)  the $\rho_E$ and $\rho_M$ are equal at $T\sim 1.1\,T_c$ for $\chi_T^u$ scheme while at $T\sim 1.7\,T_c$ for $\chi_T^L$ scheme: these temperatures are where $r_d(T)\equiv d\chi_T^{L,u}/dT$ should peak.

\begin{figure*}[!t]
\bc
\hspace{1pt}
\includegraphics[width=0.465\textwidth]{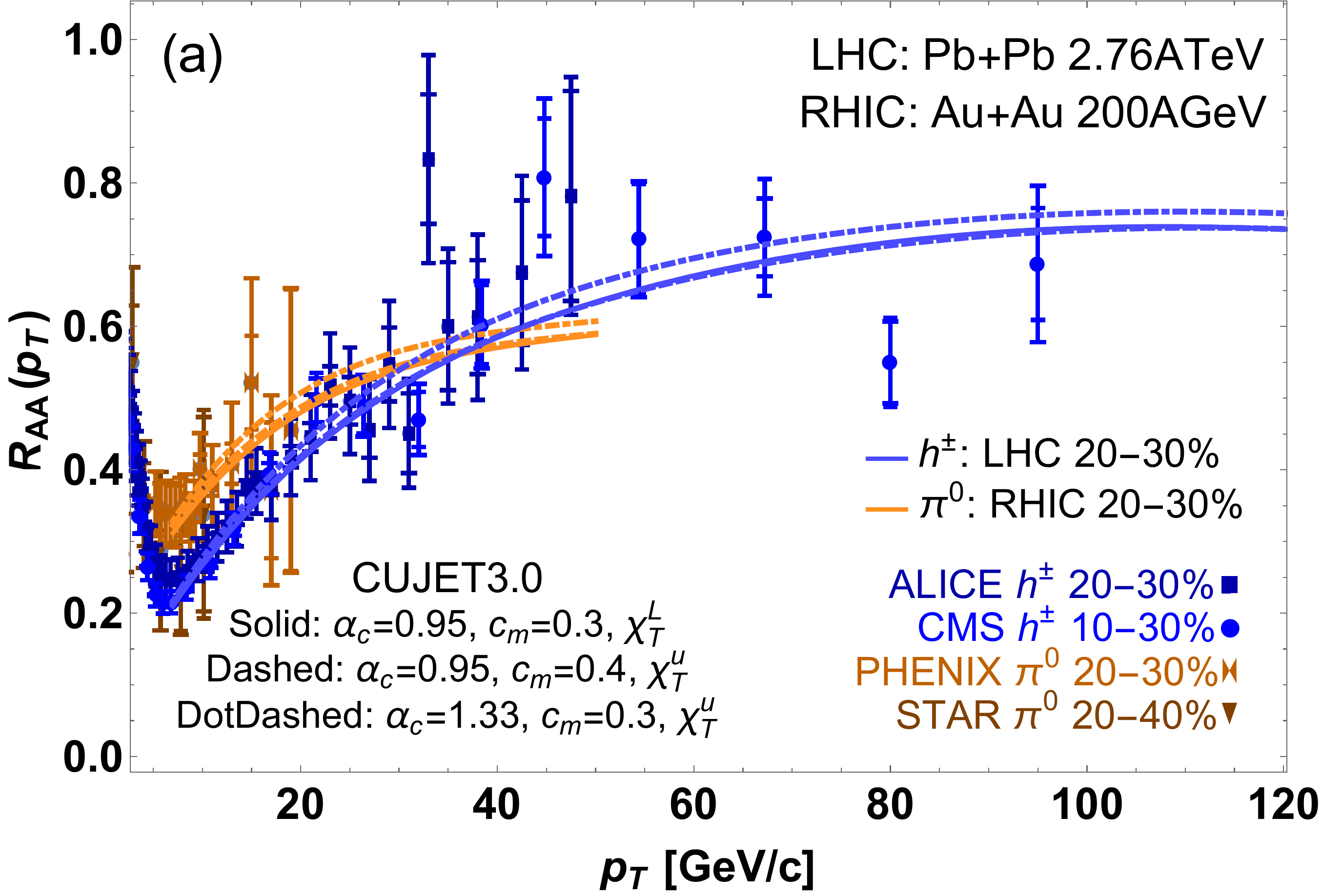}
\hspace{1pt}
\includegraphics[width=0.465\textwidth]{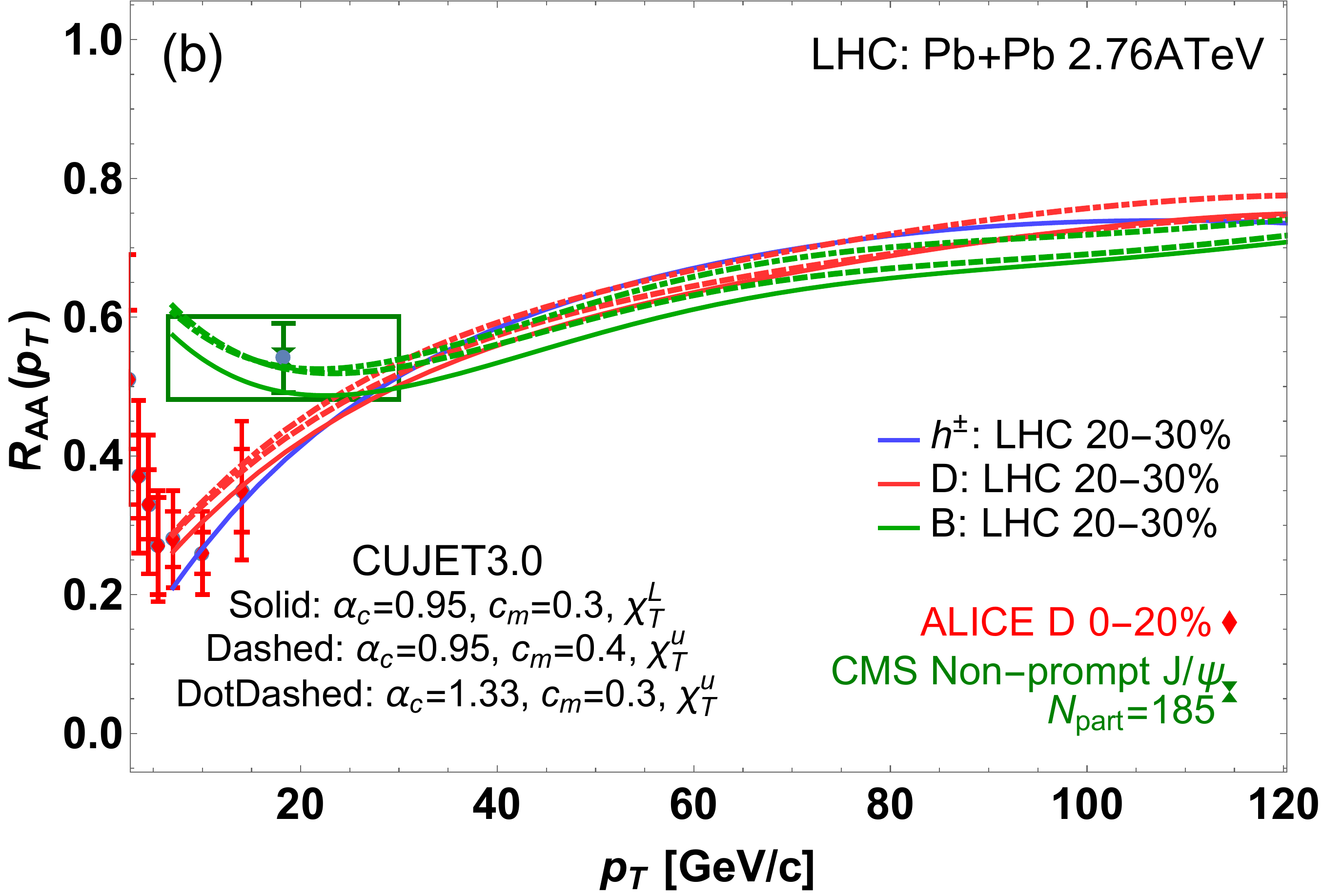}
\includegraphics[width=0.475\textwidth]{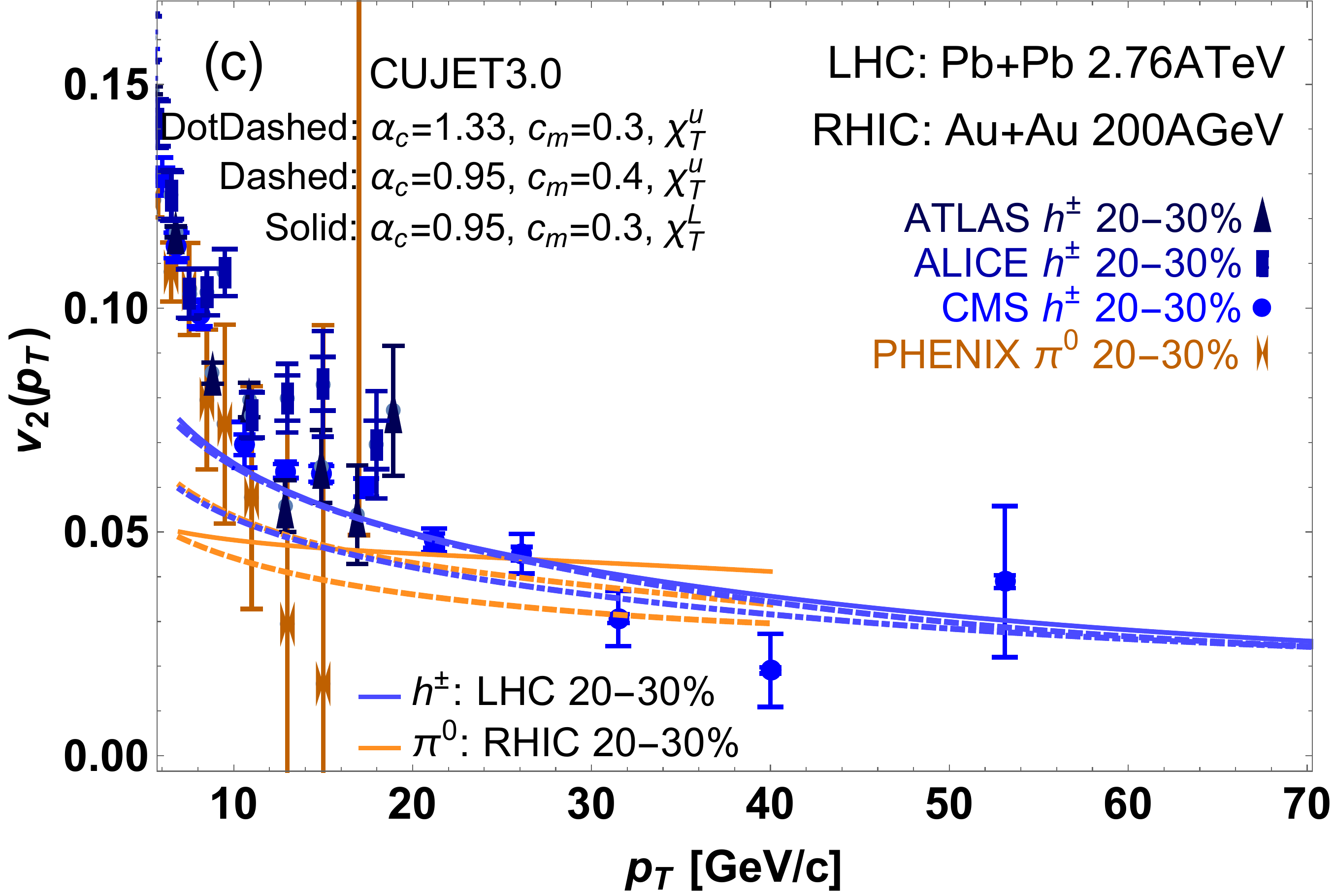}
\includegraphics[width=0.475\textwidth]{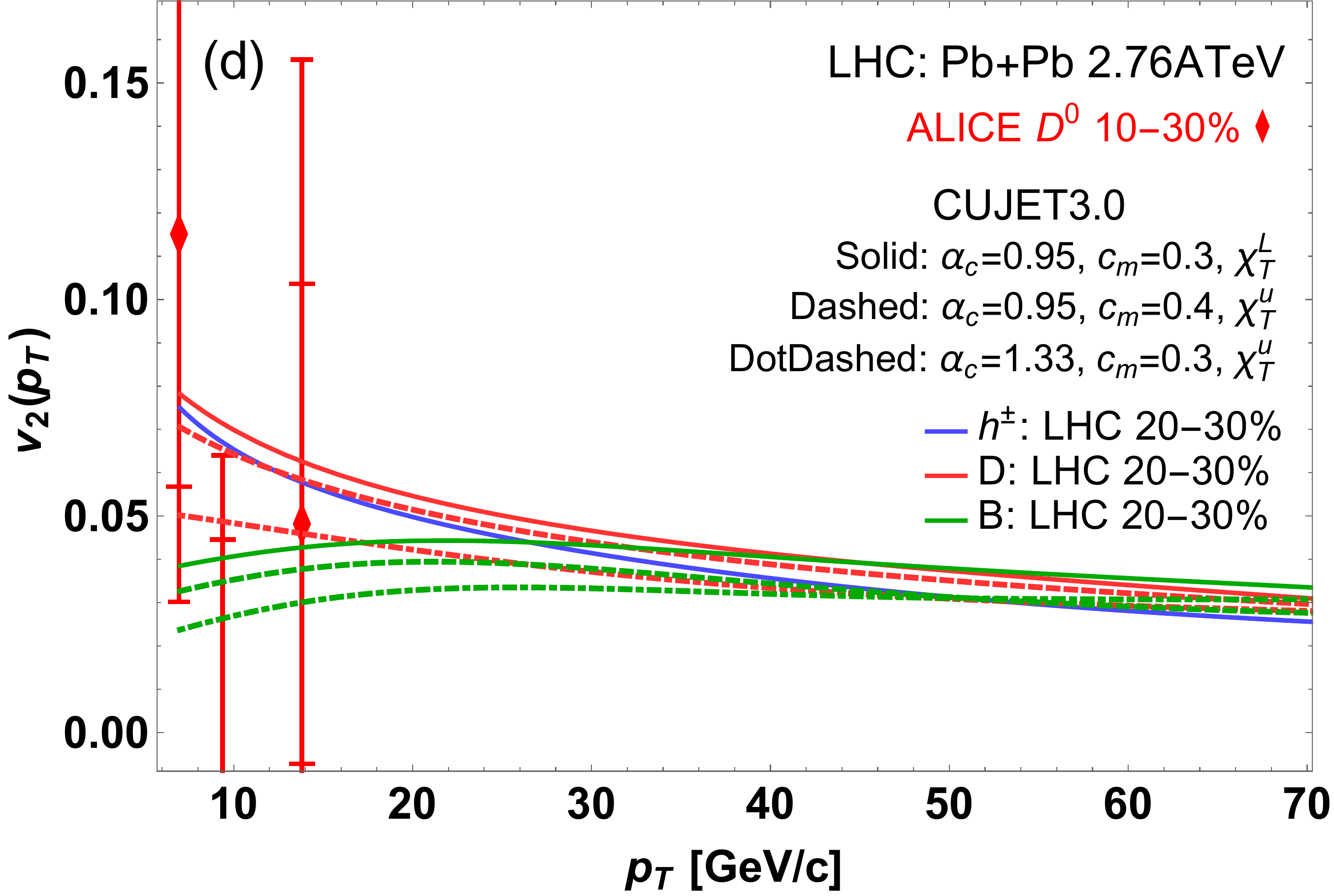}
\caption{\label{fig:FL-RAA-v2} 
(Color online)
(a) Neutral pion ($\pi^0$, brown) and charge particle ($h^\pm$, blue)'s $R_{AA}(p_T>8{\rm GeV})$ in Au+Au 200GeV and Pb+Pb 2.76TeV 20-30\% collisions, computed from CUJET3.0 with scheme (i)\eqref{SchI} $\alpha_c$=0.95, $c_m$=0.3, $\chi_T^L$ (solid), (ii)\eqref{SchII} $\alpha_c$=0.95, $c_m$=0.4, $\chi_T^u$ (dashed) and (iii)\eqref{SchIII} $\alpha_c$=1.33, $c_m$=0.3, $\chi_T^u$ (dotdashed), compared with corresponding RHIC \cite{Adare:2008qa,Adare:2010sp,Adare:2012wg,Abelev:2009wx} and LHC \cite{Abelev:2012di,Abelev:2012hxa,ATLAS:2011ah,CMS:2012aa,Chatrchyan:2012xq} measurements. The $\pi^0$ and $h^\pm$'s $v_{2}(p_T>8{\rm GeV})$ are plotted in (c). (b) The CUJET3.0 results of D meson (red) and B meson (green)'s $R_{AA}(p_T>8{\rm GeV})$ at LHC in (i)(ii)(iii) compared with available data \cite{ALICE:2012ab,Abelev:2014ipa,CMS:2012vxa}. The D and B's $v_{2}(p_T>8{\rm GeV})$ are plotted in (d).
Note that the parameters $(\alpha_c,c_m)$ are determined in scheme (i)(ii)(iii) such that the reference datum at LHC $R_{AA}^{h^\pm}(p_T=12.5{\rm GeV})\approx0.3$ as well as the lattice $\mu_M$ (\cite{Nakamura:2003pu}, c.f. Fig.~\ref{fig:FL}(b)) are reasonably fitted.
Results from all three schemes are compatible with light hadron (LH)'s $R_{AA}$; while for LH's $v_{2}$, (i) and (ii) can generate reasonable agreements with data, but (iii) underestimates the $v_{2}$. For open heavy flavors (HF), (ii) and (iii) have similar $R_{AA}$ predictions, both differ from (i); while for HF's $v_{2}$, (i), (ii), and (iii)'s prediction are all different. Such differences in the predictions for jet quenching observables  from (i)(ii)(iii) suggest that data on high $p_T$ $R_{AA}$ and $v_{2}$ can impose stringent constraints on the nonperturbative properties of the medium near $T_c$.
}
\ec
\end{figure*}

As suggested in Eq.~\eqref{f_EM}, any change in $\chi_T$ alters the electric screening mass $\mu_E$, and one expects shifts in the magnetic screening correspondingly under electric-magnetic duality. In Fig.~\ref{fig:FL}(b), lattice data of the electric and magnetic screening mass are compared with the CUJET3.0 results in three schemes:
\begin{eqnarray}
&({\rm i})& \alpha_c=0.95, c_m=0.3, \chi_T^L\,;\label{SchI}\\
&({\rm ii})& \alpha_c=0.95, c_m=0.4, \chi_T^u\,;\label{SchII}\\
&({\rm iii})& \alpha_c=1.33, c_m=0.3, \chi_T^u\,.\label{SchIII}
\end{eqnarray}
Note that the $(\alpha_c,c_m)$ parameters are chosen such that the single reference datum $R_{AA}^{h^\pm}(p_T=12.5{\rm GeV})\approx0.3$ at LHC is well-fitted, c.f. Fig.~\ref{fig:FL-RAA-v2}(a). Implicitly, $\chi_T^L$ and $\chi_T^u$ is determined by Eq.~\eqref{chiTL} and Eq.~\eqref{chiTu} respectively. All three schemes are in reasonable agreements with the lattice data. However, to be more careful, (i)(ii)(iii)'s $\mu_{E,M}$ do behave differently as temperature varies. (i)'s and (ii)'s $\mu_E/T$ and $\mu_M/T$ intersect at approximately the same temperature, i.e. $T\sim 1.5-1.6\,T_c$. Meanwhile, (iii)'s $\mu_E/T$ and $\mu_M/T$ intersect at $T\sim 1.1\,T_c$, which temperature overlaps approximately with the T where $\rho_E(T)=\rho_M(T)$ in $\chi_T^u$. Furthermore, in the near $T_c$ regime, (i)'s and (ii)'s $\mu_E-\mu_M$ are approximately equal, and both are less than (iii)'s.

Let us move to the predictions of jet quenching observables in scheme (i)(ii)(iii) within the CUJET3.0 framework, in particular, leading light hadrons' and open heavy flavors' $R_{AA}(p_T>8{\rm GeV})$ (c.f. Eq.~\eqref{RAAdef}) and $v_{2}(p_T>8{\rm GeV})$ (c.f. Eq.~\eqref{v2def}) at RHIC and LHC semi-peripheral A+A collisions are of interests. The results and corresponding data are plotted in Fig.~\ref{fig:FL-RAA-v2}.

For high $p_T$ light hadrons ($\pi^0$ at RHIC, $h^\pm$ at LHC), Fig.~\ref{fig:FL-RAA-v2}(a) shows that all three schemes can describe the $R_{AA}(p_T)$ data at RHIC ($8<p_T<20 \,{\rm GeV}$) and LHC ($8<p_T<100 \,{\rm GeV}$) simultaneously, but only (i) and (ii) are compatible with the high $p_T$ single $\pi^0$ and $h^\pm$'s $v_{2}$ as illustrated in Fig.~\ref{fig:FL-RAA-v2}(c). Since (i) and (ii) have different liberation schemes hence $r_d$'s, different absolute values of $\mu_{E,M}$, but approximately the same $\mu_E-\mu_M$ near $T_c$, this observation indicates that in boosting the $\pi^0$ or $h^\pm$'s azimuthal elliptical harmonics to be in line with data, the difference between $\mu_E$ and $\mu_M$ in the near $T_c$ regime plays a critical role. Notice that as suggested in the magnetic scenario, when cooling down to pass $T\sim 1-2 T_c$, the lightest hence the dominant DOFs in the medium shift from EQPs to MQPs, and the color screening mass is one of the indicators of this transition \cite{Liao:2006ry}. The fact that (i)(ii)'s $\mu_E(T)-\mu_M(T)$ generates a larger $v_2$ than (iii)'s $\mu_E(T)-\mu_M(T)$ who has a larger value and a lower zero point temperature $T_0$ ($T_0$ is defined as $\mu_E(T_0)=\mu_M(T_0)$) implies that, $v_2$ is sensitive to how the relative value of $\mu_E$ and $\mu_M$ inverses near $T_c$ -- the higher the $T_0$, the longer the path length that jets interact with the strongly coupled monopole dominated medium at later time of the evolution, the larger the azimuthal anisotropy. A further comment is on the absolute values of $\mu_{E,M}$ in (i)(ii), clear (ii)'s are larger. This is necessary because after T drops lower than $T_0$, MQPs dominates, (i)'s $\rho_M$ is denser than (ii)'s, to get to the same magnitude of overall leading hadron suppressions, (ii) should possess larger color screening masses, c.f. Eq.~\eqref{EMPotential}.

For open heavy flavors, specifically, high $p_T$ D and B mesons, Fig.~\ref{fig:FL-RAA-v2}(b) shows their respective $R_{AA}$ at LHC Pb+Pb $\sqrt{s_{NN}}=2.76$ TeV collisions, computed from CUJET3.0 with scheme (i)(ii)(iii). Differ from the light hadrons' $R_{AA}$ where (i)(ii)(iii) have almost identical predictions, for D and B mesons, (ii) and (iii)'s $R_{AA}$ overlap, both of which distinguish from (i)'s. One notices that (ii) and (iii) have different $\mu_{E,M}$, but the same liberation scheme $\chi_T^u$, which is distinct from (i)'s $\chi_T^L$. This implies the open heavy flavor's high $p_T$ $R_{AA}$ is critically influenced by the rate at which chromo-electric DOFs are deconfined ($r_d=d\chi_T/dT$). This connection is intrinsically embedded in the CUJET3.0 framework because the heavier masses induced dead cone effects shuffled the weights of the scattering potential \eqref{EMPotential} and the rest of the Eq.~\eqref{emEnergyLoss} in such a way that the total induced radiation became more sensitive to the deconfinement scheme $r_d$ rather than screening masses $\mu_{E,M}$ for jet quenching in sQGMP. On the other hand, Fig.~\ref{fig:FL-RAA-v2}(d) shows the prediction of open heavy flavor's $v_{2}(p_T>8{\rm GeV})$ at LHC in CUJET3.0. The $v_2$'s are all different in scheme (i), (ii), and (iii). This suggests the open charm and beauty's $v_2(p_T)$ act as good probes of the nonperturbative $(r_d,\mu_E,\mu_M)$ near $T_c$. Let us mention in passing that for the heavy quark dynamics in low $p_T$ region, the sQGMP model   also expects a strong temperature dependence of their in-medium diffusion coefficients (specifically a ``dip'' near $T_c$), which has recently been shown as an essential ingredient toward a simultaneous description of their $R_{AA}$ and $v_2$~\cite{Rapp:2009my,Das:2015ana}.

If one views the above physical connections from a different angle, the set of RHIC and LHC heavy-ion collision data on high $p_T$ light hadron (LH) and open heavy flavor (HF)'s $R_{AA}$ and $v_{2}$ will provide stringent limits on the nonperturbative properties of the QCD matter near $T_c$ in the CUJET3.0 framework. Specifically, after parameters in the model have been fixed by light quark's $R_{AA}$ data, the rate at which color DOFs are deconfined $r_d$ and the color screening masses $\mu_{E,M}$ can be stringently constrained: (1) light quark's $v_2$ regulates $\mu_{E}(T)-\mu_{M}(T)$ near $T_c$; (2) heavy quark's $R_{AA}$ determines $r_d(T)$; (3) HF's $v_{2}$ distinguishes all $r_d(T)$, $\mu_{E}(T)$ and $\mu_{M}(T)$.

\subsection{Jet transport coefficient and shear viscosity}
\label{sec:qhat-etas}

As discussed above, the high $p_T$  $R_{AA}$ and $v_{2}$ data of light and heavy quarks can provide stringent constraints on values of the nonperturbative $(r_d,\mu_E,\mu_M)$ near $T_c$. It is of great interests to further compare  how the jet and bulk transport properties differ in varied schemes (i) \eqref{SchI}, (ii) \eqref{SchII}, and (iii) \eqref{SchIII}. This will pave the way for clarifying the temperature dependence of jet quenching and shear viscous transport properties based on available high $p_T$ data in high-energy A+A collisions.

The jet transport coefficient $\hat{q}$ characterizes the averaged transverse momentum transfer squared per mean free path \cite{Burke:2013yra}. Here let us first calculate the $\hat{q}$ for a quark jet (in the fundamental representation F) with initial energy $E$, in the same way as the previous CUJET3.0 computation in \cite{Xu:2014tda}, via the following:
\begin{eqnarray}
\hat{q}_F(E,T)&=& \int_0^{6ET} d\bq_\perp^2 \frac{2\pi \bq_\perp^2}{(\bq_\perp^2+f_E^2 \mu^2)(\bq_\perp^2+f_M^2 \mu^2)} \rho(T) \nonumber\\
&\times& \left[ (C_{qq} f_q + C_{qg} f_g ) \alpha_s^2(\bq_\perp^2) + C_{qm} ( 1- f_q - f_g ) \right]\;.
\label{Effqhat}
\end{eqnarray} 
where $\rho(T)$ is the total number density, connected to the lattice pressure $p(T)$ via
\begin{eqnarray}
\rho(T) =  \xi\, p(T)/T \;, 
\label{RhoFromP}
\end{eqnarray}
with $\xi = [90\zeta(3)(16+9N_f)]/[\pi^4(16+10.5N_f)]=1.012$ 
for an $N_c=3, N_f=2.5$ ideal gas. 
The parameters $f_{q,g}$ are quasi-parton density fractions of quark (q) or gluon (g) type, in the $\chi_T^L$ and $\chi_T^u$ scheme, they are respectively
\begin{eqnarray}
f_q&=& c_q L(T) ,\;f_g=c_g L(T)^2,\; {\rm if}\;\chi_T^L\;; \nonumber \\ 
f_q&=& c_q \tilde{\chi}_2^u(T) ,\;f_g=c_g L(T)^2,\; {\rm if}\;\chi_T^u  \;.
\label{FracScheme} 
\end{eqnarray}  
The $c_{q,g}$ and $L(T)$ are the same as in Eq.~\eqref{chiTL} and \eqref{PolyakovLoop}. 
The magnetically charged quasi-particle density fraction is hence $f_m(T)=1-f_q(T)-f_g(T)$. The color factors in Eq.\eqref{Effqhat} are given by 
\begin{eqnarray}
C_{qq} &=&  \frac{4}{9} , \; C_{gg}=C_{mm}= \frac{9}{4}  , \;  \nonumber\\
C_{qg} &=&  C_{gq} = C_{qm} = C_{mq} = 1 \;.
\label{Effqhat1}
\end{eqnarray}

\begin{figure*}[!t]
\bc
\includegraphics[width=0.475\textwidth]{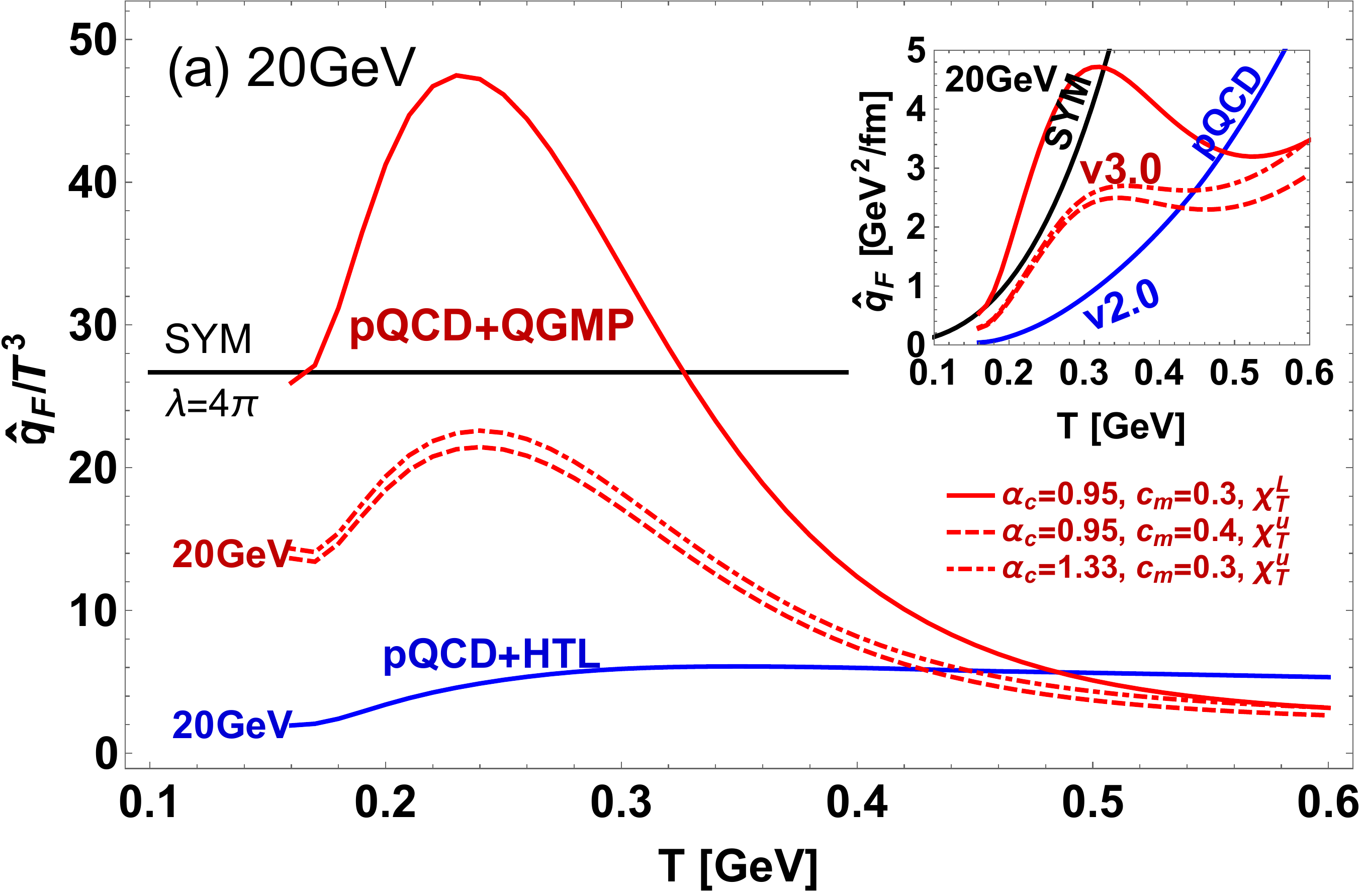}
\includegraphics[width=0.475\textwidth]{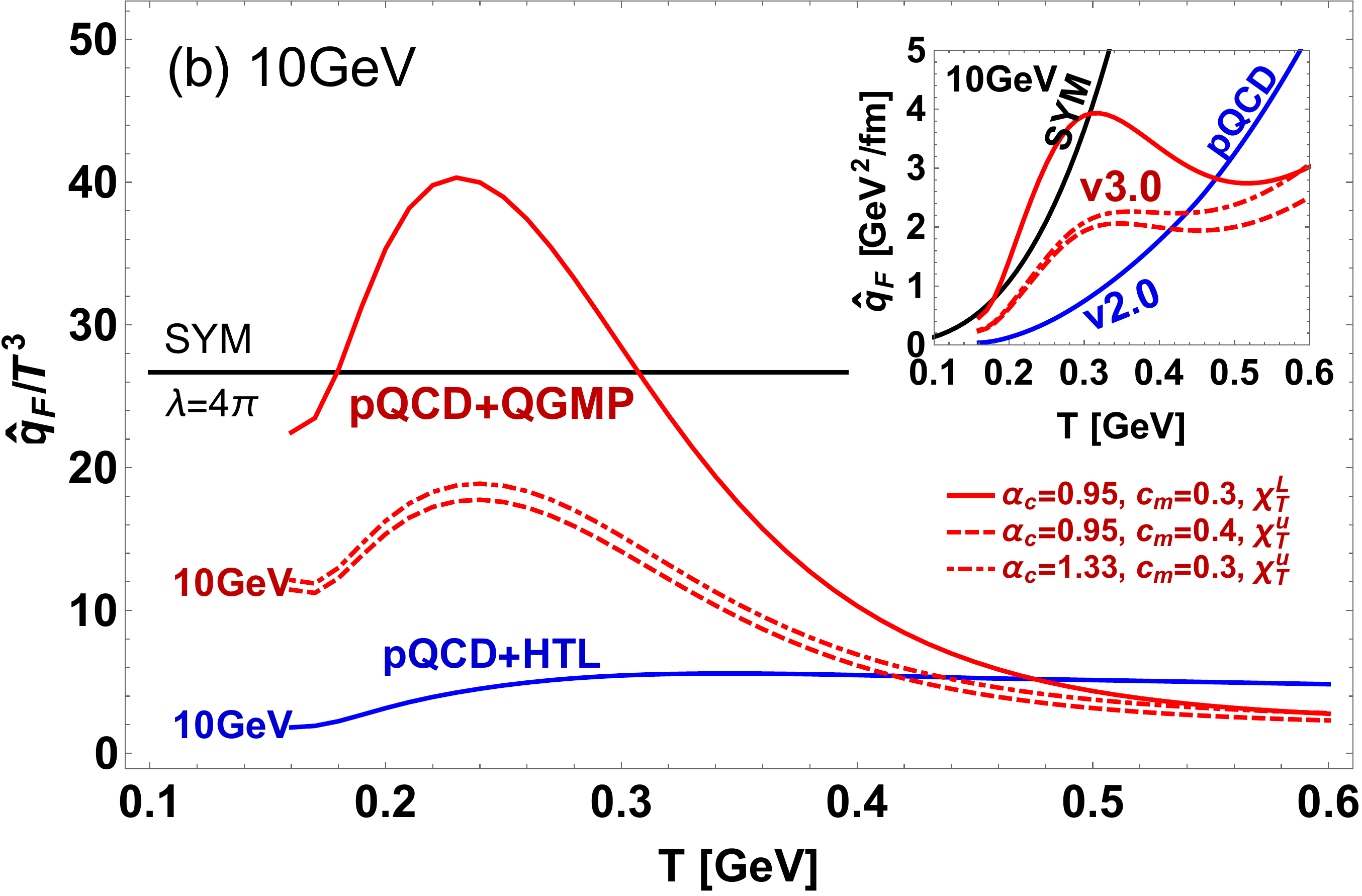}
\hspace{1pt}
\includegraphics[width=0.475\textwidth]{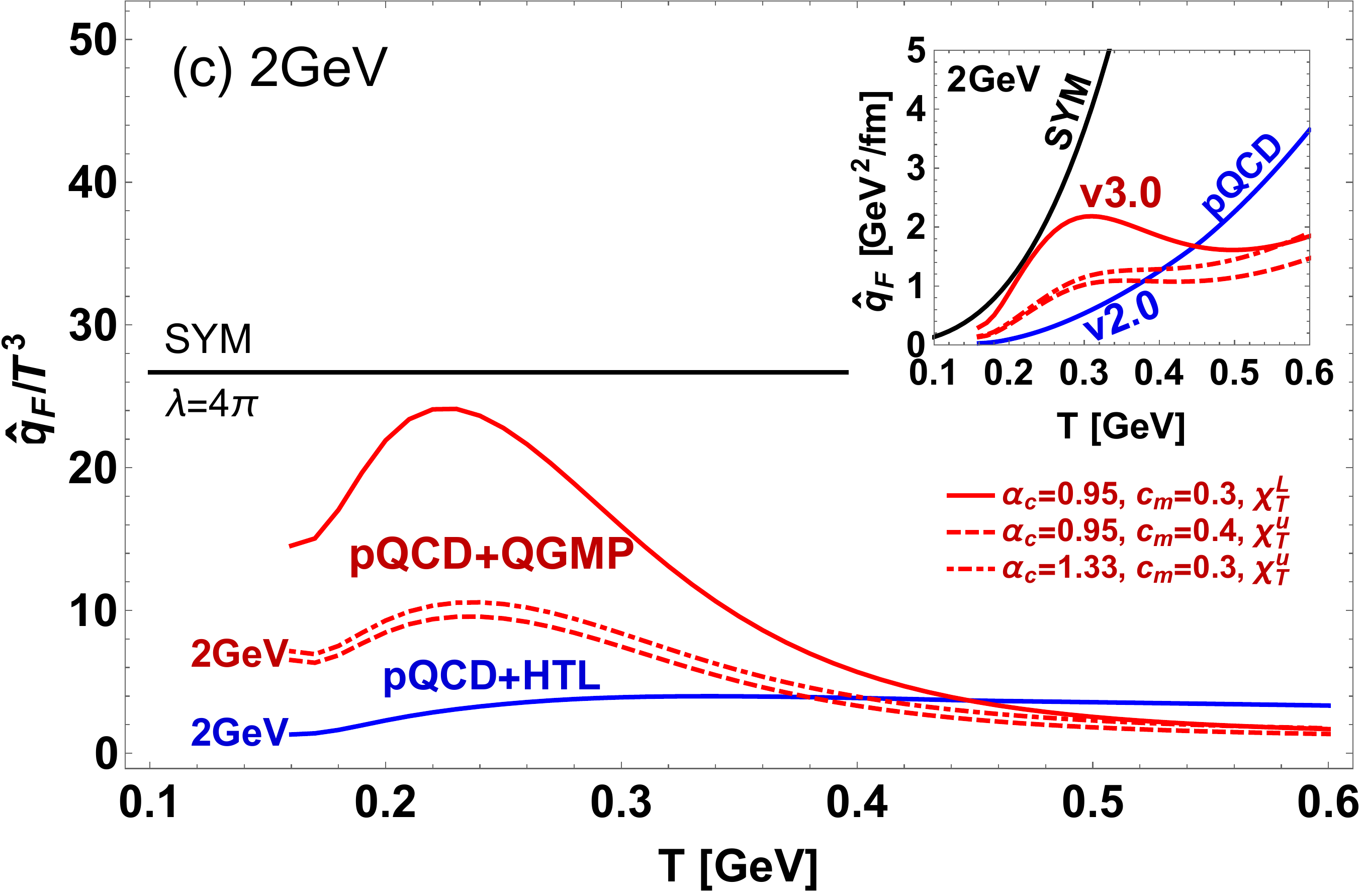}
\includegraphics[width=0.485\textwidth]{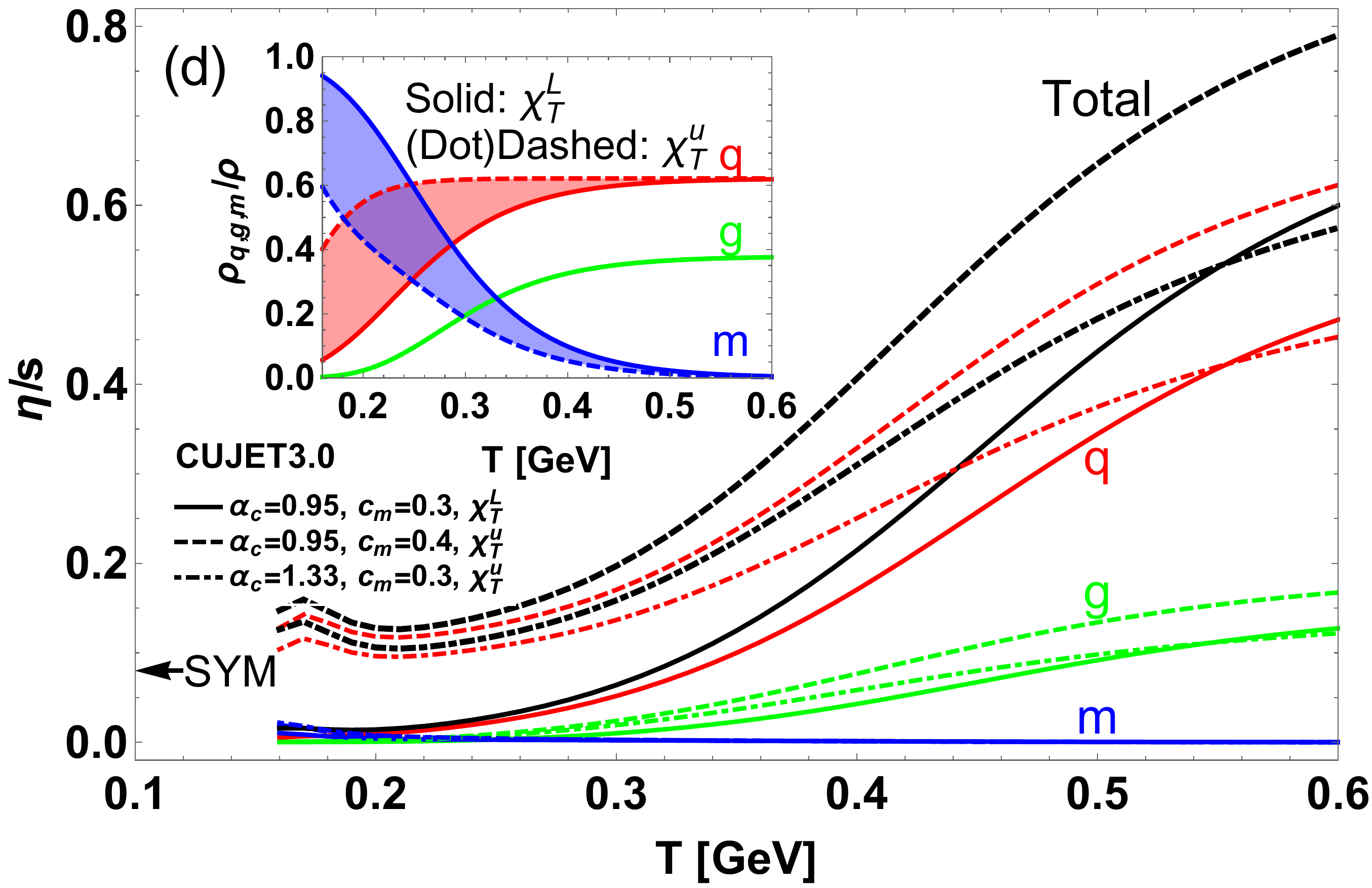}
\caption{\label{fig:FL-qhat-etas} 
(Color online)
The temperature dependence of the dimensionless jet transport coefficient $\hat{q}/T^3$ for a light quark jet (F) with initial energy $E=$ (a) 20GeV, (b) 10GeV, (c) 2GeV in the CUJET3.0 framework (Red) with the three schemes: (i) \eqref{SchI} (solid), (ii) \eqref{SchII} (dashed), and (iii) \eqref{SchIII} (dotdashed). The CUJET2.0 $\hat{q}_F/T^3$ with $(\alpha_{max},f_E,f_M)=(0.39,1,0)$ (Blue) and the $\mathcal{N}=4$ Super Yang-Mills (SYM) $\hat{q}_{SYM}/T^3=\frac{\pi^{3/2}\Gamma(\frac{3}{4})}{\Gamma(\frac{5}{4})}\sqrt{\lambda}$ \cite{Liu:2006ug} with 't Hooft coupling $\lambda=4\pi$ (Black) are plotted for comparisons. The insets show the absolute $\hat{q}_F$ in CUJET3.0, CUJET2.0 and SYM. Note that the (ii) and (iii)'s $\hat{q}$ are similar, and both are smaller than (i)'s. (d) The shear viscosity to entropy density ratio $\eta/s$ estimated in the kinetic theory using the $\hat{q}$ extrapolation Eq.~\eqref{Effetas1} in CUJET3.0 with scheme (i) (solid) (ii) (dashed) (iii) (dotdashed), for quasi-patron type q (quark, red), g (gluon, green), and m (monopole, blue). The total $\eta/s$ is plotted with black curves. The inset shows quasi-particle number density fraction of q, g, m in the liberation scheme $\chi_T^L$ and $\chi_T^u$. Note that in the near $T_c$ regime, in the $\chi_T^u$ scheme,  the total $\eta/s$ is dominated by q, while in the $\chi_T^L$ ``slow'' quark liberation scheme the total $\eta/s$ is dominated by m. In addition, there is a clear $\eta/s\sim 0.12$ minimum at $T\sim210$ MeV in (ii) and (iii) which utilize the same $\chi_T^u$ ``fast'' quark liberation  scheme. This $(\eta/s)_{min}$ is larger and phenomenologically more favorable than that in the ``slow'' quark liberation scheme.
}
\ec
\end{figure*}

In the CUJET3.0 framework, once the jet transport coefficient $\hat{q}$ has been computed, the shear viscosity to entropy density ratio $\eta/s$ can be calculated based on kinetic theory in a weakly-coupled quasi-particle picture, as proposed in~\cite{Danielewicz:1984ww,Hirano:2005wx,Majumder:2007zh}. 
An estimate of $\eta/s$ can be derived as 
\begin{eqnarray}
\eta/s &=& \frac{1}{s}\, \frac{4}{15} 
\sum_{a} \rho_a \langle p\rangle_a \lambda_a^{\perp} \nonumber\\
&=&\frac{4T}{5s}  \sum_a \rho_a \left(\sum_b \rho_b \int_0^{\langle \mathcal{S}_{ab}  \rangle /2}dq_\perp^2 
\frac{4q_\perp^2}{\langle \mathcal{S}_{ab} \rangle }\frac{d\sigma_{ab}}{dq_\perp^2}\right)^{-1} \nonumber\\
&=&\frac{18T^3}{5s}  \sum_a \rho_a/{\hat{q}}_a(T,E=3T)\;\;.
\label{Effetas1} 
\end{eqnarray}
Note that the $\hat{q}(T,E)$ is extrapolated down to thermal energy scales $E\sim 3T$. The $\rho_a(T)$ is the quasi-parton density of type $a=q,g,m$. The mean thermal Mandelstam variable $\langle \mathcal{S}_{ab} \rangle \sim 18T^2$. The entropy density $s(T)$ is interpolated from lattice calculations \cite{Bazavov:2009zn}. The $\hat{q}_{a=q}\equiv \hat{q}_F$ is calculated as in Eq.~\eqref{Effqhat}. The $\hat{q}_{a=g,m}$ for are computed via
\begin{eqnarray}
\hat{q}_{a=g}(E,T)&=& \int_0^{6ET} d\bq_\perp^2 \frac{2\pi \bq_\perp^2}{(\bq_\perp^2+f_E^2 \mu^2)(\bq_\perp^2+f_M^2 \mu^2)} \rho(T) \nonumber\\
&\times& \left[ (C_{gq} f_q + C_{gg} f_g ) \alpha_s^2(\bq_\perp^2) + C_{gm} ( 1- f_q - f_g ) \right]\;,\label{qhatG}\\
\hat{q}_{a=m}(E,T)&=& \int_0^{6ET} d\bq_\perp^2 \frac{2\pi \bq_\perp^2}{(\bq_\perp^2+f_E^2 \mu^2)(\bq_\perp^2+f_M^2 \mu^2)} \rho(T) \nonumber\\
&\times& \left[ (C_{mq} f_q + C_{mg} f_g ) +  C_{mm} ( 1- f_q - f_g )/\alpha_s^{2}(\bq_\perp^2) \right]\;.\label{qhatM}
\end{eqnarray} 
Clearly the $\eta/s$ of the system is dominated by the ingredient which has the largest $\rho_a/\hat{q}_a$.

Fig.~\ref{fig:FL-qhat-etas}(a)(b)(c) shows the dimensionless jet transport coefficient $\hat{q}/T^3$ in CUJET3.0 for a quark jet with initial energy E = 20, 10, 2 GeV respectively, compared with the CUJET2.0 result \cite{Xu:2014ica} and the AdS/CFT limit \cite{Liu:2006ug}. The $\hat{q}$ in scheme (i)(ii)(iii) are plotted. Compared with (i) which has $\chi_T^L$ ``slow'' quark liberation, the $\hat{q}$ in (ii) and (iii) which have $\chi_T^u$ ``fast'' quark liberation scheme are significantly smaller. This may  be understood as follows: in the $\chi_T^u$ scheme, as temperature rises, the chromo-electric DOFs are excited faster than those in the $\chi_T^L$ scheme, and leads to a smaller fraction of magnetically charged quasi-particles in the near $T_c$ regime. Since chromo-magnetic monopoles are the key contributors to the enhancement of jet opacity, c.f. appendix \ref{appx:hybrid}, less monopoles thus result in a diminished $\hat{q}$ in the $\chi_T^u$ scheme compared with the one in the $\chi_T^L$ scheme. Interestingly, (ii) and (iii)'s $\hat{q}$ still get close to the Super Yang-Mills (SYM) limit near $T_c$. Note that (ii) and (iii)'s $\mu_{E,M}$ behave very differently as shown in Fig.~\ref{fig:FL}(b), then a crucial observation one can draw is that among the nonperturbative $(r_d,\mu_E,\mu_M)$, the high energy jet transport property which is determined by the quenching parameter $\hat{q}(T)$, is sensitive to $r_d$, i.e. the rate at which confined colors are excited near $T_c$. Apart from such sensitivity, the near-$T_c$ enhancement of jet-medium interaction is a very robust feature in all schemes and is deeply tied with the nonconformal, nonperturbative dynamics near the transition temperature. It may be worth mentioning that a number of studies based on holographic QCD models~\cite{Li:2014hja,Rougemont:2015wca} that build in near-$T_c$ conformal-breaking effects, have universally found similar near-$T_c$ enhancement of the $\hat{q}/T^3$.  

A  surprising aspect of the comparison between CUJET2.0 and CUJET3.0 in Fig.~\ref{fig:FL-qhat-etas}, is that both models describe equally well the azimuthally averaged $R_{AA}$ data (which characterize overall suppression), in spite of their rather different functional forms of $\hat{q}_F(T)$. The two models differ from each other most significantly in the near-Tc regime:  the $\hat{q}_F/T^3$ of CUJET3.0 is much larger than that of CUJET2.0 for $T\sim (1-2)T_c$.   Above $T\sim 3T_c$ and by $T\sim 6T_c$, the $\hat{q}_F/T^3$ of CUEJT2.0 is $\sim 50\%$ larger than that of CUJET3.0. The overall energy loss is controlled by $\hat{q}_F$ (rather than $\hat{q}_F/T^3$) and therefore more dominated by contributions from  the high temperature QGP in the fireball. This explains   why both CUJET2.0 and CUJET3.0 are able to fit the overall $R_{AA}$. The azimuthal anisotropy $v_2$, on the other hand, is more sensitive to the late time contributions to energy loss coming from the lower temperature $T\sim (1-2)T_c$ part of the fireball. While CUJET2.0 fails to describe $v_2$, the CUJET3.0 successfully describes the $v_2$ data precisely by virtue of the strongly enhanced near-$T_c$ contributions due to the emergent color magnetic monopole degrees of freedom in the sQGMP. The contrast between the CUJET2.0 and CUJET3.0 demonstrates again the importance of simultaneous descriptions for both $R_{AA}$ and $v_2$ data in order to differentiate energy loss models. 

The fact that {\bf remarkably different $\hat{q}(T)$ dependence could be consistent with the same $R_{AA}$ data, demonstrates clearly  the inadequacy of focusing on the jet path averaged  quantity $\left\langle\hat{q}\right\rangle$ as the only relevant medium property to characterize jet energy loss. }Evidently while the $\left\langle\hat{q}\right\rangle$ captures the important transverse  ``kick'' factor, there are other essential factors like the actual chromo electric and magnetic composition of the plasma, the screening masses  and the running couplings at multiple scales which all strongly influence jet energy loss and imprint their effects  beyond just in the $\left\langle\hat{q}\right\rangle$.

In Fig.~\ref{fig:FL-qhat-etas}(d) the $\hat{q}$ extrapolated $\eta/s$ in scheme (i)(ii)(iii) following Eq.~\eqref{Effetas1} are plotted. Note that when $T<T_c$, as $T$ keeps cooling down $\eta/s$ rises due to the hadron resonance gas (HRG) contributions, as computed in \cite{NoronhaHostler:2008ju,Niemi:2011ix,NoronhaHostler:2012ug,Christiansen:2014ypa}. In terms of the total $\eta/s$ near $T_c$, both (ii) and (iii) has a clear minimum $(\eta/s)_{min}\sim 0.12$ at $T\sim 210$ MeV, while (i) has a $(\eta/s)_{min}\sim 0.02$ which is under the $\eta/s\sim 0.08$ quantum bound. This suggests (1) the liberation scheme of color DOFs, i.e. $r_d$, significantly influences the lower bound of $\eta/s$; (2) $(\eta/s)_{min}$ is insensitive to the screening masses $\mu_{E,M}$. If one strictly assumes $\eta/s(T=T_c)=0.08$, then the $r_d(T)$ should be in between $d\chi_T^L/dT$ and $d\chi_T^u/dT$. On the other hand, despite a $\Delta\eta/s\sim 0.15$ difference in the absolute magnitude, as temperature increases, (i) and (ii)'s $\eta/s$ rise at about the same rate, i.e. approximately identical $d(\eta/s)/dT$, and both are larger than the one in (iii). Notice that (i) and (ii) have different $r_d$, $\mu_{E}$ and $\mu_{M}$ but similar $\mu_{E}-\mu_{M}$ near $T_c$. This implies $d(\eta/s)/dT$ is sensitive to $\mu_{E}-\mu_{M}$ but is only limitedly affected by $r_d$ and the absolute values of $\mu_{E,M}$.

\begin{figure*}[!t]
\bc
\includegraphics[width=0.475\textwidth]{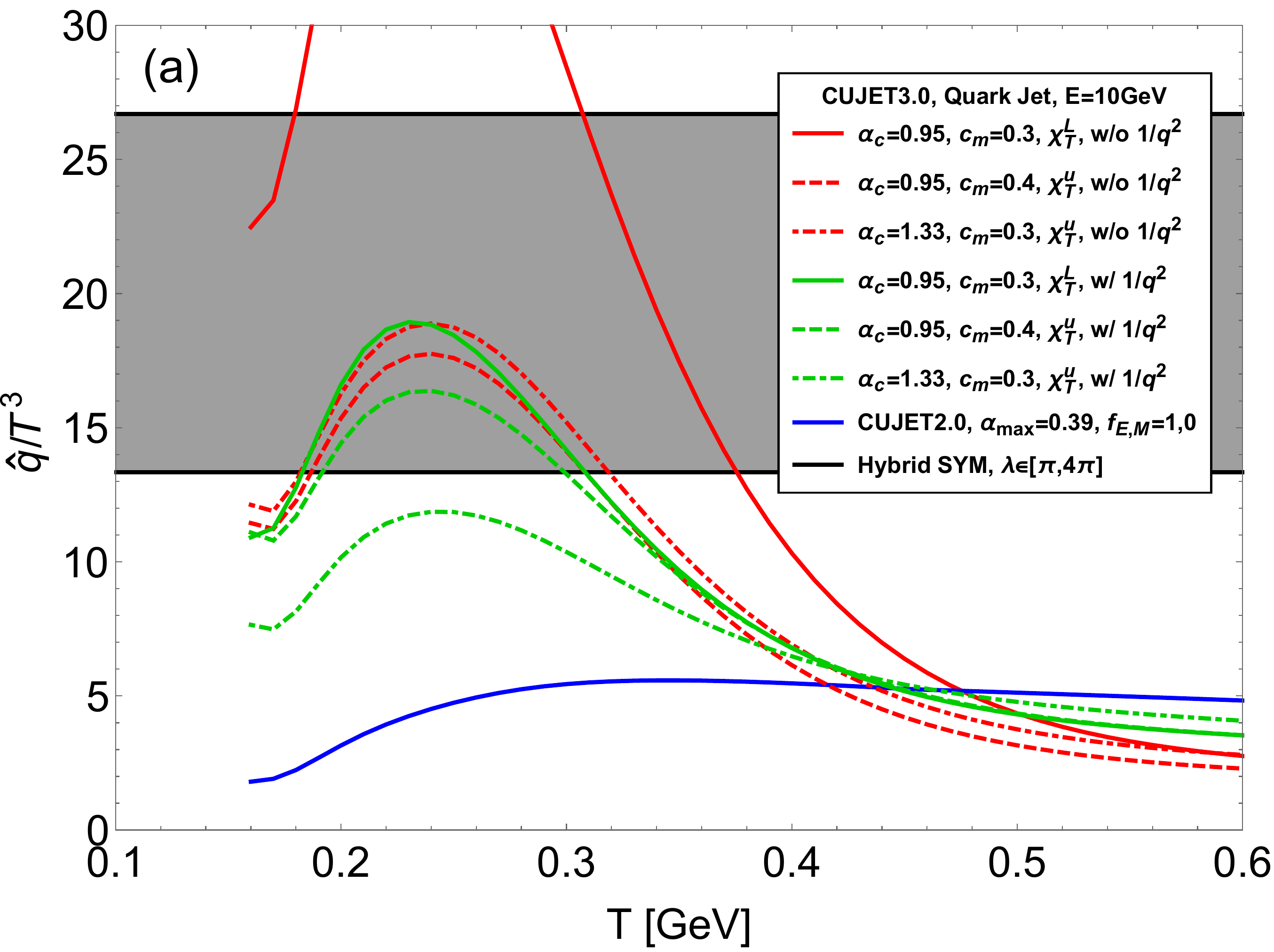}
\includegraphics[width=0.475\textwidth]{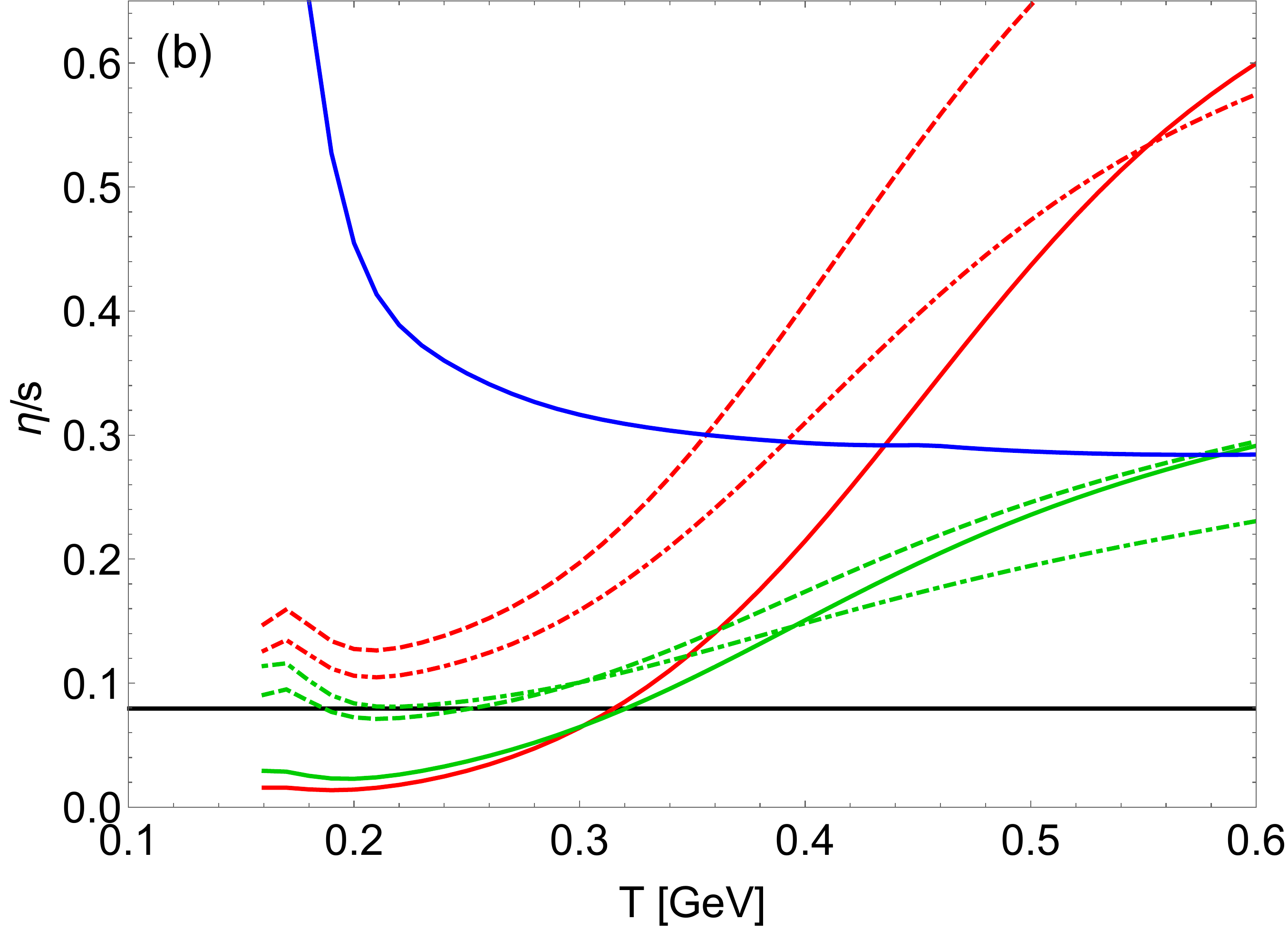}
\caption{\label{fig:Q2} 
(Color online)
(a) The temperature dependence of the dimensionless jet transport coefficient $\hat{q}/T^3$ for a light quark jet with initial energy $E=10$ GeV in the CUJET3.0 framework (Red) with scheme: (i) \eqref{SchI} (solid), (ii) \eqref{SchII} (dashed), and (iii) \eqref{SchIII} (dotdashed), compared with corresponding   $\hat{q}'$ (Green) as defined in Eq.~\eqref{Techqhat}. The CUJET2.0 $\hat{q}_F/T^3$ with $(\alpha_{max},f_E,f_M)=(0.39,1,0)$ (Blue) and the $\mathcal{N}=4$ SYM $\hat{q}_{SYM}/T^3=\frac{\pi^{3/2}\Gamma(\frac{3}{4})}{\Gamma(\frac{5}{4})}\sqrt{\lambda}$ \cite{Liu:2006ug} with 't Hooft coupling $\lambda\in[\pi,4\pi]$ (Black shaded) are plotted as references. (b) The shear viscosity to entropy density ratio $\eta/s$ estimated in the kinetic theory using extrapolation Eq.~\eqref{Effetas1} from $\hat{q}$'s in (a). Note that there is a clear $\eta/s$ minimum at $T\sim210$ MeV in the CUJET3.0 framework regardless of the schemes been chosen. The corresponding $(\eta/s)'$ (determined from $\hat{q}'$) converges to the pQCD weakly-coupled QGP limit at high temperature in (i)(ii)(iii) as expected. The $(\eta/s)_{min}$ in the fast liberations always sit above the quantum bound while in the Polyakov liberation it does not. In the near $T_c$ regime within the fast liberation schemes, the relative magnitude of $\eta/s$'s does not follow the naive inverse of the quark $\hat{q}_F$'s. This is because the computation of the $(\eta/s)'$ receives  enhanced contributions from softer scales that have stronger electric couplings, and consequently suppressing the transverse mean free path. 
}
\ec
\end{figure*}

\subsubsection{Alternative determination of jet transport coefficient}

The $\hat{q}$ computation above has followed the  previous CUJET3.0 prescription~\cite{Xu:2014tda} as given in Eq.~\eqref{Effqhat}, where the scattering kernel for the jet transport coefficient $\hat{q}$ is symmetric under inter-exchange of $E$ and $M$ in accord with   E-M duality considerations. There is however a subtle ambiguity: the form of scattering potential in Eq.~\eqref{Effqhat} differs from the scattering potential in the generalized kernel Eq.~\eqref{EMPotential} (as given in the second line of Eq.~\eqref{emEnergyLoss}) that is used in the actual CUJET3.0 modeling. The $1/\bq_\perp^2$ factors, present in Eq.~\eqref{emEnergyLoss} while absent in Eq.~\eqref{Effqhat},   increase the weight of soft momentum transfers in the computation of $\hat{q}$. It is important to examine the results for $\hat{q}$ and $\eta/s$ determined from the following alternative $\hat{q}'$ measure of the CUJET3.0 quenching field, and compare them with the computation from Eq.~\eqref{Effqhat}: 
\begin{eqnarray}
\hat{q}'_F(E,T)= & \int_0^{6ET} & d\bq_\perp^2  \frac{ 2\pi }{(\bq_\perp^2 + f_E^2 \mu^2(\bz))(\bq_\perp^2 + f_M^2 \mu^2(\bz))} \rho(T) \nonumber\\
&\times& \Big\{ \left[C_{qq} f_q + C_{qg} f_g \right]\cdot\left[  \alpha_s^2(\bq_\perp^2)\right]\cdot\left [f_E^2 \bq_\perp^2+ {f_E^2 f_M^2 \mu^2(\bz)} \right ] + \nonumber\\
&  &\;\;\; \left[ C_{qm} ( 1- f_q - f_g )\right]\cdot \left[ 1\right]\cdot \left[f_M^2{\bq_\perp^2} + {f_E^2 f_M^2 \mu^2(\bz)}\right]  \Big\}\;.
\label{Techqhat}
\end{eqnarray} 
The ``prime'' generalizations of quenching parameters of gluon and monopole jets follow straightforwardly. By substituting Eq.~\eqref{Techqhat} into Eq.~\eqref{Effetas1}, one can compute the corresponding  $(\eta/s)'$ in the quasi-particle picture according to kinetic theory. Fig.\ref{fig:Q2}(a) shows the temperature dependence of both $\hat{q}/T^3$ and $\hat{q}'/T^3$ for a light quark jet with initial energy $E=10$ GeV in the CUJET3.0 framework with scheme: (i) \eqref{SchI}, (ii) \eqref{SchII}, and (iii) \eqref{SchIII}. Fig.\ref{fig:Q2}(b) shows the corresponding comparison of $\eta/s$ and $(\eta/s)'$. There is a clear $\eta/s$ minimum at $T\sim210$ MeV in the CUJET3.0 framework in both ways of determining the quenching parameter. The  $(\eta/s)'$ nicely converges to the weakly-coupled HTL QGP limit at very high temperature $T>500$ MeV in all (i)(ii)(iii) schemes, as expected from Eq.~\eqref{emEnergyLoss} in the $\chi_T\rightarrow 1$ limit. Interestingly, for both estimates of $\eta/s$, the $(\eta/s)_{min}$ in the ``fast'' quark liberation schemes stay above the quantum bound while in the ``slow'' quark liberation scheme it does not. The general relations between $[$liberation schemes $+$ screening masses$]$ and   $[$$(\eta/s)_{min}$ $+$ $d(\eta/s)/dT$$]$ that one could infer from Fig.~\ref{fig:FL-qhat-etas} do not alter significantly for the $(\eta/s)'$ results. Within the fast liberation schemes, in the near $T_c$ regime, the relative magnitude of $\eta/s$'s do not follow the naive inverse of the quark $\hat{q}_F$'s. This is understandable since the computation of $(\eta/s)'$ from Eq.~\eqref{Techqhat} puts more weights on softer scales that have stronger electric couplings, given $\alpha_E\alpha_M =1$ at all scales. Thus the important EM scattering channel is not affected while the transverse mean free path will be suppressed due to larger EE scattering channel cross sections. Consequently one gets smaller $(\eta/s)'$ values as  compared with the $\eta/s$ values.

Overall, the lesson that one learns from Fig.~\ref{fig:FL-qhat-etas} is that $r_d$ and $\mu_E-\mu_M$ determines $(\eta/s)_{min}$ and $d(\eta/s)/dT$ respectively. Combined with the observations that one draws from Fig.~\ref{fig:FL-RAA-v2}, we can arrive at the following: \\
(1) If data of high $p_T$ light hadron (LH) and open heavy flavor (HF)'s $R_{AA}$ and $v_{2}$ at RHIC and LHC with sufficiently small uncertainties become available, then in the CUJET3.0 framework, after one has constrained the model parameters with LH's $R_{AA}$ and lattice calculations, one can use (1) LH's $v_2$ to estimate $\mu_{E}(T)-\mu_{M}(T)$ near $T_c$; (2) HF's $R_{AA}$ to constrain $r_d(T)$; (3) HF's $v_{2}$ to limit the absolute values of $\mu_{E}$ and $\mu_{M}$ as well as to further constrain $r_d(T)$. Even if the $(r_d(T),\mu_{E}(T),\mu_{M}(T))$ are not completely fixed in CUJET3.0, insights on $\hat{q}(T)$ and $\eta/s(T)$ can be drawn within this model because for jet energy loss in sQGMP, (a) $\hat{q}(T)$ is regulated by $r_d(T)$; (b) $d(\eta/s)/dT$ is shaped by $\mu_{E}(T)-\mu_{M}(T)$ near $T_c$; (c) $(\eta/s)_{min}$ is bounded by $r_d(T)$. \\
(2) In addition, the investigation of how the near $T_c$ physics affects the temperature dependence of the bulk viscosity $\zeta/s(T)$ as well as the role that $\zeta/s(T)$ plays on the experimental observables at RHIC and LHC is a topic of significant interests, there have been studies from the soft hydro sector \cite{Ryu:2015vwa}. Exploring the importance of $\zeta/s(T)$ from the hard jet quenching sector within CUJET3.0 will be explored elsewhere.

\subsection{Theoretical uncertainties related with monopole density constraints}

Thus far we have concentrated on using the total lattice QCD pressure, $p(T)$,
to constrain the chromo-magnetic-monopole (cmm) density assuming an ideal gas of chromo-electric-charged (cec) and cmm
degrees of freedom that leads to
\begin{equation}
\rho_m(T)= \xi_p p(T)/T - \rho_q(T) -\rho_g(T)\equiv \rho_m^{(PS)}(T)
\;\; .
\label{PS} \end{equation}
Where $\xi_p=1.012$ for a $N_c=3, N_f=2.5$ Stefan-Boltzmann gas as in Eq.~\eqref{RhoFromP}. We refer to Eq.~\eqref{PS} as the Pressure Scheme (PS) to fix the partial pressure
of magnetic monopoles from the total pressure minus the suppressed 
semi-QGP densities.

Such Pressure Scheme may ``suffer'' from  the potential or ``bag'' $B(T)$ contribution to thermodynamics whereby pressure
 $p= Ts/4 -B$ and energy density $\epsilon=3Ts/4+B$. In this case one would then have nonzero ``trace anomaly'' $\epsilon-3p=4B$ and indeed  lattice QCD data have shown the existence of such a ``bag'' contribution. In this regard, the entropy density $s=(\epsilon+p)/T$, free from any ``bag'' terms, may serve as a useful ``counting'' scheme for quasiparticle densities. We therefore also introduce an independent Entropy Scheme (ES) for determining the total number density via $\rho(T)=\xi_s s(T)/4$, where $\xi_s=\xi_p/4=0.253$ for a $N_c=3, N_f=2.5$ Stefan-Boltzmann gas, and examine the corresponding uncertainty in our modeling. 
 
\begin{figure*}[!t]
\bc
\includegraphics[width=0.475\textwidth]{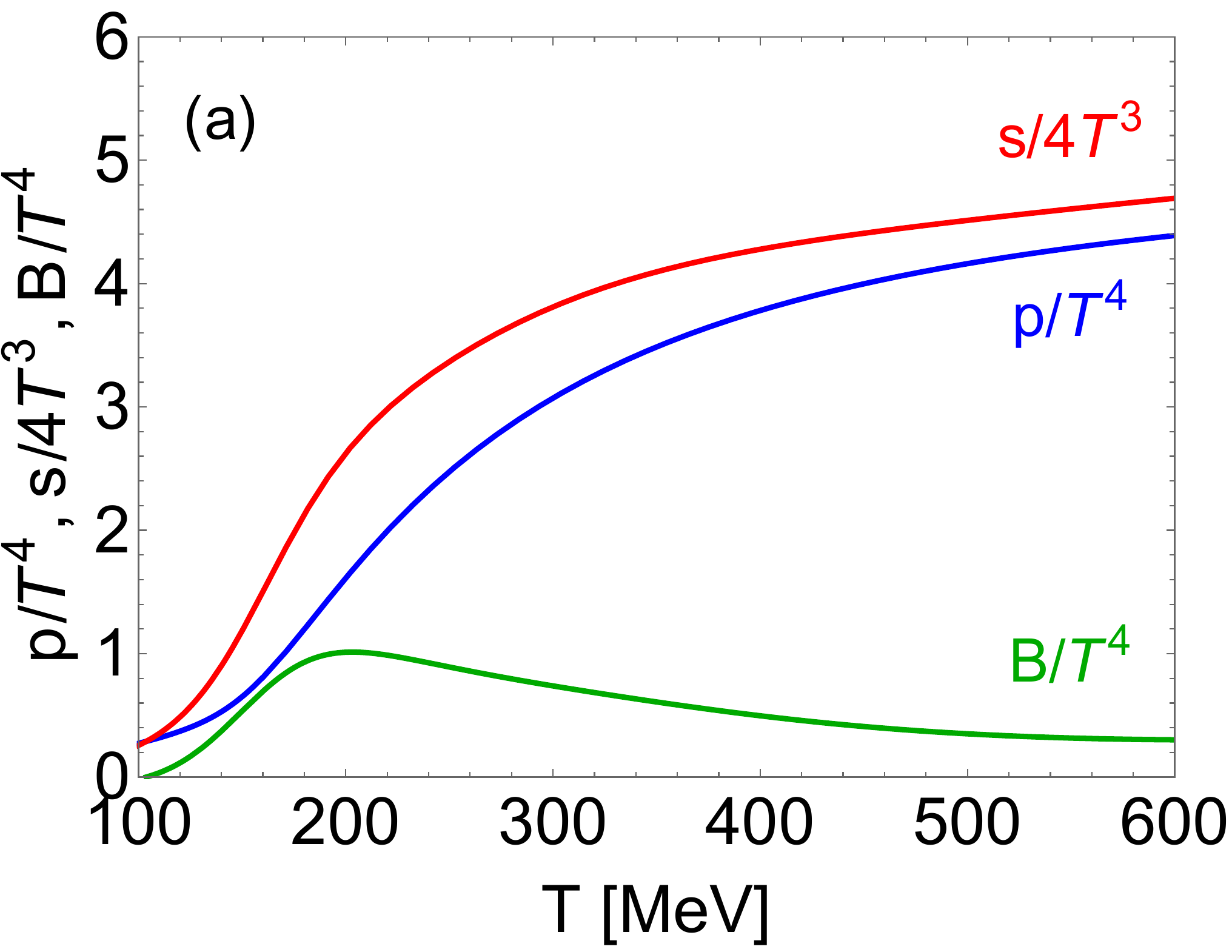}
\includegraphics[width=0.475\textwidth]{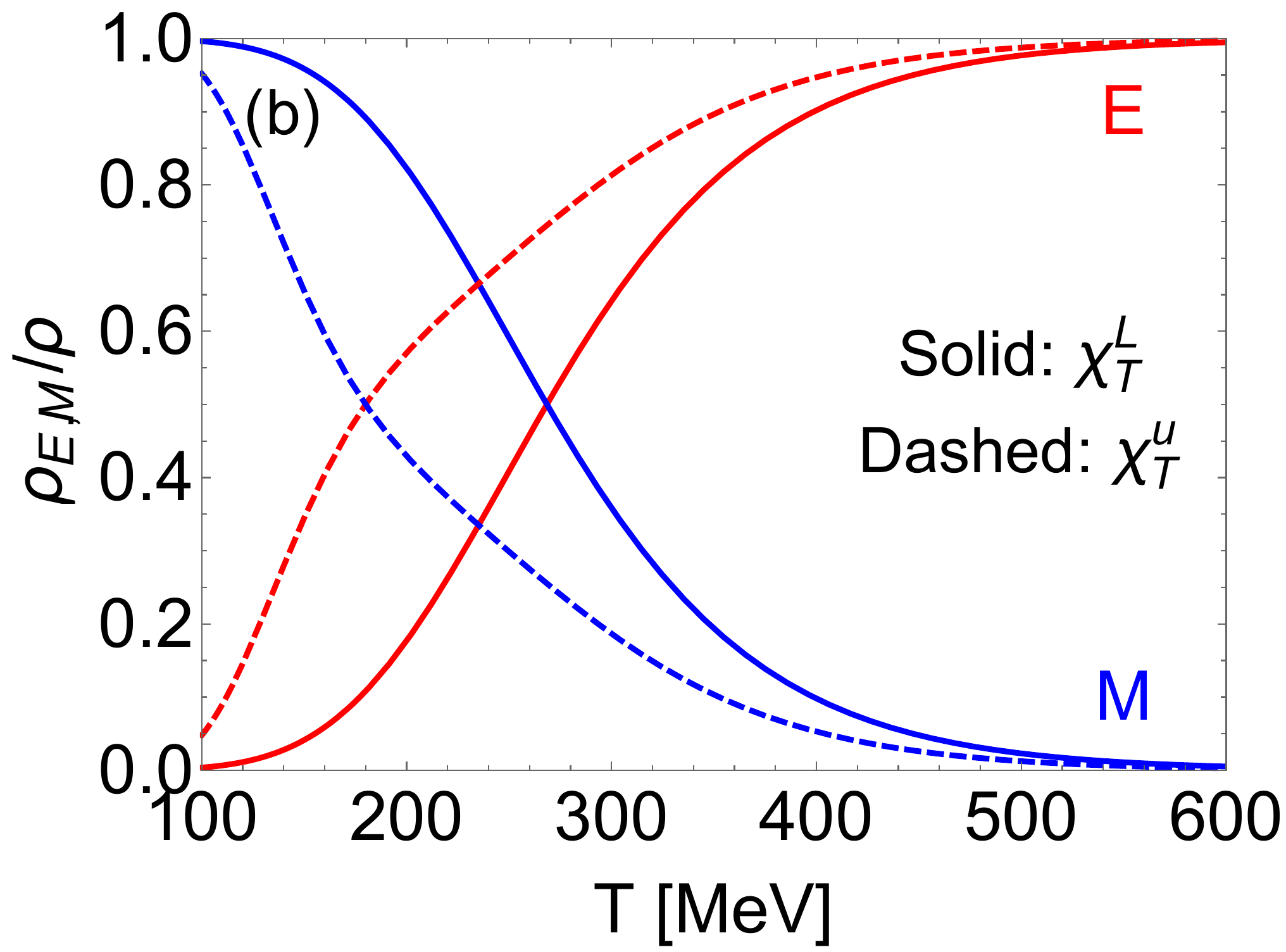}
\includegraphics[width=0.475\textwidth]{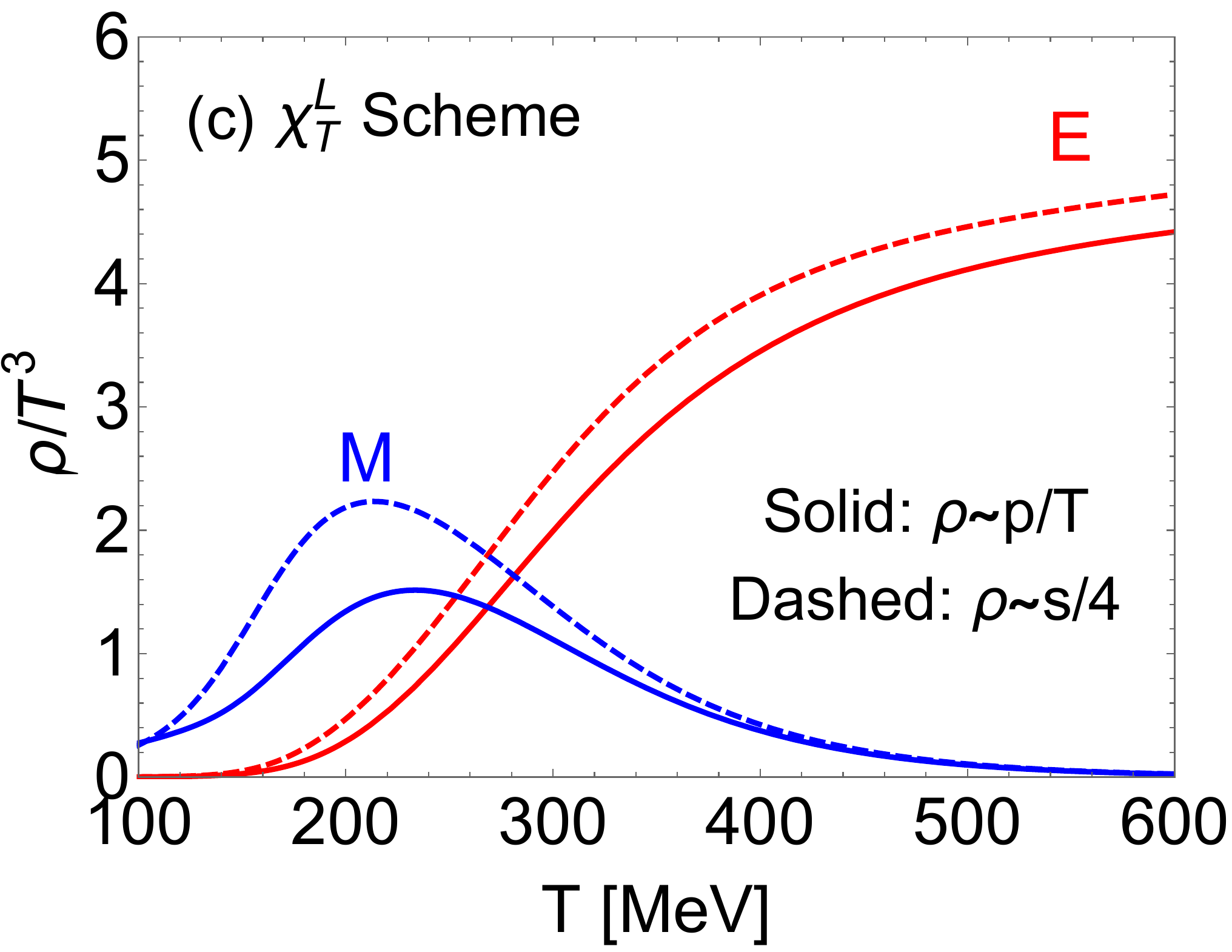}
\includegraphics[width=0.475\textwidth]{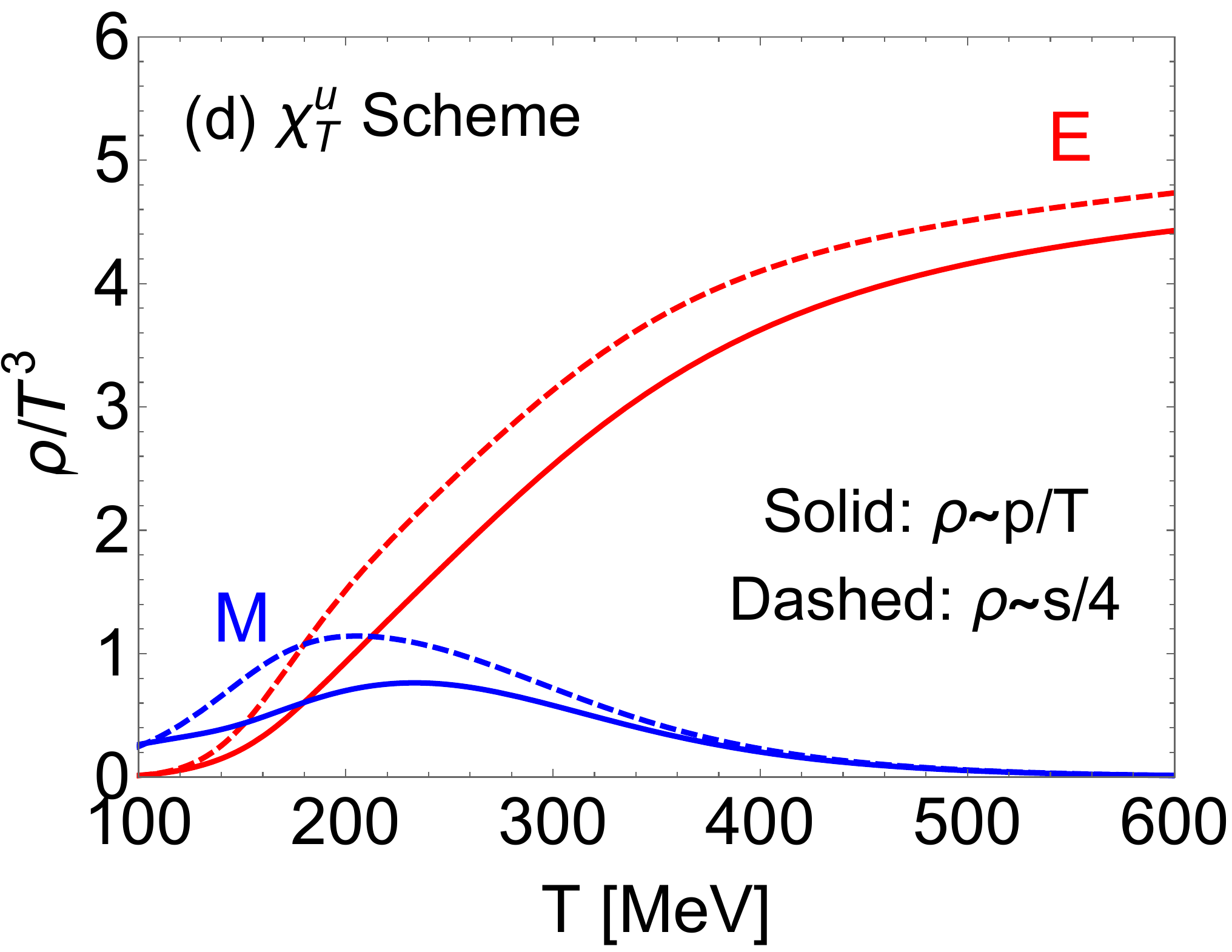}
\caption{\label{fig:PSES} 
(Color online)
(a) The effective ideal quasiparticle density, $\rho/T^3=\xi_p P/T^4$, in the Pressure Scheme (PS, Blue) is compared with effective density,  $\rho/T^3=\xi_p S/4T^3$, in the Entropy Scheme (ES, Red) based on fits to lattice data from HotQCD Collaboration \cite{Bazavov:2014pvz}. The difference is due to an interaction 'bag'' pressure $-B(T)/T^4$ (Green) that encodes  the QCD 
conformal anomaly $\epsilon-3p\ne 0$. (b) The density fraction of the electric (E, red) and magnetic (M, blue) degrees of freedom in the $\chi_T^L$ (solid, Eq.~\eqref{chiTL}) and $\chi_T^u$ (dashed, Eq.~\eqref{chiTu}) liberation scheme. The dimensionless E and M density $\rho/T^3$ in the two schemes are shown in (c)$\chi_T^L$ and (d)$\chi_T^u$ respectively, where both the $\rho/T^3$ in the PS (solid) and ES (dashed) are plotted. In both liberation schemes, the $\rho_m$ in the ES near $T_c$ is around twice the $\rho_m$ in the PS.
}
\ec
\end{figure*}

\begin{figure*}[!t]
\bc
\includegraphics[width=0.45\textwidth]{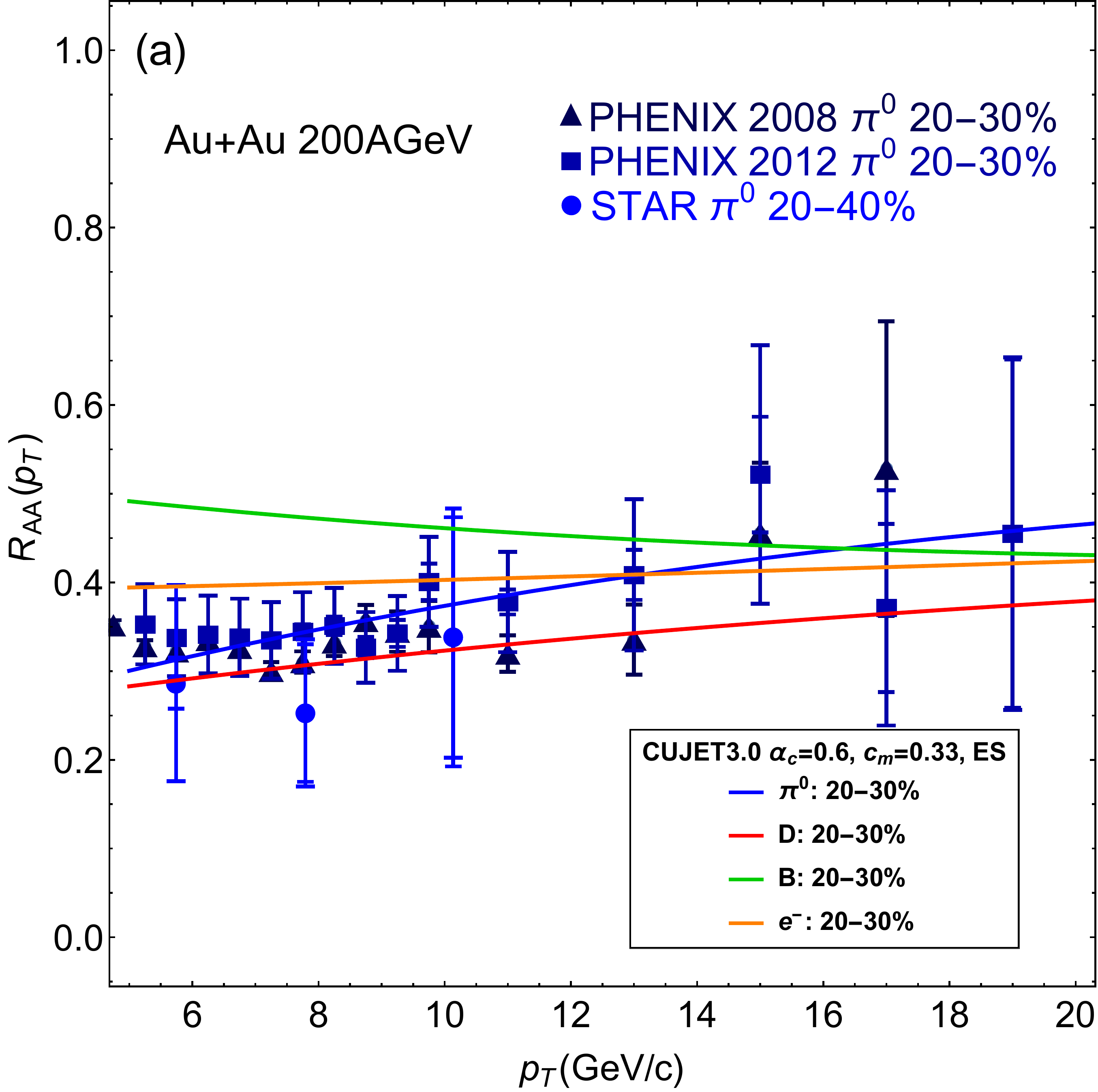}
\includegraphics[width=0.45\textwidth]{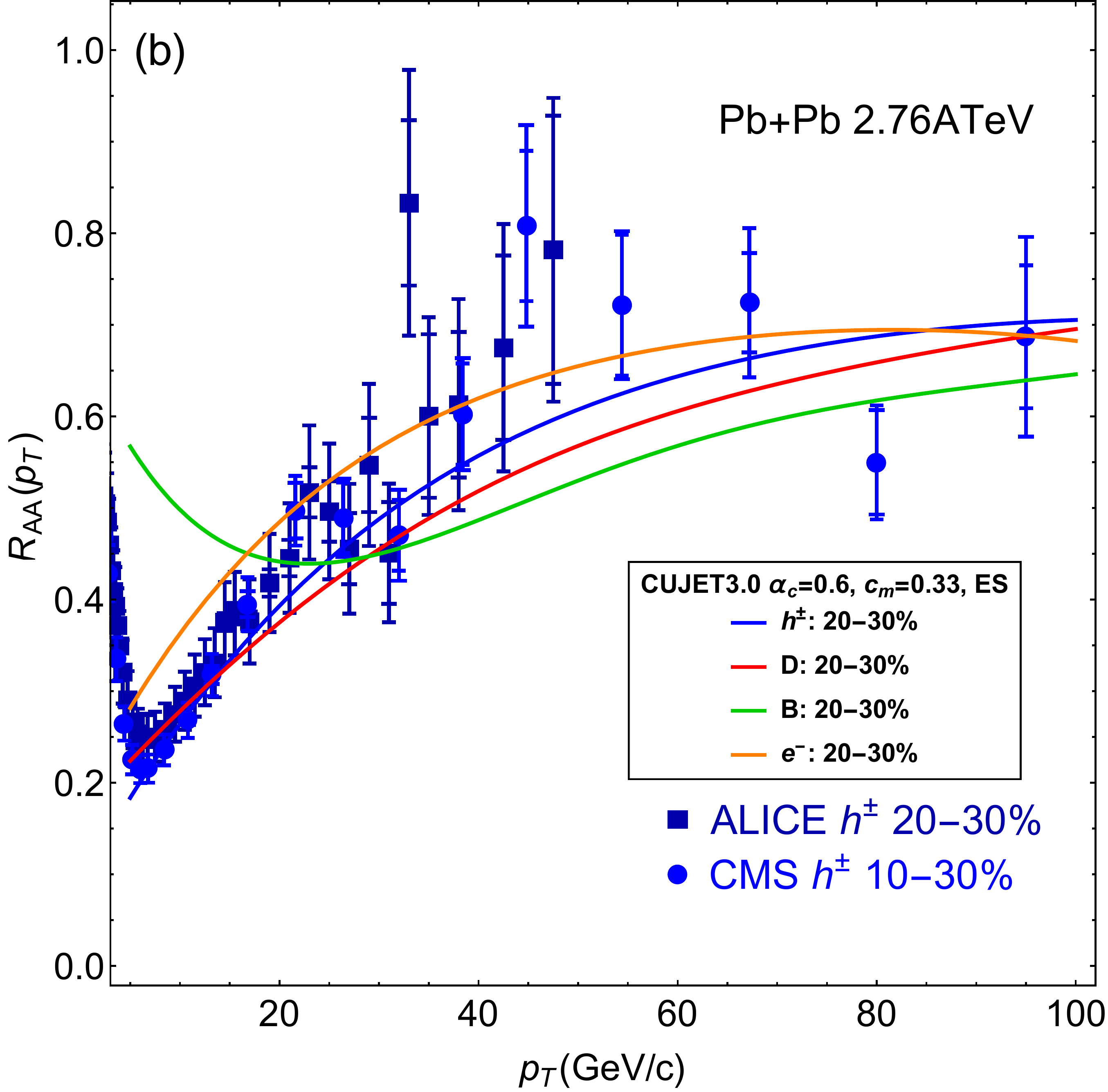}
\includegraphics[width=0.45\textwidth]{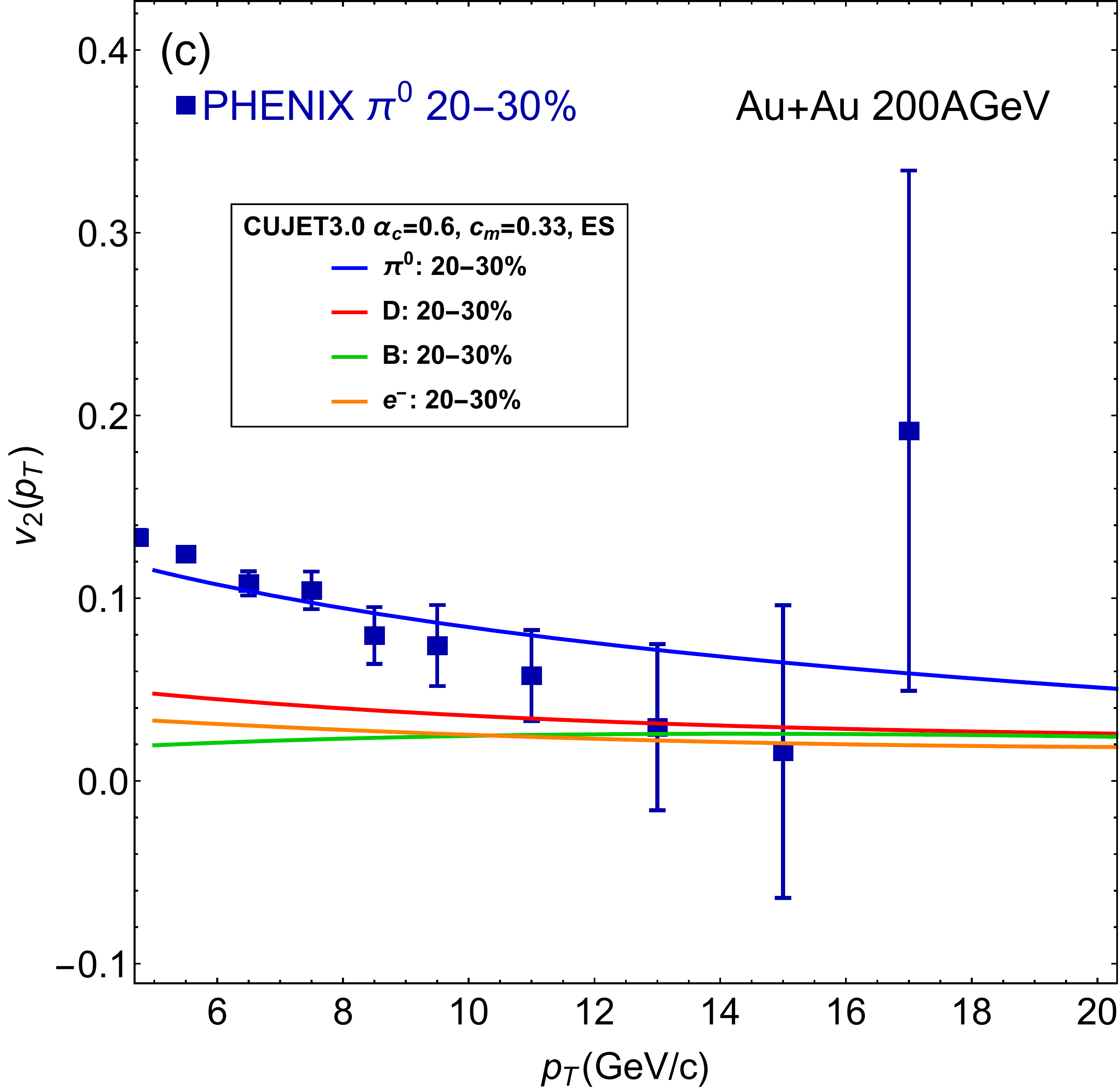}
\includegraphics[width=0.45\textwidth]{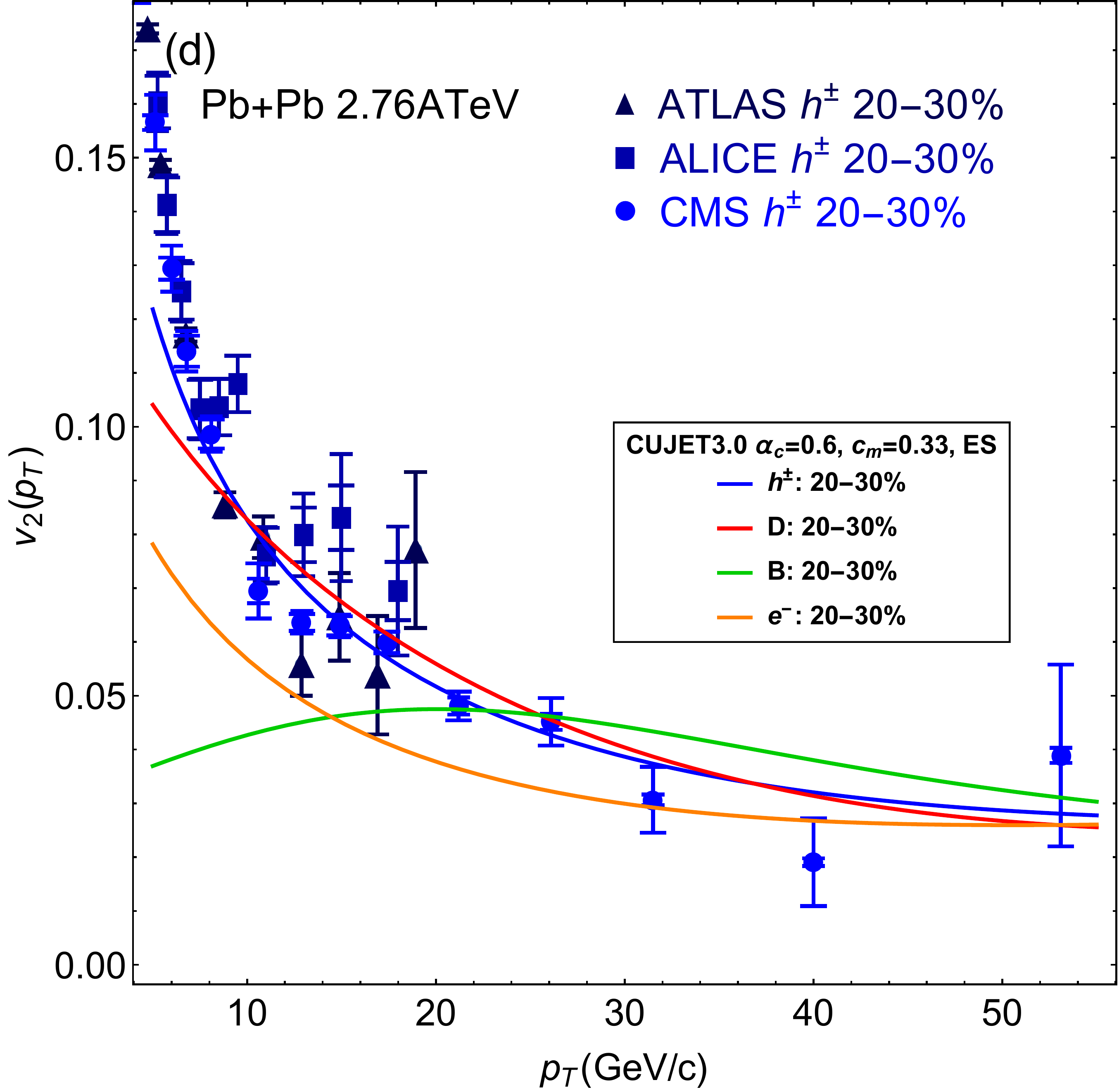}
\caption{\label{fig:ES} 
(Color online)
In the CUJET3.0 model with Entropy Scheme (ES) and $\chi_T^L$ liberation, the $\alpha_c$ and $c_m$ is adjusted to 0.6 and 0.33 to fit to the reference datum at LHC $R_{AA}^{h^\pm}(p_T=12.5{\rm GeV})\approx0.3$ as well as the lattice $\mu_M$ (\cite{Nakamura:2003pu}, c.f. Fig.~\ref{fig:FL}(b)). With this parameter setup, the $\pi^0/h^\pm$'s high $p_T$ $R_{AA}$ and $v_{2}$ at RHIC Au+Au 200GeV and LHC Pb+Pb 2.76TeV 20-30\% collisions are in perfect agreements with data \cite{Adare:2008qa,Adare:2010sp,Adare:2012wg,Abelev:2009wx,Abelev:2012di,Abelev:2012hxa,ATLAS:2011ah,CMS:2012aa,Chatrchyan:2012xq}. The result of prompt D meson, B meson, and heavy flavor decay $e-$ results in the ES scheme is plotted in red, green, and orange, respectively.
}
\ec
\end{figure*}

\begin{figure*}[!t]
\bc
\includegraphics[width=0.45\textwidth]{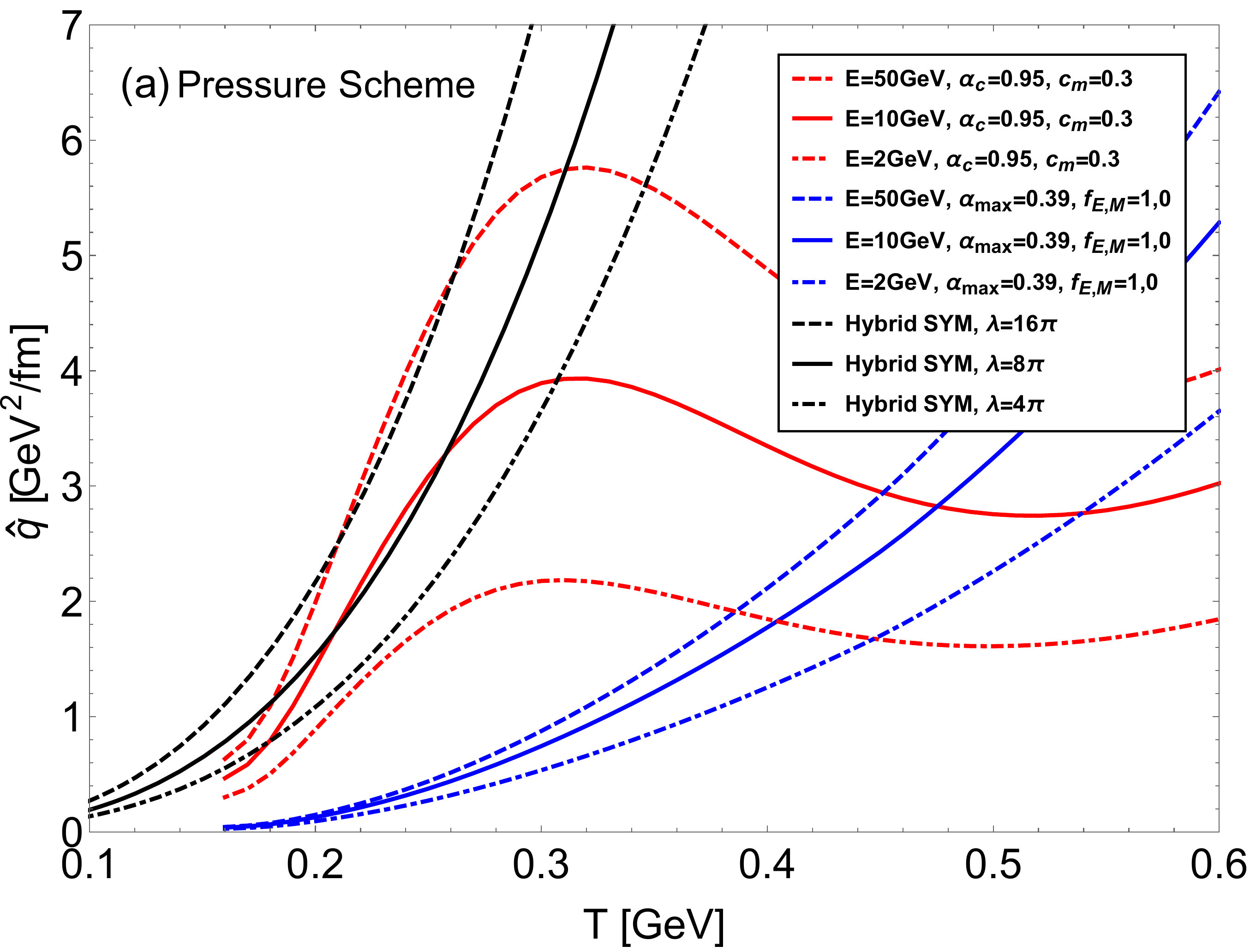}
\includegraphics[width=0.45\textwidth]{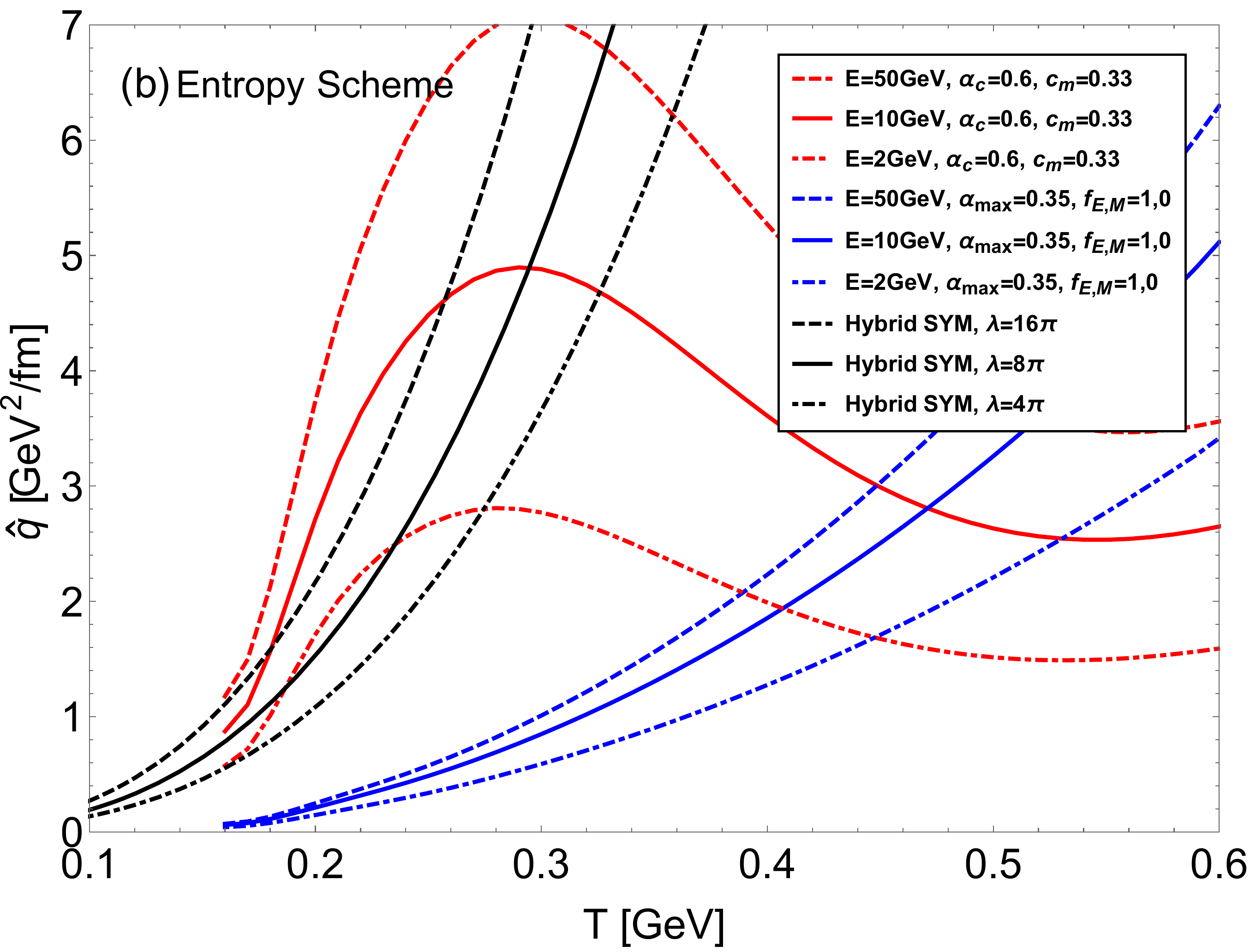}
\includegraphics[width=0.45\textwidth]{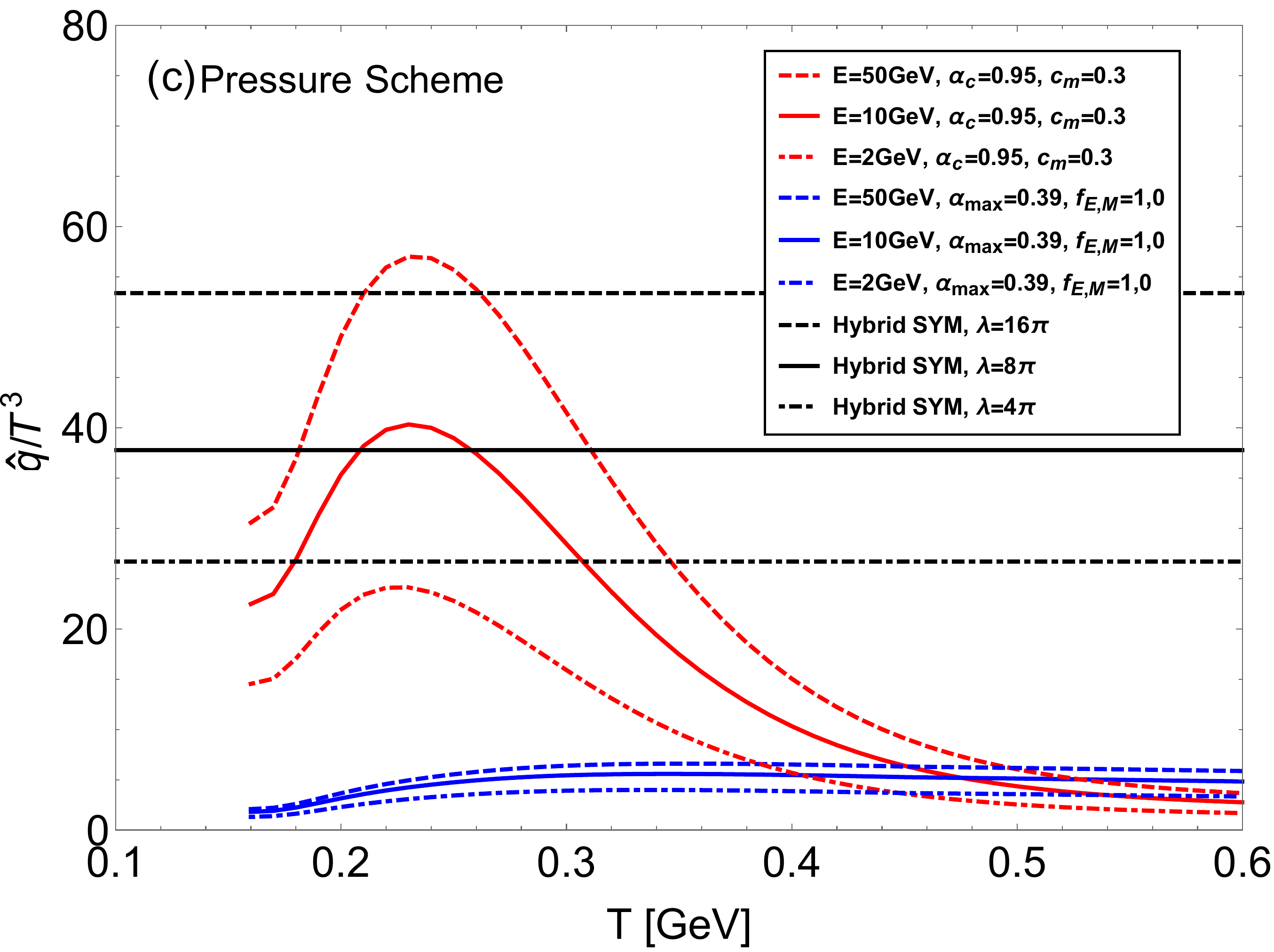}
\includegraphics[width=0.45\textwidth]{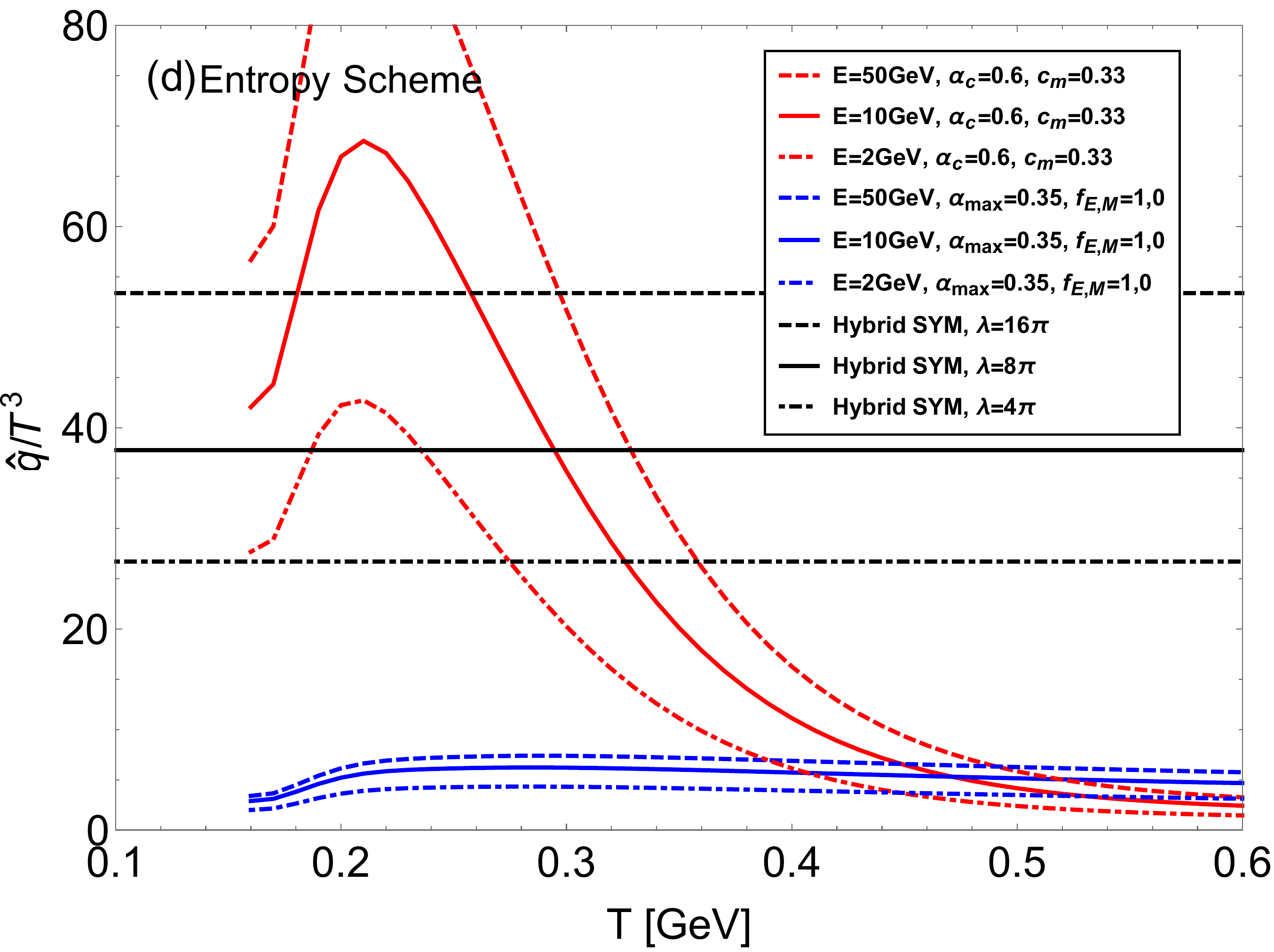}
\caption{\label{fig:PvsS} 
(Color online). (a) The $\hat{q}(T)$ for a hard quark with initial energy $E=$ 2 GeV (dotdashed), 10 GeV (solid), 50 GeV (dashed) computed according to Eq.~\eqref{Effqhat} in the Pressure scheme, for CUJET3.0 (red), CUJET2.0 (blue) and $\mathcal{N}=4$ SYM (black). The dimensionless $\hat{q}(T)/T^3$ is plotted in (c). (b)(d) The counterpart of (a)(c) in the Entropy scheme. Note that $(\alpha_c,c_m)$ in CUJET3.0 and $(\alpha_{max},f_E,f_M)$ in CUJET2.0 has been readjusted to $(0.6,0.33)$ and $(0.35,1,0)$ respectively fit to the LHC reference datum (cf. Fig.~\ref{fig:FL-RAA-v2}). The $\hat{q}$ in the ES near $T_c$ is $\sim 50\% $ larger than in the PS due to the ``bag'' contribution.
}
\ec
\end{figure*}

As can be seen from Fig.\ref{fig:PSES}, in the Entropy Scheme (ES) the total quasiparticle density is higher than in the Pressure Scheme (PS) due to the bag constant: 
\begin{equation}
\rho_m^{(ES)}(T)\equiv \xi_s s(T) - \rho_q(T) -\rho_g(T)= \rho_m^{(PS)}(T) + \xi_p B(T)/T
\;\; .
\label{ES} \end{equation}
Choosing the ES vs the PS scheme to fix the monopole density would increase the monopole density near $T_c$ by a factor of 2 and increases the $\hat{q}$ by approximately 50\% near $T_c$. 
To fit the same reference path averaged $R_{AA}$ we would need to adjust the $(\alpha_c,c_m)$ in CUJET3.0 and the $(\alpha_{max},f_E,f_M)$ in CUJET2.0 somewhat respectively. Fig.\ref{fig:PvsS} shows the absolute $\hat{q}$ and the dimensionless $\hat{q}/T^3$ in the two schemes. After readjusted $(\alpha_c,c_m)$ in CUJET3.0 and $(\alpha_{max},f_E,f_M)$ in CUJET2.0 to $(0.6,0.33)$ (as shown in Fig.~\ref{fig:ES}) and $(0.35,1,0)$ respectively, the $\hat{q}$ near $T_c$ in ES is around 50\% larger than in the PS.

\begin{figure*}[!t]
\bc
\includegraphics[width=0.475\textwidth]{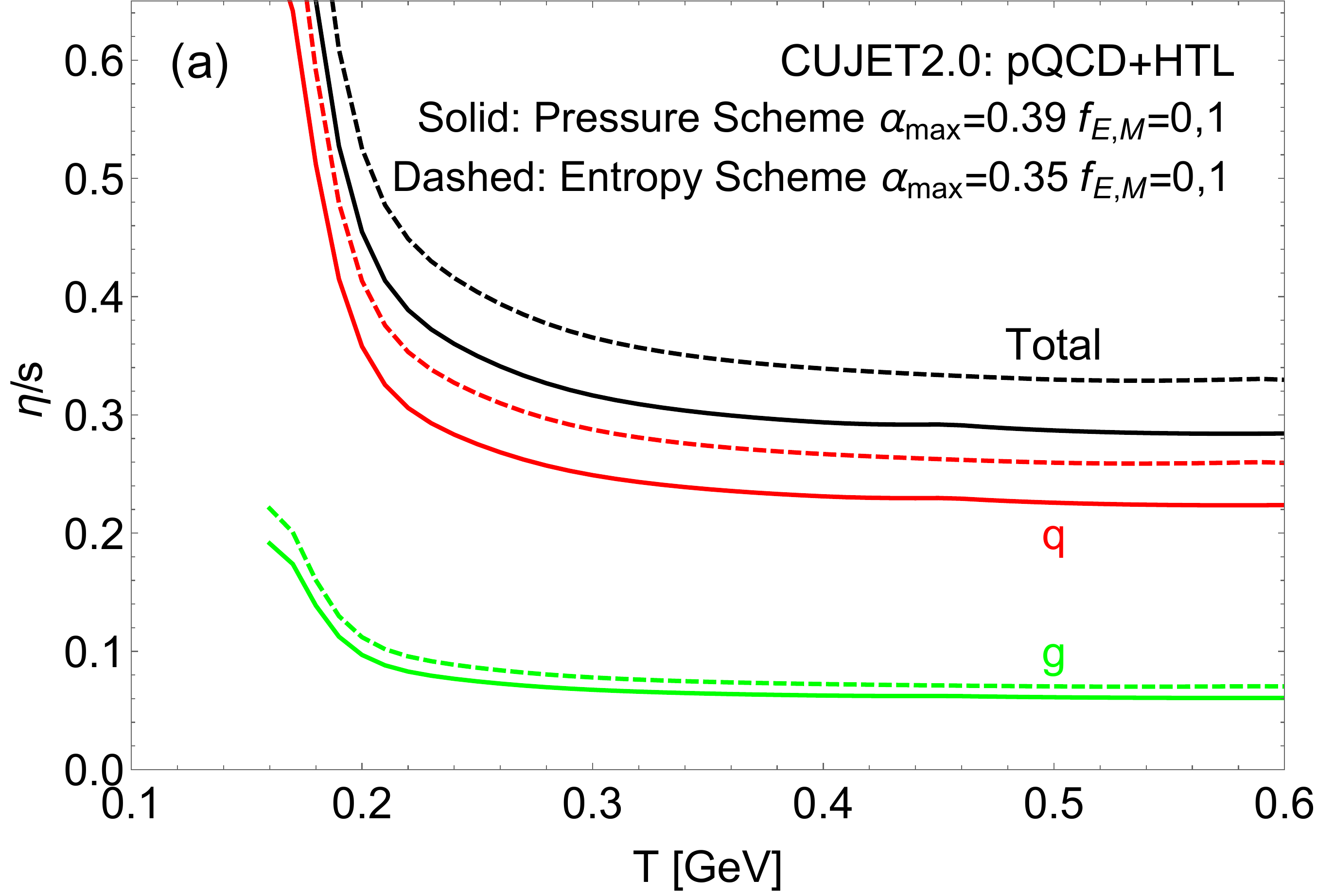}
\includegraphics[width=0.475\textwidth]{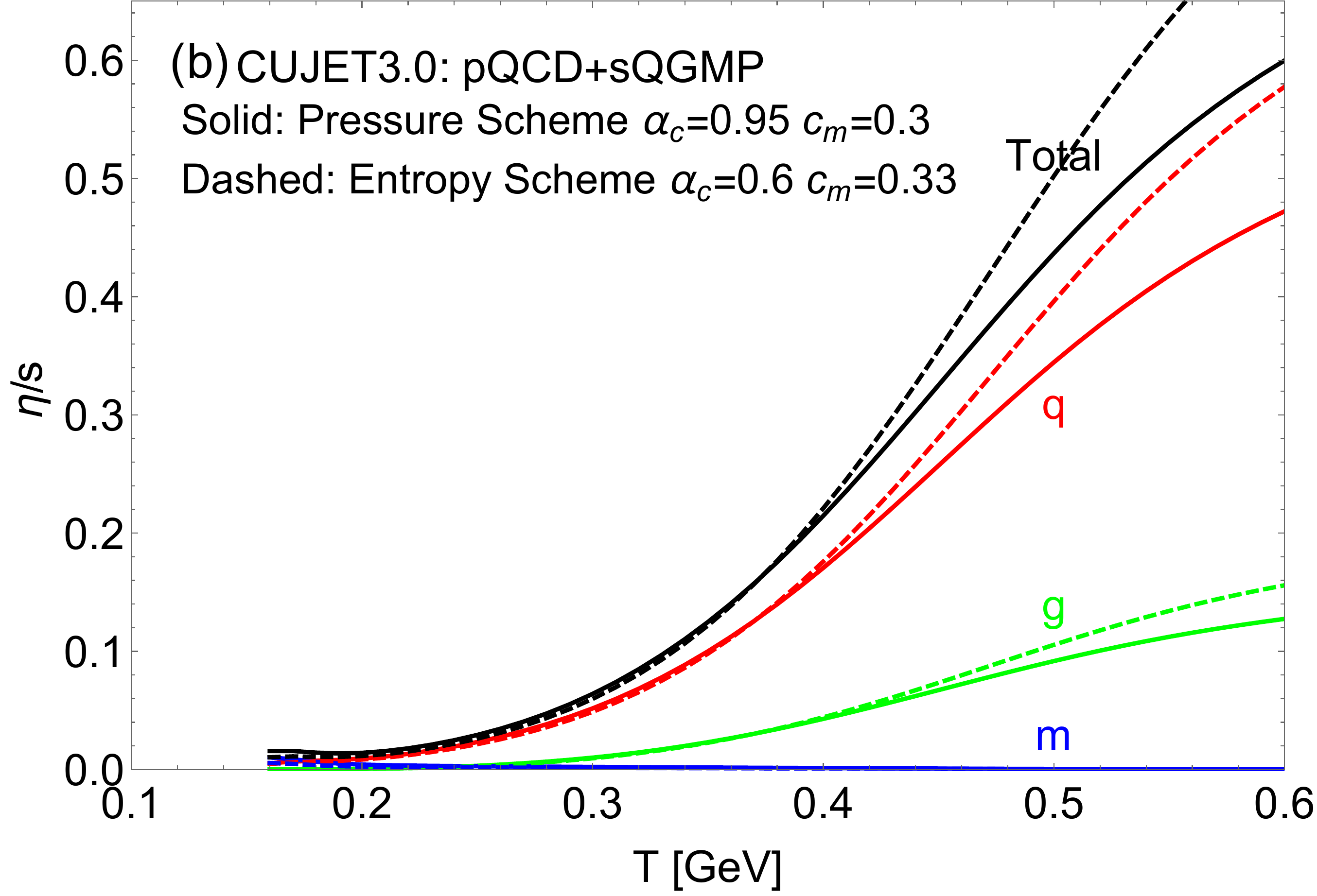}
\caption{\label{fig:2vs3} 
(Color online). (a) The $\eta/s$ in CUJET2.0 for quasi-quarks (q, red), quasi-gluons (g, green), and its total value (black) computed from inversing the $\hat{q}/T^3$ according to Eq.~\eqref{Effetas1}. The solid lines correspond to the PS scheme, while the dashed lines correspond to the ES scheme. (b) The counterpart of (a) in CUJET3.0. Note that the addition of the monopole (m, blue) like quasi-particle degrees of freedom in sQGMP does not alter the overall $\eta/s$ significantly since the strong magnetic coupling shrinks the transverse mean free path for monopoles and suppresses the shear viscosity contributions from monopoles. Since the sQGMP is dominated by monopole degrees of freedom near $T_c$, the total $\eta/s$ in the PS and ES scheme then should naturally converge to the same value. 
}
\ec
\end{figure*}

The $\eta/s$ in CUJET2.0 and 3.0 computed from inversing the $\hat{q}/T^3$ according to Eq.~\eqref{Effetas1} are plotted in Fig.\ref{fig:2vs3}. One sees that the $\eta/s$ in the ES scheme hardly deviates from its value in the PR scheme. This is understood since $\eta/s$ is dominated by the ``free'' 
quasi-quark degrees of freedom. The monopole fluid is almost  viscous free
in either scheme as one has already seen in Fig.\ref{fig:FL-qhat-etas}.  

From this consideration we see that requiring the consistency of hard and soft probes can only determine
a lower bound on the monopole density near $T_c$. Our default PR scheme
is above that lower bound. We leave the search for that lower bound
to a future study.

\section{Conclusions}
\label{sec:conclusion}

We have presented a detailed study of the jet energy loss in semi-Quark-Gluon-Monopole-Plasmas (sQGMP), within the new CUJET3.0 framework  of jet quenching in bulk constrained
  (VISH2+1D) viscous hydrodynamic backgrounds  by
  extending the perturbative QCD based (CUJET2.0) model  
  to include possible non-perturbative chromodynamical features of the QCD
  confinement phase transition near $T_c\approx 160$ MeV.  
  We test the robustness and consistency
  of this new model by
  comparing predictions of the leading hadron nuclear modification
  factor, $R_{AA}(p_T>10{\rm GeV/c},\sqrt{s})$, and its azimuthal elliptic
  asymmetry $v_2(p_T>10{\rm GeV/c},\sqrt{s})$ with available data from nuclear
  collisions at RHIC($\sqrt{s}=0.2$ ATeV) and LHC($\sqrt{s}$=2.76
  ATeV).  
    The sQGMP model
  depends on two parameters: (1) the value of the QCD running coupling
  $\alpha_c \approx 0.95-1.33$ at low $Q<T_c$ and (2) the ratio $c_m=
  g\mu_E/\mu_M$ of nonperturbative electric to magnetic screening
  scales. We study three specific cases, $c_m=0, 0.3,0.4$.  The value
  of $\alpha_c$ is fixed for each case by fitting a single reference
  datum, $R^{ch}_{PbPb}(p_T=12{\rm GeV/c})\approx 0.3$ at LHC at 20-30\%
  centrality.  Consistency with all available data is then tested
  comparing predictions of CUJET3.0 on $R^h_{AA}(p_T )$ {\em and}
  $v^h_2(p_T)$, for $h=\pi,D,B$ at both RHIC and LHC.  The emergent
  chromomagnetic degrees of freedom in the sQGMP model near $T_c$ are shown to
  solve efficiently the long standing $R_{AA}$ vs $v_2$ puzzle by
  leading to a broad maximum of the jet quenching parameter
  $\hat{q}(E,T)/T^3$ between $(1-2)T_c$. In addition and most
  remarkably, by extrapolation of the sQGMP $\hat{q}$
  down to thermal energy $E\sim 3 T$ scales  one finds the 
  shear viscosity to entropy density ratio $\eta/s \approx T^3/\hat{q} \sim
  1/4\pi$ to be near the unitary bound, in the critical $(1-2) T_c$
  transition temperature range, which is consistent with viscous hydrodynamic
  fits to bulk azimuthal harmonics $v_n$ at $p_T<2 {\rm GeV/c}$.  
  A main goal of the present study is to test the robustness of such phenomenological success in CUJET3.0 with respect to a number of theoretical uncertainties in the model implementation.

A key theoretical uncertainty of this model is the rate at which color electric degrees of freedom are liberated near the deconfinement transition temperature $T_c$. We have investigated three very different schemes of color liberation and found that such rate  does not alter the model's agreements with high $p_T$ single light hadrons (LH)' $R_{AA}$. However, this rate ($r_d\equiv d\chi_T/dT$) considerably influences the temperature dependence of the electric screening mass. It is found that the light hadrons' $v_{2}(p_T>10{\rm GeV})$ is an observable that is sensitive to the difference between the electric (E) and magnetic (M) screening mass ($\mu_{E,M}$) near $T_c$. For open heavy flavors (HF), $r_d$ regulates their high $p_T$ $R_{AA}$'s in CUJET3.0, and their $v_{2}$'s are affected by both the screening masses and the $r_d$. In terms of the jet quenching parameter $\hat{q}$, it is influenced by $r_d$, but is insensitive to the screening masses. On the other hand, for the shear viscosity to entropy density ratio $\eta/s$, $(\eta/s)_{min}$ and $d(\eta/s)/dT$ would vary strongly  if $r_d$ and $\mu_E-\mu_M$ changes.

Therefore, for jet quenching in sQGMP from the CUJET3.0 framework, after model parameters are constrained by data of LH's $R_{AA}$ and relevant lattice calculations, the rate of deconfinement $r_d$ and the screening masses $\mu_{E,M}$ affect jet fragments observables in different ways: (1) $\mu_{E}-\mu_{M}$ near $T_c$ influences LH's $v_2$; (2) $r_d$ influences HF's $R_{AA}$; (3) $r_d$ and absolute values of $\mu_{E,M}$ influence HF's $v_{2}$. On the other hand, the CUJET3.0's jet transport coefficient $\hat{q}$ and the shear viscosity $\eta/s$ are affected by $r_d$ and $\mu_{E,M}$ differently: (a) $\hat{q}(T)$ is constrained by $r_d(T)$; (b) $d(\eta/s)/dT$ is constrained by $\mu_{E}(T)-\mu_{M}(T)$ near $T_c$; (c) $(\eta/s)_{\rm min}$ is constrained by $r_d(T)$. Given our findings of these dependences, it is expected that  such model uncertainties can be significantly reduced by experimental input such as future high precision data  for  LH and HF's  $R_{AA}$ as well as  $v_{2}$ at both RHIC and LHC. For example, the HF's $R_{AA}$ limits $r_d(T)$ therefore constrains $\hat{q}(T)$ and $(\eta/s)_{\rm min}$, LH's $v_{2}$ limits $\mu_{E}-\mu_{M}$ near $T_c$ therefore constrains $d(\eta/s)/dT$, while the two can be combined to infer the functional shape of $\eta/s(T)$; in addition, HF's $v_{2}$ can be utilized to constrain the absolute magnitude of $\mu_E(T)$ and $\mu_M(T)$.

Additional discussions on three aspects of the present model have also been included  in the appendices. The first is about different relativistic corrections from viscous hydrodynamical flows which are found not to  affect  either $R_{AA}$ or $v_{2}$ for light hadrons. This is as expected since under eikonal approximation, the number of parton-medium scatterings is fixed in any frame once the initial jet production coordinate and the azimuthal jet propagation angle are specified. 
The second aims to identify which one of the various ingredients in CUJET3.0  makes the most significant  contribution toward obtaining a successful description of the high $p_T$ azimuthal elliptical harmonics data. We find that  the emergent chromo magnetic monopoles play the decisive role.  As long as monopoles are present in the near $T_c$ regime, $v_{2}$ is insensitive to the detailed form of the running coupling $\alpha_s(Q^2)$, provided that lattice screening masses can be reasonably described. The last one is about the path length ($L$) dependence of the light/heavy quark energy loss, which is an informative dynamical feature of jet quenching models. It is found that  both CUJET3.0 and CUJET2.0 converge to be around the pQCD/elastic limit at high temperatures $T\gtrsim400$ MeV. However as $T$ drops, starting from $T\sim 300$ MeV, chromo-magnetic monopoles gradually dominate the medium. Consequently we find that the energy loss's dependence on $L$ starts to deviate from pQCD/elastic toward the AdS/CFT-like strong coupling behavior. 
It is noteworthy that   the ``dead cone'' suppression is not altered by the nonperturbative sQGMP near $T_c$. It is also found that the modification on the $L$ dependence of light and heavy quark energy loss caused by the transition from high-T QGP to near-$T_c$ sQGMP are quite similar.

In summary, with our detailed investigations of the jet quenching phenomena in the sQGMP model, we conclude that the phenomenological consequences of the sQGMP, for both hard and soft probes in heavy ion collisions, stay very robust with respect to certain important systematic theoretical uncertainties. The phenomenological success includes simultaneous descriptions of  all available  high $p_T$  $R_{AA}$ and $v_2$ data at both RHIC and  LHC, as well as providing an intuitive dynamical mechanism that may explain how the shear viscosity to entropy density ratio could approach the $1/4\pi$ unitary
  bound in the vicinity of $T_c$ as required 
to explain the bulk low $p_T<2$ GeV
  ``perfect fluidity''. The sQGMP model therefore provides a first model that consistently accounts for both hard and soft transport properties  of the new state of QCD matter discovered in high energy  nuclear collisions at RHIC and LHC.

We end by emphasizing again the fact that  remarkably different $\hat{q}(T)$ dependence could be consistent with the same $R_{AA}$ data, demonstrates clearly  the inadequacy of focusing on the jet path averaged  quantity $\left\langle\hat{q}\right\rangle$ as the only relevant medium property to characterize jet energy loss. Evidently while the $\left\langle\hat{q}\right\rangle$ captures the important transverse  ``kick'' factor, there are other essential factors like the actual chromo electric and magnetic composition of the plasma, the screening masses  and the running couplings at multiple scales which all strongly influence jet energy loss and imprint their effects  beyond just in the $\left\langle\hat{q}\right\rangle$. It is of significant interest and importance to thoroughly exploit these multiple  facets of jet physics and the opportunities they offer for progressing toward an ultimate understanding of the microscopic making of the sQGP.

\begin{acknowledgments}
We thank especially Peter Petreczky for critical and insightful discussions. JX is grateful to Gabriel Denicol, Dima Kharzeev, Zhe Liu, Rob Pisarski, Chun Shen and Xin-Nian Wang for helpful conversations. The research of JX and MG is supported by U.S. DOE Nuclear Science Grants No. DE-FG02-93ER40764.
The research of JL is supported by the National Science Foundation (Grant No. PHY-1352368). JL also acknowledges partial support from the RIKEN BNL Research Center.
\end{acknowledgments}
\\

\appendix

\section{Relativistic flow corrections to jet energy loss}
\label{appx:flow}

Generally speaking, in pQCD based energy loss models, the non-Abelian bremsstrahlung amplitudes are derived assuming a static QGP medium \cite{Baier:2000mf,Gyulassy:2003mc,Kovner:2003zj,Jacobs:2004qv,Armesto:2011ht}. With a hydrodynamically evolving background, flows move at relativistic velocities in certain cells at certain time; if a light-like high-$p_T$ parton happens to pass through a unit cell flowing close to the speed of light, since the radiative formulas are applicable only in the co-moving frame, then one should boost back to the lab framework for correct predictions of jet quenching observables. Therefore, modifications on a pQCD energy loss theory are necessary if it is coupled to a hydrodynamically expanding medium.

In \cite{Baier:2006pt,Liu:2006he}, the relativistic flow corrections to $\hat{q}$ are calculated using weak and strong coupling approaches. The authors in those papers derived that in existence of hydrodynamical flows, the $\hat{q}$ should  be multiplied by a factor
\begin{eqnarray}
\Gamma(\vec{z})=\frac{u_f^\mu p_\mu}{p_0}=\gamma_{f}(\vec{z})\left[ 1-\vec{\beta}_{j}(\vec{z})\cdot\vec{\beta}_{f}(\vec{z})\right] \;.
\end{eqnarray}
Where $\vec{z}$ and $p^\mu=(p_0,\vec{p})$ is the coordinate and the four momentum of the high-$p_T$ parton in the lab frame, respectively. Note that $\vec{\beta}_j=\vec{p}/p_0$. The $u_f^\mu=\gamma_f(1,\vec{\beta}_f)$ is the flow four velocity.

On the other hand, from naive considerations of the jet opacity and relativistic boosts, one can get
\begin{eqnarray}
\frac{L}{\lambda}\longrightarrow\frac{L'}{\lambda'}&=&\rho'\cdot L'\cdot\sigma \nonumber \\
&=& \left\lbrace \frac{\rho}{\gamma_{f}}\right\rbrace \cdot \left\lbrace L \gamma_{f}(\vec{z}) \left[ 1-\vec{\beta}_{j}(\vec{z})\cdot\vec{\beta}_{f}(\vec{z})\right]\right\rbrace \cdot\sigma\nonumber \\
&=& \left[ 1-\vec{\beta}_{j}(\vec{z})\cdot\vec{\beta}_{f}(\vec{z})\right] \cdot \frac{L}{\lambda}\;.
\end{eqnarray}
Therefore, to systematically study the relativistic corrections to the energy loss kernel hence jet quenching observables, we compare three schemes: (1) $\Gamma=\gamma_{f}(1-\vec{\beta}_{j}\cdot\vec{\beta}_{f})$; (2) $\Gamma=1-\vec{\beta}_{j}\cdot\vec{\beta}_{f}$; (3) $\Gamma=1$.

\begin{figure*}[!t]
\bc
\includegraphics[width=0.475\textwidth]{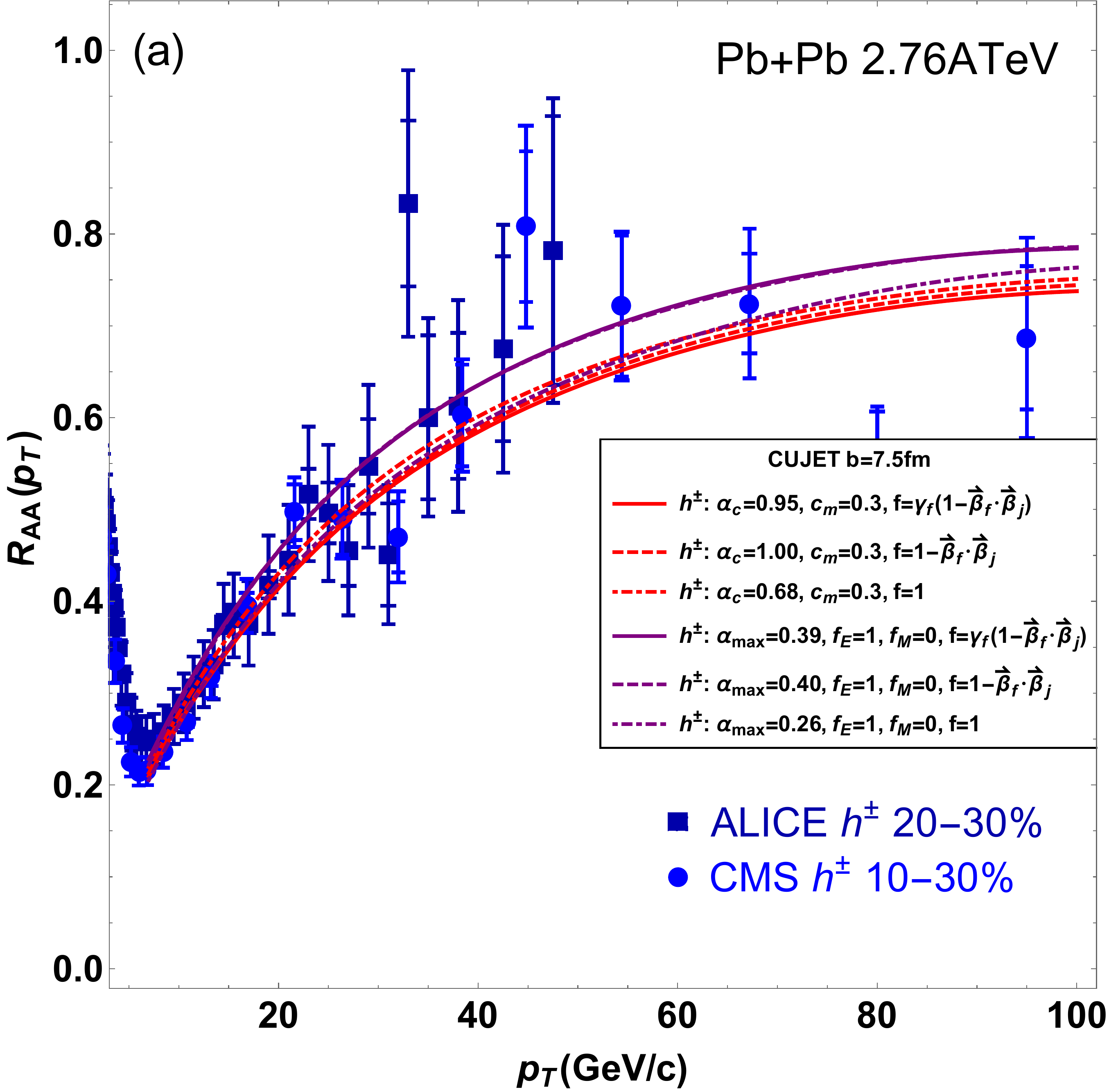}
\includegraphics[width=0.48\textwidth]{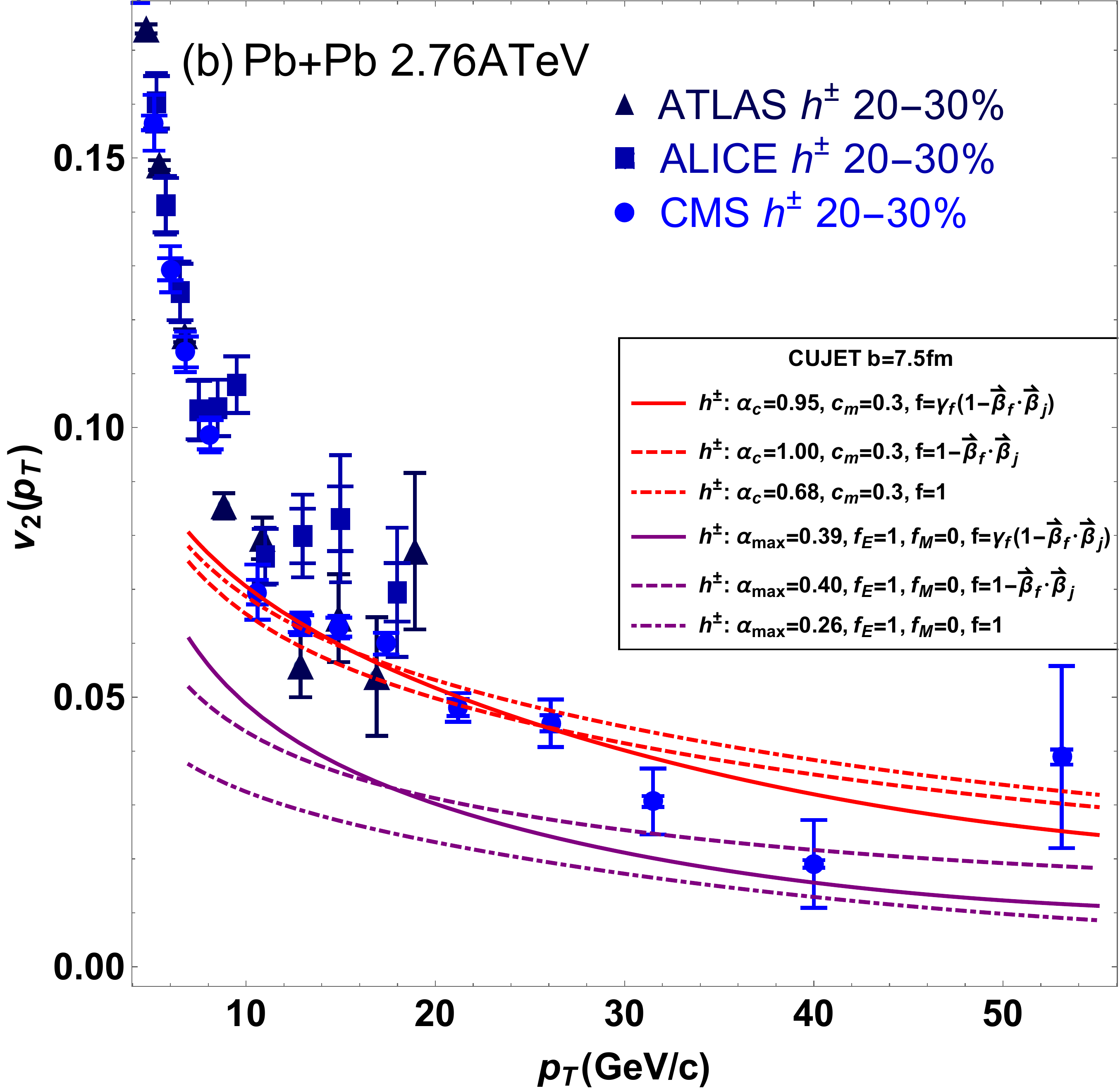}
\caption{\label{fig:rel-flow}
(Color online)
Charged particles' (a) $R_{AA}$ and (b) $v_{2}$ in LHC Pb+Pb $\sqrt{s_{NN}}$=2.76TeV semi-peripheral collisions, computed from CUJET2.0 HTL $f_E=1$, $f_M=0$ (purple) and CUJET3.0 $c_m=0.3$ (red) with relativistic flow corrections (1) $\Gamma=\gamma_{f}(1-\vec{\beta}_{j}\cdot\vec{\beta}_{f})$ (solid) \cite{Baier:2006pt,Liu:2006he}; (2) $\Gamma=1-\vec{\beta}_{j}\cdot\vec{\beta}_{f}$ (dashed); (3) $\Gamma=1$ (dotdashed) to the energy loss kernel, compared with relevant data \cite{Abelev:2012di,ATLAS:2011ah,Chatrchyan:2012xq,Abelev:2012hxa,CMS:2012aa} (blue). The parameters $\alpha_{max}$ (v2.0) and $\alpha_{c}$ (v3.0) are adjusted to fit to the $R_{AA}^{h^\pm}(p_T=12.5{\rm GeV})\approx 0.3$ reference datum. Note that both $R_{AA}$ and $v_{2}$ do not distinguish the different flow corrections at a measurable level. This can be partially understood as the number of parton-medium scatterings is fixed for a given jet path in any frame once the initial production coordinate and azimuthal propagation angle have been specified. 
		}
\ec
\end{figure*}

Fig.~\ref{fig:rel-flow}(a)(b) shows the results of high $p_T$ charged particles' $R_{AA}(p_T)$ and $v_{2}(p_T)$ at LHC Pb+Pb $\sqrt{s_{NN}}=2.76$ TeV 20-30\% centrality collisions using scheme (1)(2)(3) within CUJET3.0 and CUJET2.0, compared with available data \cite{Abelev:2012di,ATLAS:2011ah,Chatrchyan:2012xq,Abelev:2012hxa,CMS:2012aa}. While all the six combinations generate good agreements with the $R_{AA}$, only the CUJET3.0 results are consistent with the $v_{2}$. It is because the nonperturbative sQGMP in the CUJET3.0 framework strongly enhances the strength of parton-medium interaction near $T_c$, effectively increases late time jet energy loss, and boosts the underestimated $v_2$ to be in line with data \cite{Xu:2014tda}.

Surprisingly, as shown in Fig.~\ref{fig:rel-flow}(b), the relativistic flow correction scheme (1)(2)(3) do not alter the $v_{2}$ significantly in both CUJET3.0 and CUJET2.0, despite very different values of $\alpha_c$ and $\alpha_{max}$ (note that $\alpha_c$'s and $\alpha_{max}$'s are adjusted to fit reasonably the reference datum $R_{AA}^{h^\pm}(p_T=12.5{\rm GeV})\approx 0.3$). Under eikonal approximation, a jet path is fixed once the initial jet production coordinate and the azimuthal jet propagation angle are specified. Therefore, the robustness of the $R_{AA}$ and $v_2$ in response to relativistic flow corrections can be understood as along the fixed jet path, the number of parton-medium scatterings is invariant in any frame. To be more careful, one would argue that the $v_2$ shifts by $\sim0.01$ from scheme (3) $\Gamma=1$ to (1) $\Gamma=\gamma_{f}(1-\vec{\beta}_{j}\cdot\vec{\beta}_{f})$, which is not ``unchanged''. In fact, this magnitude of variation is consistent with the conclusion of \cite{Baier:2006pt}, where the authors argue that the corrections on ordinary jet quenching observables because of the hydro flow are too small to be measured hence can be neglected at current stage. 

\section{What contributes most significantly to the strongly enhanced jet opacity in sQGMP?}
\label{appx:hybrid}

In CUJET3.0, the dimensionless jet quenching parameter $\hat{q}/T^3$ is strongly enhanced in the near $T_c$ regime, and several factors may contribute to this enhancement: the enlargement of $\alpha_s\rightarrow \alpha_c$ as $T\rightarrow T_c^+$ in Eq.~\eqref{TcEnhancement}, the separation of the electric and magnetic quasi-particle density fraction according to Eq.~\eqref{chiTL}, and the magnetic screening mass regulator $c_m$ in Eq.~\eqref{f_EM}. A critical question to ask is: which factor contributes most significantly to the enhancement of the jet opacity in sQGMP? In order to answer this, a practical solution is to hybridize the CUJET2.0 running coupling scheme (denote it as $\alpha_{max}$, Eq.~\eqref{AlphaRunMax}) with the CUJET3.0 energy loss kernel (denoted it as QGMP, Eq.~\eqref{emEnergyLoss}); and hybridize the CUJET3.0 running coupling scheme (denote it as $\alpha_{c}$, Eq.~\eqref{TcEnhancement}) with the CUJET2.0 energy loss kernel (denot it as HTL, Eq.~\eqref{rcDGLV}); then compare the predictions of jet quenching observables, in particular, high $p_T$ $R_{AA}$ and $v_{2}$ from the four models: $[\alpha_{c}/\alpha_{max}] + [{\rm QGMP/HTL}]$.

\begin{figure*}[!t]
\bc
\includegraphics[width=0.475\textwidth]{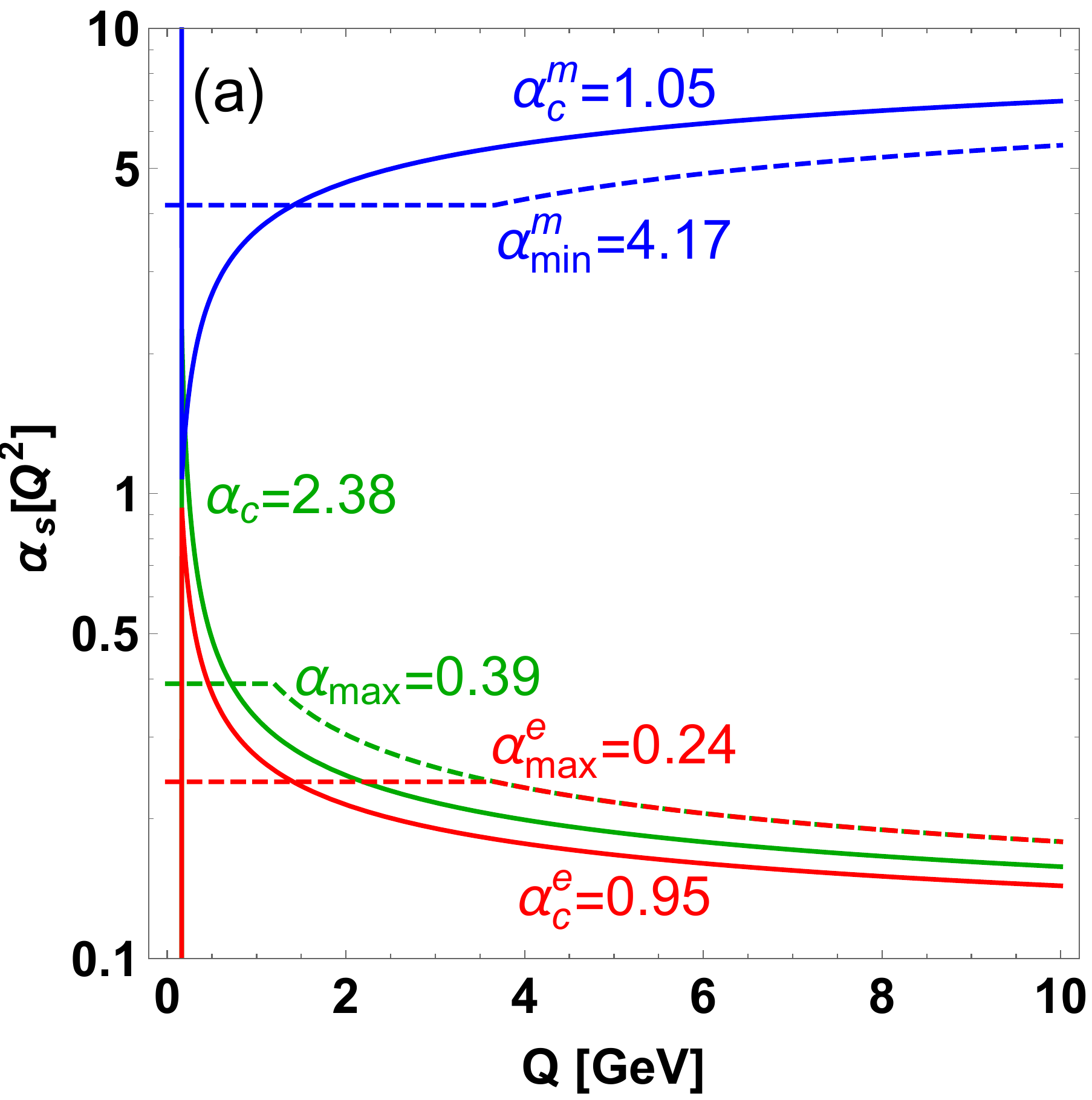}
\includegraphics[width=0.455\textwidth]{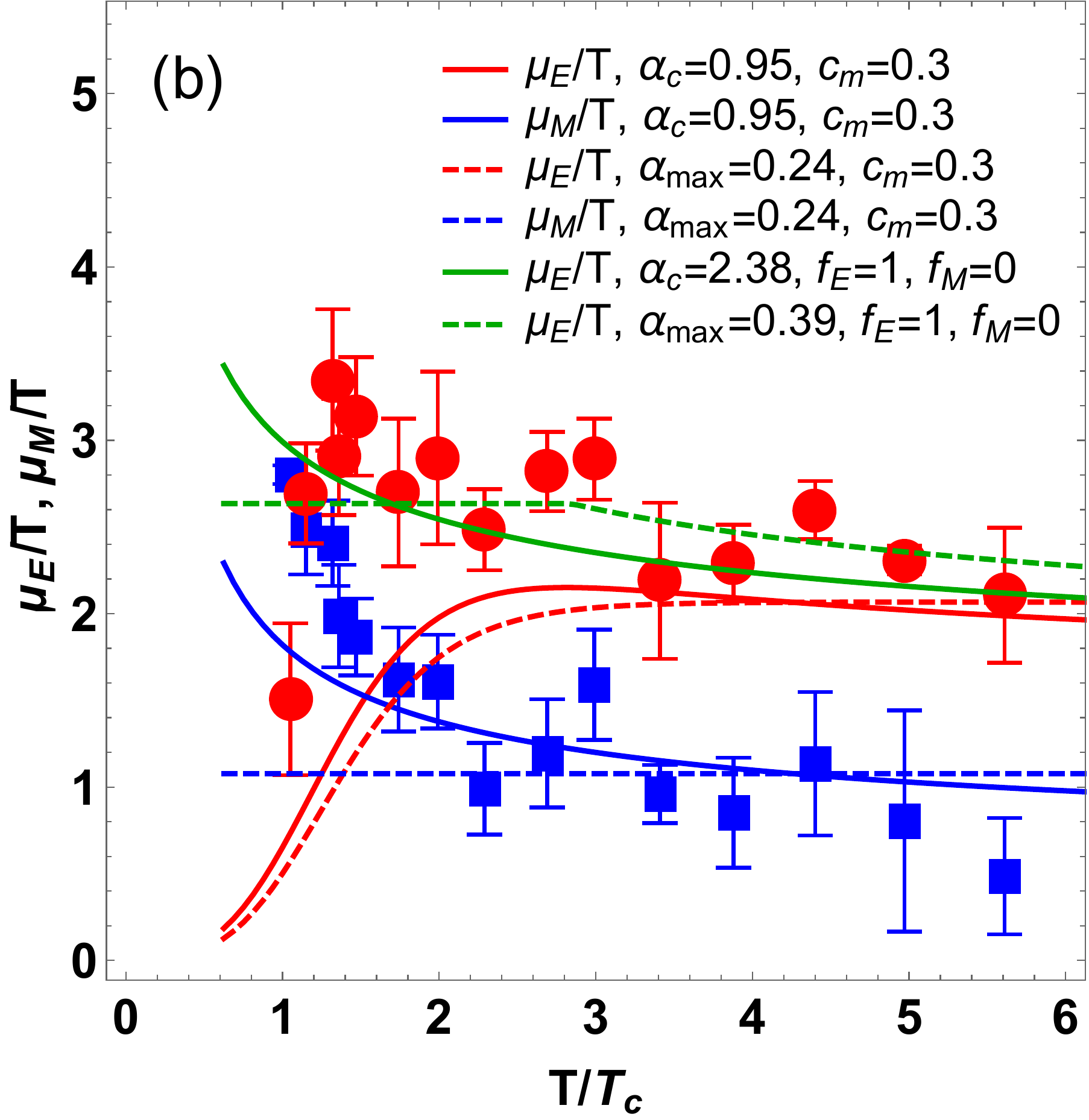}
\caption{\label{fig:hybrid}
(Color online) 
		(a) The running strong coupling $\alpha_s(Q^2)$ in four different models, $[\alpha_{c}/\alpha_{max}] + [{\rm QGMP/HTL}]$. Notice that $\alpha_{max}$ (dashed) and $\alpha_{c}$ (solid) is parametrized as in Eq.~\eqref{AlphaRunMax} and Eq.~\eqref{TcEnhancement} respectively. Note in CUJET3.0 energy loss kernel, i.e. Eq.~\eqref{emEnergyLoss} [QGMP], both chromo-electric (red) and chromo-magnetic (blue) coupling exists, while in CUJET2.0 energy loss kernel, i.e. Eq.~\eqref{rcDGLV} [HTL], only the electric coupling (green) is present. (b) The electric screening mass ($\mu_E$) and magnetic screening mass ($\mu_M$) as temperature varies in the four models, compared with lattice data (E, red; M, blue; \cite{Nakamura:2003pu}). Note in plotting the $[\alpha_{c}/\alpha_{max}] + [{\rm QGMP/HTL}]$ results, the curve styles are the same as in panel (a). In [$\alpha_{c}$], $T_c=160~{\rm MeV}$; In [$\alpha_{max}$], $\Lambda_{QCD}=200~{\rm MeV}$. Note that in [HTL] the magnetic screening mass is zero because $f_M=0$ \cite{Djordjevic:2008iz}.
		}
\ec
\end{figure*}

\begin{figure*}[!t]
\bc
\includegraphics[width=0.475\textwidth]{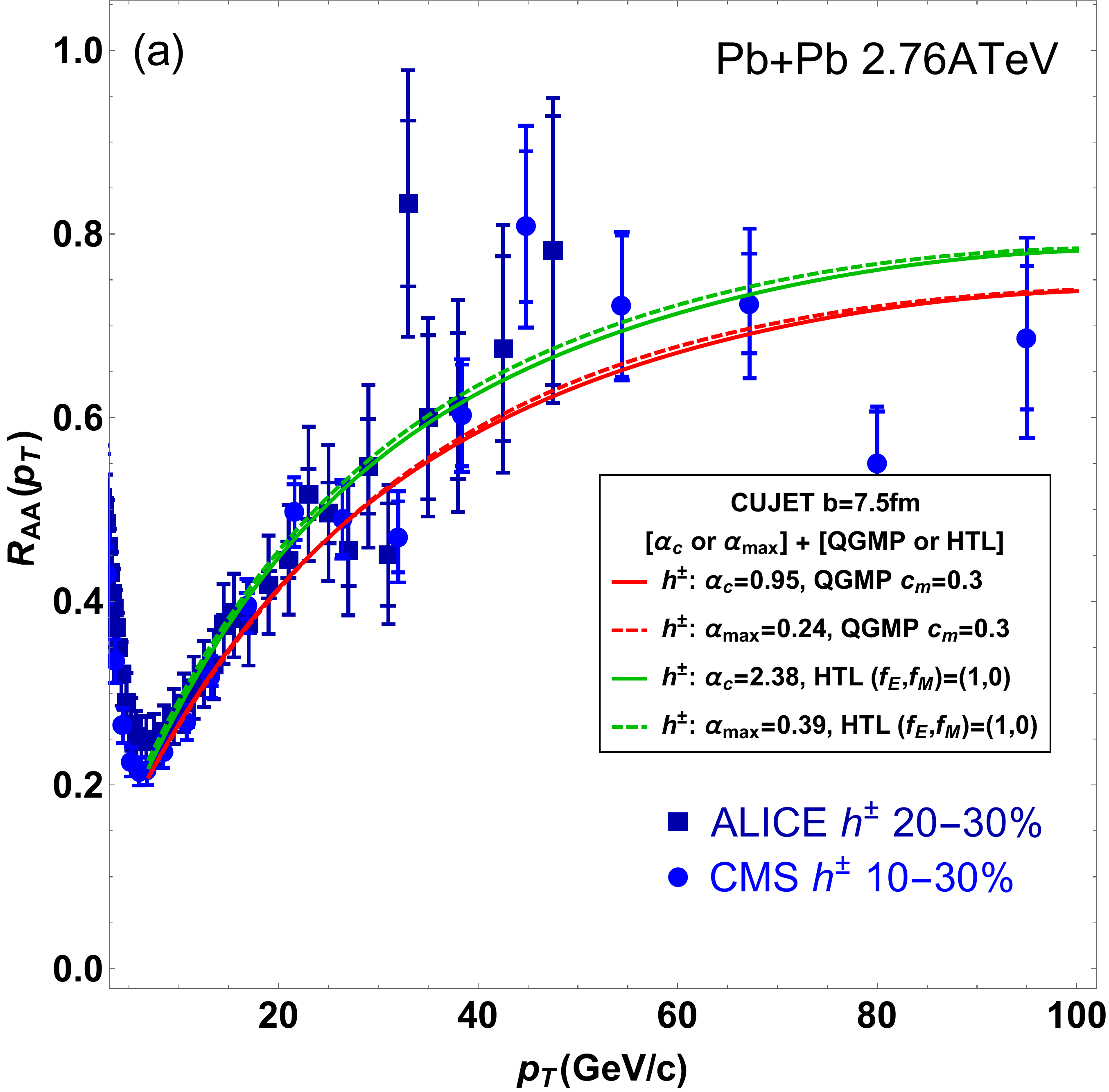}
\includegraphics[width=0.48\textwidth]{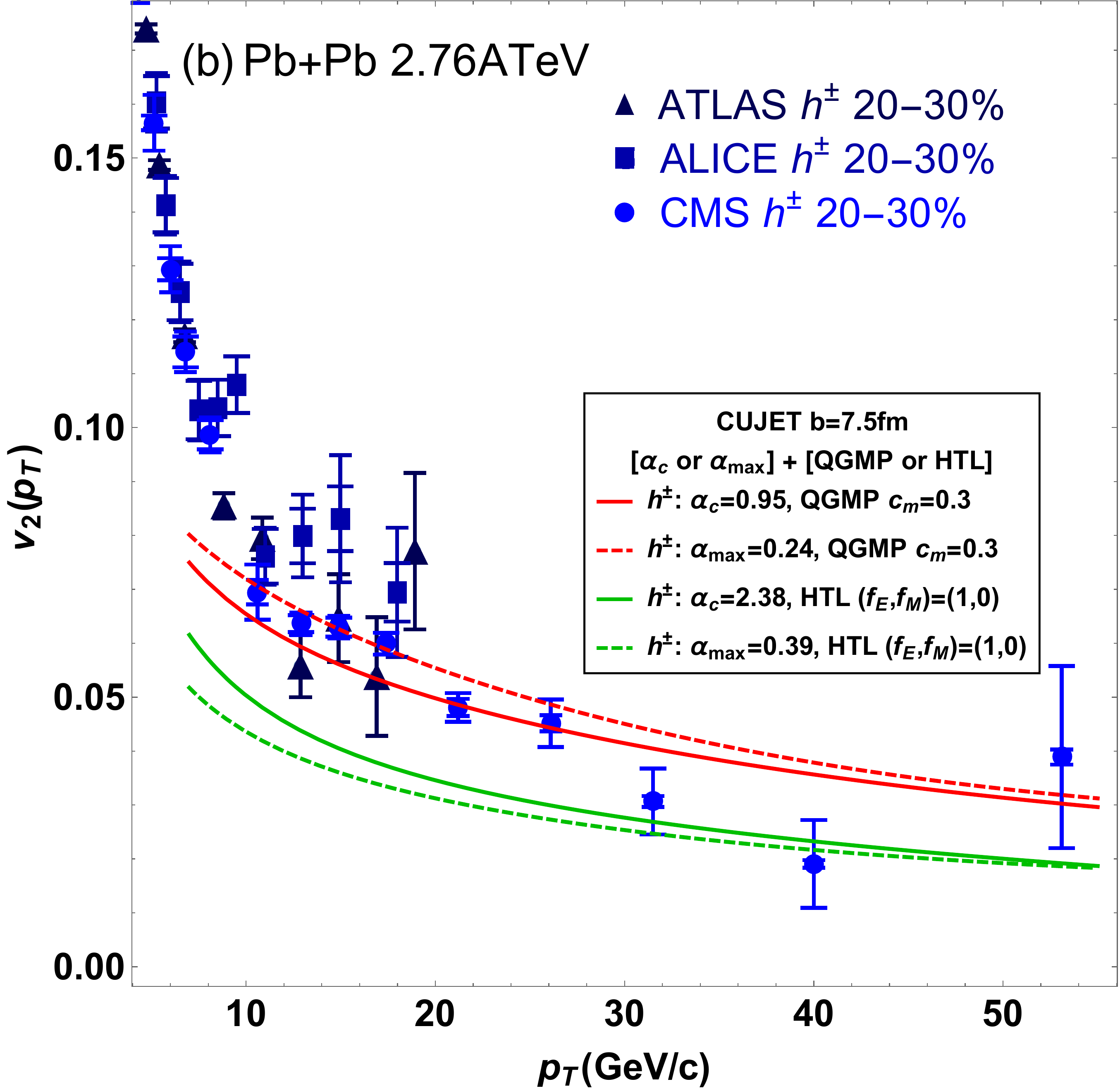}
\caption{\label{fig:hybrid-RAA-v2} 
(Color online)
		Charged particles' (a) $R_{AA}$ and (b) $v_{2}$ in LHC Pb+Pb $\sqrt{s_{NN}}$=2.76TeV semi-peripheral collisions, computed from CUJET2.0 [HTL] $f_E=1$, $f_M=0$ (green, Eq.~\eqref{rcDGLV}) and CUJET3.0 [QGMP] $c_m=0.3$ (red, Eq.~\eqref{emEnergyLoss}) with the [$\alpha_{max}$] (dashed, Eq.~\eqref{AlphaRunMax}) and [$\alpha_{c}$] (solid, Eq.~\eqref{TcEnhancement}) running coupling scheme, compared with available data \cite{Abelev:2012di,ATLAS:2011ah,Chatrchyan:2012xq,Abelev:2012hxa,CMS:2012aa} (blue). The parameter $\alpha_{max}$ and $\alpha_{c}$ are adjusted to fit to the $R_{AA}^{h^\pm}(p_T=12.5~{\rm GeV/c})\approx 0.3$ reference point. Note that while all four models can explain the high $p_T$ $R_{AA}$, only [$\alpha_{c}$]+[QGMP] and [$\alpha_{max}$]+[QGMP] can explain the high $p_T$ $v_{2}$. This suggests the emergence of chromo-magnetic monopoles as $T\rightarrow T_c^+$ contributes most significantly to the strongly enhanced $\hat{q}/T^3$ near $T_c$ and generates the simultaneous description of high $p_T$ light hadrons' $R_{AA}$ and $v_{2}$.
		}
\ec
\end{figure*}

Fig.~\ref{fig:hybrid}(a) shows the running coupling used in the four models, note that the parameters are fixed by fitting the LHC $R_{AA}^{h^\pm}(p_T=12.5{\rm GeV})\approx 0.3$ reference datum, c.f. Fig.~\ref{fig:hybrid-RAA-v2}(a). Note that the [$\alpha_{max}$]+[QGMP] model has a weaker chromo-electric coupling in the near $T_c$ regime but a stronger one at large $Q>2$ GeV than the [$\alpha_{c}$]+[QGMP] model. This model also has a smaller $\alpha_{max}^e=0.24$ than the $\alpha_{max}=0.39$ in the [$\alpha_{c}$]+[HTL] model. It is as expected since [$\alpha_{max}$]+[QGMP] has an extra monopole fraction with extremely strong couplings.

Fig.~\ref{fig:hybrid}(b) shows the electric and magnetic screening masses in the four models compared with lattice data \cite{Nakamura:2003pu}. Both [$\alpha_{max}$]+[QGMP] and [$\alpha_{c}$]+[QGMP] can describe both $\mu_{E}$ and  $\mu_{M}$ reasonably well. Both [$\alpha_{max}$]+[HTL] and [$\alpha_{c}$]+[HTL] are in agreements with $\mu_{E}$, but they have $\mu_M=0$ because $f_M=0$ in [HTL] \cite{Djordjevic:2008iz}.

Fig.~\ref{fig:hybrid-RAA-v2} show the results of high $p_T$ light hadrons' (a) $R_{AA}$ and (b) $v_{2}$ in the four models, i.e. [$\alpha_{c}$ or $\alpha_{max}$] + [QGMP or HTL], at LHC Pb+Pb $\sqrt{s_{NN}}=2.76$ TeV 20-30\% collisions, compared with available data \cite{Abelev:2012di,ATLAS:2011ah,Chatrchyan:2012xq,Abelev:2012hxa,CMS:2012aa}. Note the model parameters in the four models are fixed as:
\begin{eqnarray}
&&\rm[\alpha_{c}]+[QGMP]:\;\alpha_{c}=0.95,\;c_m=0.3\;; \\
&&\rm[\alpha_{max}]+[QGMP]:\;\alpha_{max}=0.24,\;c_m=0.3\;; \\
&&\rm[\alpha_{c}]+[HTL]:\;\alpha_{c}=2.38,\;f_E=1.0,\;f_M=0.0\;; \\
&&\rm[\alpha_{max}]+[HTL]:\;\alpha_{max}=0.39,\;f_E=1.0,\;f_M=0.0\;.
\end{eqnarray}
While all four models are compatible with the $R_{AA}$ data, only [$\alpha_{c}$ or $\alpha_{max}$] + [QGMP] can fit to the high $p_T$ charged particles' $v_2$. Noted that the boost in azimuthal elliptical harmonics is contributed mainly by the enhancement of the jet opacity near $T_c$; [$\alpha_{c}$]+[QGMP] and [$\alpha_{max}$]+[QGMP] have different running coupling schemes as well as different $\mu_M$'s in the near $T_c$ regime but share the same CUJET3.0 energy loss kernel; we can therefore conclude that the dividing of electric (E) and magnetic (M) fractions according to Eq.~\eqref{chiTL} results in the transition of the nonperturbative medium from E dominate to M dominate as temperature goes towards $T_c^+$, and contributes most significantly to the strongly enhanced parton-medium interaction near $T_c$ for the jet energy loss in semi-Quark-Gluon-Monopole-Plasmas.

\section{Path length dependence of parton energy loss in sQGMP}
\label{sec:PL}

CUJET3.0 is a jet quenching framework that hybridizes the perturbative dynamical DGLV opacity expansion theory, the TG elastic energy loss, and the nonperturbative sQGMP near $T_c$. It is consistent with high $p_T$ $R_{AA}$ and $v_2$ data at RHC and LHC simultaneously, and intrinsically builds a connection between high energy jet quenching and bulk perfect fluidity \cite{Xu:2014tda}. Beyond this phenomenological success, a crucial question to ask is how does the path length dependence of jet energy loss look like in this hybrid model. From addressing this, one can gain insights into e.g. at what temperature does the nonperturbative physics enter jet quenching, why different $\hat{q}(T)$'s lead to the same suppression factor in CUJET3.0 and CUJET2.0, whether or not the beyond leading order effects change light and heavy quark energy loss identically, etc.

To obtain useful insights about the path length dependence of jet energy loss in general scenarios, one can take a step back to look at the so-called ``abc'' model~\cite{Betz:2013caa,Betz:2014cza} where the parton energy loss is simplified as a power law of the energy E, the path length L, and the local temperature T:
\begin{eqnarray}
\frac{dE}{dL}=-\kappa E^a L^b T^{2-a+b}\;.
\label{abc}
\end{eqnarray}
Depending on underlying energy loss mechanism, the $b$ may take quite different values  (see discussions in e.g. \cite{Betz:2013caa,Betz:2014cza}: for collisional processes dominated energy loss, $b\approx 0$; at leading order (LO) in pQCD, for non-Abelian bremsstrahlung processes dominated energy loss, $b\approx 1$; as the coupling strength $\alpha_s$ becomes extremely strong (as well as $N_c\rightarrow \infty$) and AdS/CFT correspondence is applicable, holographic jet quenching generally has $b\approx 2$. In the following subsections, after we have interpolated $(\Delta E/E)(L)$ at various temperatures, we will further extract out a ``b'' factor from $b=d\log(\Delta E/E)/d\log(L/L_0)-1$. Different from the abc model with ``global'' power law dependence for  the jet energy loss, our extracted ``b'' factor would be a sort of ``local index'' but it nevertheless is an informative indicator that can help achieve deeper understandings about the energy loss dynamics  encoded in the computed $(\Delta E/E)(L)$ from the CUJET3.0.

\subsection{Light Quark}
\label{sec:PL-light}

In the radiative sector, there have been next-to-leading order (NLO) pQCD calculations for energy loss assuming massless projectile partons \cite{Liou:2013qya,Kang:2013raa,Iancu:2014kga,Blaizot:2014bha}, and they all suggest a double logarithmic path length dependence of the jet quenching parameter, i.e.
\begin{eqnarray}
\hat{q}\propto \log \tilde{L} (1+\log^2 \tilde{L})\;.
\end{eqnarray}
Where $\tilde{L}\equiv L/L_0$ and $L_0$ is a proper ultraviolet cutoff. Generally speaking, the differentiate jet energy loss has $dE/dL\propto \hat{q}L$. Let us assume $\log \tilde{L}$ varies much slower than $ \tilde{L}$, after straightforward integrations and simplifications, one arrives at
\begin{eqnarray}
\log(\Delta E/E) \propto 2 \log \tilde{L} + \log \log \tilde{L} + \log (1 +\log^2 \tilde{L} ) + {\rm const} \;.
\label{NLOergloss}
\end{eqnarray}
This form is of cause not general enough because of the $\log \tilde{L}$ approximation and it is derived at NLO in pQCD. Nevertheless, motivated by Eq.~\eqref{NLOergloss}, we will use
\begin{eqnarray}
\log\frac{\Delta E}{E}=A+B\log\frac{L}{L_0}+\log(1+C\log\frac{L}{L_0})+\log(1+D\log^2\frac{L}{L_0}) \;,
\label{dEdLvsL}
\end{eqnarray}
to interpolate the path length dependence of radiative jet energy loss in sQGMP within the CUJET3.0 framework. On the other hand, combining the above with Eq.~\eqref{abc}, one can easily get
\begin{eqnarray}
b&\equiv&\frac{d\log(\Delta E/E)}{d\log(L/L_0)} -1 \label{powerb}\\
&\stackrel{\mathclap{\normalfont\mbox{\tiny rad}}}{=}& B-1+\frac{C}{1+C\log(L/L_0)}+\frac{D}{1+D\log^2(L/L_0)} \label{powerb1}\\
&\rightarrow& B - 1\;(L\rightarrow \infty) \label{powerb2}\;.
\end{eqnarray}

\begin{figure*}[!t]
\bc
\hspace{0.5pt}
\includegraphics[width=0.98\textwidth]{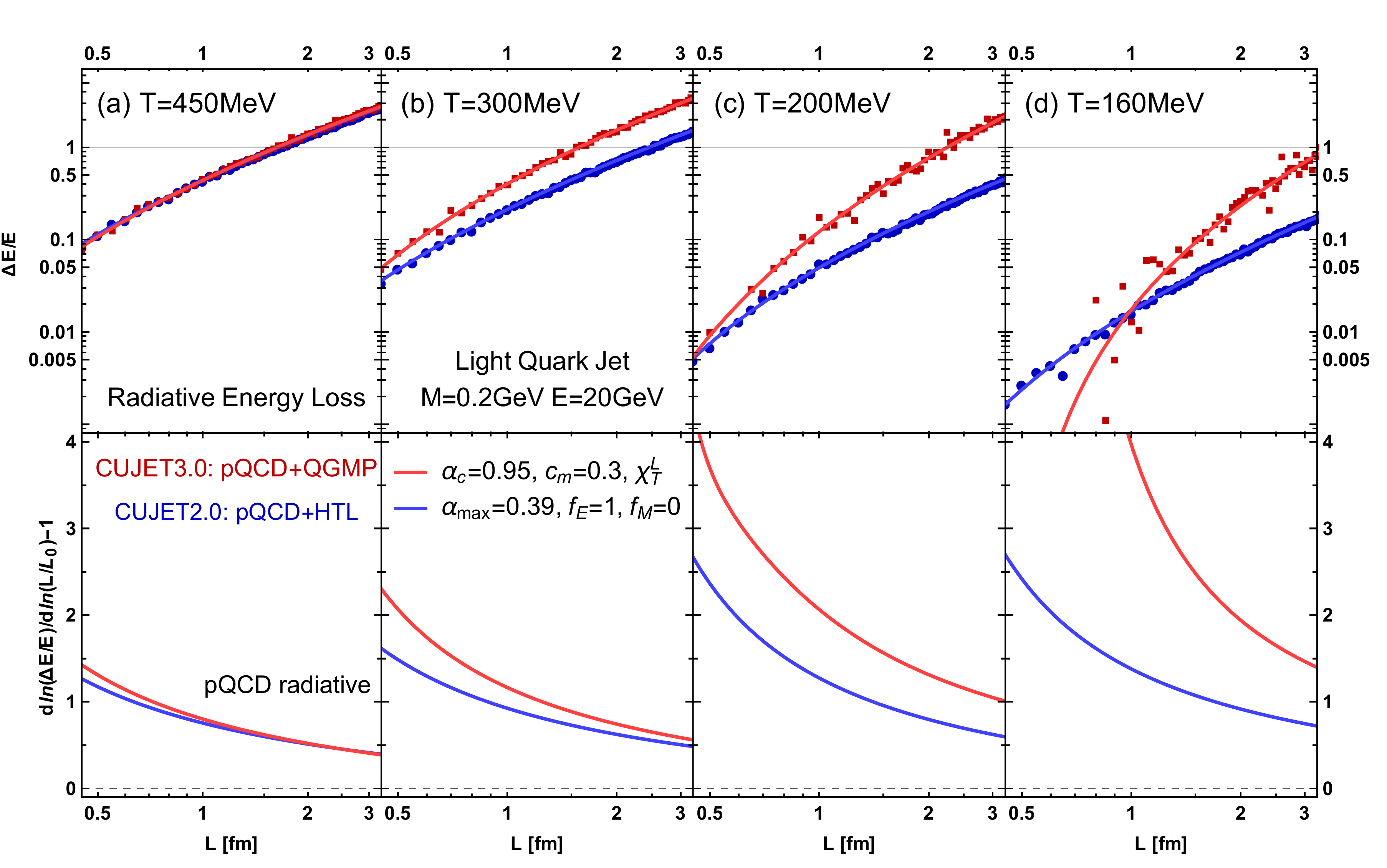}
\caption{\label{fig:PL-light} 
(Color online)
(Upper panels)
The radiative energy loss ratio $\Delta E/E$ of a light quark jet ($M=200$ MeV) with initial energy $E=20$ GeV propagating through a brick plasma with various thicknesses $L$ at temperature $T=$ (a) 450, (b) 300, (c) 200, (d) 160 MeV, in the CUJET3.0 $\alpha_c=0.95$, $c_m=0.3$, $\chi_T^L$ model (red) and in the CUJET2.0 $\alpha_{max}=0.39$, $f_E=1$, $f_E=0$ model (blue). When interpolating the $\Delta E/E(L)$'s, the NLO pQCD motivated Eq.~\eqref{dEdLvsL} is used.
Notice that $\Delta E /E$ exceeds 1 at some large $L$ for a certain $T$, this is due to the fact that in the computation $E$ is not dynamically updated in each unit cell.
Since with a smooth viscous hydro evolution background, the hard parton will stay at a certain temperature for limited time, the relevant energy loss informations are mostly at short $L$.
As temperature decreases, CUJET3.0's $\Delta E/E(L)$ and the stopping distance $L_1$ (defined in Eq.~\eqref{L_1}) respectively gets steeper and larger compared with CUJET2.0's. At low and intermediate T, $(\Delta E/E)_{v3.0}<(\Delta E/E)_{v2.0}$ in the small $L$ regime.
(Lower panels)
The path length $L$ dependence of the power $b$ in Eq.~\eqref{powerb} at different temperatures. Note that $b$ = 0, 1, 2 is approximately the elastic, pQCD and AdS limit respectively. At high temperature $T\sim400$ MeV, the $b(L)$ of CUJET3.0 and CUJET2.0 converge to around the pQCD limit. As temperature cools down, when $T\sim300$ MeV, CUJET3.0's $b(L)$ start becoming larger than CUJET2.0's. This signals the transition from E to M dominant as well as from weak to strong coupling for the bulk. In the near $T_c$ regime, the $b(L)$ in the CUJET3.0 framework is higher than LO pQCD, and is close to the AdS limit. This implies the model ingredients in CUJET3.0 do effectively bring in nonperturbative dynamics into the original pQCD/DGLV energy loss kernel.
}
\ec
\end{figure*}

The upper panels of Fig.~\ref{fig:PL-light} show the path length dependence of the radiative energy loss $\Delta E/E$ of a high-$p_T$ light quark (mass $M=200$ MeV) with initial energy $E=20$ GeV traversing a brick plasma (fixed density) at varied temperatures in CUJET3.0 and in CUJET2.0. The (a)(b)(c)(d) corresponds to jet quenching in the brick medium with temperature $T=450$, $300$, $200$, $160$ MeV respectively. In the computation, Eq.~\eqref{emEnergyLoss} and Eq.~\eqref{rcDGLV} is used for CUJET3.0 and CUJET2.0 respectively, and the Monte-Carlo integration is iterated 1,000,000 times to enforce convergence. The brick size $L$ is increased from 0.45 fm to 3.05 fm with 0.5 fm intervals. We fit the $\Delta E/E$ vs $L$ in both CUJET3.0 and CUJET2.0 with Eq.~\eqref{dEdLvsL}.
Note the phenomenon that $\Delta E/E$ exceeds 1 for some large size bricks is purely technical, since in the computation $E$ is not dynamically updated in Eq.~\eqref{emEnergyLoss} and \eqref{rcDGLV} as hard partons propagate through and lose energies. Nevertheless, the $\Delta E/E(L)$ at small $L$'s, the slope of $\Delta E/E(L)$, the thickness where $\Delta E/E=1$, and the relative information about parton energy loss at the four different temperatures are all meaningful.

For the CUJET2.0 model with pQCD+HTL, as temperature decreases, $d(\Delta E/E)/dL$ is altered significantly. Define the stopping distance $L_1$ as
\begin{eqnarray}
L_1: \Delta E/E|_{L=L_1}=1\;.
\label{L_1}
\end{eqnarray}
It monotonically increases in this picture. This is as expected, since in the CUJET2.0, bricks with lower temperatures have less opacities, and it takes a longer path for a jet to lose all the energy. However, for the $L_1$ in CUJET3.0, though it monotonically increases, its increasing rate is much less than CUJET2.0's. More importantly, the CUJET3.0's $d(\Delta E/E)/dL$ significantly grows as the temperature decreases below $T=300$ MeV. All these pointed to the fact that, as the temperature approaches $T_c$, chromo-magnetic monopoles gradually dominate the medium, since $\alpha_M \gg \alpha_E$, the jet-medium interaction may weaken at a reduced rate or it may be enhanced (telling from Fig.~\ref{fig:FL-qhat-etas}(a), this is the correct picture) despite the decrease of the quasi-particle number density.

To this point, there is a critical question that one should address regarding the jet energy loss in sQGMP: if one compares the CUJET3.0 and CUJET2.0's $\hat{q}/T^3$ in Fig.~\ref{fig:FL-qhat-etas}(a)(b), the former is always above the latter in the temperature range of $T<450$ MeV, then why both of them can reasonably describe the high $p_T$ light hadron's $R_{AA}$ at RHIC and LHC? The upper panels of Fig.~\ref{fig:PL-light} give one the answer: at $T=400$ MeV, CUJET3.0 and CUJET2.0's $\Delta E/E(L)$ almost overlap; as the temperature cools down, because of the transition of the nonperturbative medium from EQPs dominate to MQPs dominate,  the CUJET3.0's $\Delta E/E(L)$ becomes steeper and steeper than the CUJET2.0's, while the former's $L_1$ becomes less and less than the latter's; these effects lead to $(\Delta E/E)_{v3.0}<(\Delta E/E)_{v2.0}$ at small $L<L_{eq}$, where $L_{eq}$ is defined as
\begin{eqnarray}
L_{eq}: (\Delta E/E)_{v3.0}|_{L=L_{eq}}=(\Delta E/E)_{v2.0}|_{L=L_{eq}}\;,
\label{L_eq}
\end{eqnarray}
and this $L_{eq}$ keeps enlarging as T decreases. For hadron suppressions in a hydrodynamically evolving smooth medium in A+A collisions, along a given jet path (let the initial production point be at the origin), if one sequentially divides it into sections with average temperature of 450, 300, 200, 160 MeV and mark the traveling time in each section as $l_{450}$, $l_{300}$, $l_{200}$, $l_{160}$, in a general estimation, the ordering is $l_{450} > l_{300} > l_{200} > l_{160}$, and $l_{160}\ll 1$ fm while $l_{200}\lesssim 0.5$ fm. Define
\begin{eqnarray}
\Delta_{E}(\bar{T}=\frac{T_i+T_{i+1}}{2})\equiv \frac{1}{L_>-L_<} \int_{T(L_<)=T_i}^{T(L_>)=T_{i+1}} dL \left[ (\Delta E)_{v3.0} - (\Delta E)_{v2.0}\right] \;.
\end{eqnarray}
Although $\Delta_{E}(T=450, 300\,{\rm MeV})$ are most likely positive, $\Delta_{E}(T=200, 160\,{\rm MeV})$ can be nontrivially negative and thus compensate the formers. This cancellation results in the overall $\sum_{T_j}\Delta_{E}\approx 0$, meaning similar averaged $R_{AA}$ predictions for CUJET3.0 and CUJET2.0.

The lower panels of Fig.~\ref{fig:PL-light} show the extracted power b ($b\equiv d\log(\Delta E/E)/d\log(L/L_0)-1$, c.f. Eq.~\eqref{powerb}) versus the brick thickness $L$. A first observation is that, at high temperature, the CUJET3.0's $b(L)$ is almost identical to CUJET2.0's; but as T goes down, starting from $T\sim300$ MeV, the former gets larger than the latter. This is a clear signal showing that the chromo-magnetic monopoles begin emerging and bringing up strong coupling effects from $T\sim300\,{\rm MeV}\approx1.8\,T_c$. Let $L_2$ be the path length that satisfies
\begin{eqnarray}
b(L=L_2)=1\;.
\end{eqnarray}
In both CUJET2.0 and CUJET3.0, $L_2$ increases as the temperature decreases, this is understood as the opacity $L/\lambda=\rho\sigma L\sim T^3 L$, if $T$ drops, a larger $L$ is required to keep the same opacity level and hence the similar antenna structure. Meanwhile, the CUJET2.0's $db/d\tilde{L}$ does not undergo significant shifts as temperature varies, but this is not the case for CUJET3.0. Define
\begin{eqnarray}
b_{1.5}\equiv b\,(L=1.5\,{\rm fm})\;.
\end{eqnarray}
In CUJET2.0, $b_{1.5}$ rises from $\sim 0.6$ at 450 MeV to $\sim 1.2$ at 160 MeV, which is roughly consistent with the LO pQCD expectation of $b=1$. In CUJET3.0, the $b_{1.5}$ rises from $\sim 0.6$ at 450 MeV to $\sim 2.5$ at 160 MeV, suggests that the sQGMP introduces some nonperturbative effects into the DGLV opacity series and effectively causes the resummation of higher orders in the full QCD amplitude.

\subsection{Heavy Flavor}
\label{sec:PL-heavy}

\begin{figure*}[!t]
\bc
\hspace{0.5pt}
\includegraphics[width=0.98\textwidth]{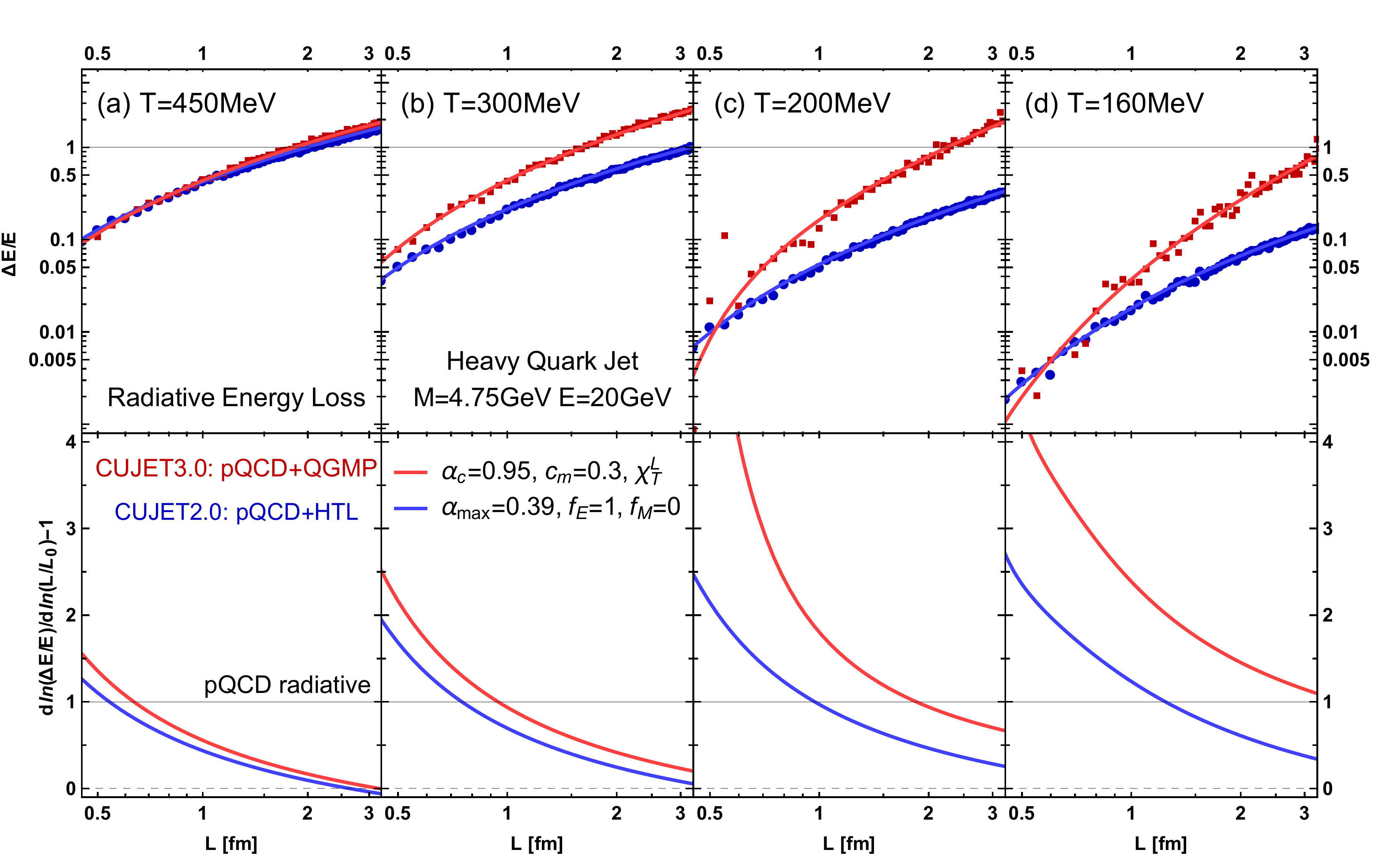}
\caption{\label{fig:PL-heavy} 
(Color online)
		The radiative energy loss ratio $\Delta E/E$ and the power $b$ (c.f. Eq.~\eqref{powerb}) of a high-$p_T$ heavy quark ($M=4.75$ GeV) with initial energy $E=20$ GeV traversing a brick plasma with thickness L at various temperatures in CUJET3.0 and in CUJET2.0. All marks and computational details are the same as in Fig.~\ref{fig:PL-light}. Note that the heavy quark's $d(\Delta E/E)/dL$ and $b(L)$ are smaller than the light quark's as expected from the dead cone suppression. At high $T\sim$ 450 MeV, both CUJET3.0 and CUJET2.0 converge at around the linear elastic energy loss limit. As $T$ drops towards $T_c$, beginning from $T\sim300$ MeV, the CUJET3.0's $b(L)$ starts deviating from CUJET2.0's because of the emergence of chromo-magnetic monopoles; the former's $\Delta E/E(L)$ gets steeper than the latter's, while $L_1$ (c.f. Eq.~\eqref{L_1}) gets shorter. All these alternations for the heavy quark suppression are similar to those for the light quark, and the magnitude of the $b(L)$ deviation for the two different flavors are almost identical. This suggests that the nonperturbative effects in the near-$T_c$ sQGMP modify the energy loss kernel of light and heavy quarks in a very similar way.
		}
\ec
\end{figure*}

Let us now turn   to the path length dependence of the heavy quark energy loss in the sQGMP.
The upper panels of Fig.~\ref{fig:PL-heavy} show the $\Delta E/E(L)$ of a heavy quark jet (mass $M=4.75$ GeV) with initial energy $E=20$ GeV transversing a brick plasma at $T=$ 450, 300, 200, 160 MeV, in CUJET3.0 and in CUJET2.0. Except for the jet mass $M$, all technical details in this computation are the same as the in the one for the light quark energy loss. Compared with Fig.~\ref{fig:PL-light}, one notice that slope of $\Delta E/E(L)$ in both CUJET3.0 and CUJET2.0 are more gentle for the heavy quark than for the light quark, and the $L_1$ (c.f. Eq.~\eqref{L_1}) grows faster when cooling down. This clearly indicates the dead cone effects suppress the induced radiation regardless of whether or not the sQGMP is present. As the temperature gets lower, for heavy quarks, CUJET3.0's $\Delta E/E(L)$ and $L_1$ also becomes steeper and smaller than CUJET2.0's. This phenomenon has the same physical origin as for light quarks discussed in section \ref{sec:PL-light}, i.e. a transition from EQPs to MQPs dominate. Interestingly, in the near $T_c$ regime, the $L_{eq}$ (c.f. Eq.~\eqref{L_eq}) for heavy quarks is smaller than for light quarks. Based on the discussions in section \ref{sec:PL-light}, this will lead to the high $p_T$ leading B meson $R_{AA}$ predictions from CUJET3.0 being slightly lower than CUJET2.0. In fact, this is case from the comparison of the $R^B_{AA}$ in \cite{Xu:2014tda} and in \cite{Xu:2014ica}.

The lower panels of Fig.~\ref{fig:PL-heavy} show the extracted power $b$ (c.f. Eq.~\eqref{powerb}) versus the medium thickness $L$. Notice that for some temperatures at large $L$, the $b$ becomes less than 1, nevertheless this can be neglected since in these regimes the $\Delta E/E$ has became larger than 1, which is unphysical. In CUJET2.0, the $b_{1.5}$ (c.f. section \ref{sec:PL-light}) rises from $\sim 0.3$ at 450 MeV to $\sim 0.6$ at 160 MeV, this is weaker than the LO pQCD radiative energy loss $b=1$, and approaches the elastic limit $b=0$. Compared with the energy loss for light quarks, the dead cone suppression is significant for heavy quarks. In CUJET3.0, the $b_{1.5}$ rises from $\sim 0.4$ at 450 MeV to $\sim 1.6$ at 160 MeV, this suggests the strong coupling effects hence high order resummations also enter the energy loss kernel for the heavy quark.

At high $T\sim450$ MeV, the CUJET3.0's and CUJET2.0's $b(L)$ overlap, as $T$ drops, beginning from $T\sim300$ MeV, the former starts to deviate from the latter, suggesting the commencement of monopoles taking control of the medium transport properties. This initiating temperature $T_{initial}\sim300$ MeV for the heavy quark coincides exactly with the $T_{initial}$ for the light quark; meanwhile, the magnitude of the deviation in $b(L)$ between CUJET3.0 and CUJET2.0 for the two different flavors resemble each other; these observations imply that the nonperturbative effects brought up by the sQGMP near $T_c$ influences the light quark jet quenching and the heavy quark jet quenching in approximately the same manner within CUJET3.0.

\bibliographystyle{JHEP}
\bibliography{CUJET3.0}

\end{document}